\documentclass{article}

\usepackage{arxiv} 
\usepackage[utf8]{inputenc}
\usepackage[T1]{fontenc}

\usepackage[hidelinks]{hyperref}

\usepackage{graphicx}
\usepackage{amsmath, amsthm, amssymb, mathtools}
\usepackage{fullpage}
\usepackage{curves}
\usepackage{url}
\usepackage{paralist}
\usepackage{xspace}
\usepackage{pythonhighlight}

\usepackage{tikz}
\usetikzlibrary{matrix,positioning,patterns,arrows,shapes,calc}
\usepackage{listings}
\usepackage{array,makecell}
\usepackage{adjustbox}
\usepackage{multicol}
\usepackage{pgfplots,pgfplotstable}\pgfplotsset{compat=1.18}
\usepackage{tcolorbox}
\usepackage{caption,subcaption}
\usepackage{pifont}
\usepackage{listings, textcomp, color, verbatim}
\lstdefinelanguage{cpplang}{
	language=C++,                      
	basicstyle=\ttfamily\footnotesize,              
	keywordstyle=\color{purple}\bf,        
	commentstyle=\color{green!35!black}\slshape, 
	xleftmargin=1pt, xrightmargin=1pt, 
	frame=single,                        
}
\lstnewenvironment{cpp}{\lstset{language=cpplang}}{}

\definecolor{nvgreen}{HTML}{76b900}

\theoremstyle{definition}
\newtheorem{definition}{Definition}[section]

\theoremstyle{remark}

\newcommand{\CuTe}{\textsc{CuTe}\xspace}

\newcommand{\Q}{\mathbb{Q}}
\newcommand{\Z}{\mathbb{Z}}

\newcommand{\N}{\mathbb{N}}

\newcommand{\D}{\mathcal{D}}
\renewcommand{\i}{\imath}

\newcommand{\abs}[1]{\left| {#1} \right|}

\newcommand{\ceil}[1]{\left\lceil {#1} \right\rceil}
\newcommand{\floor}[1]{\left\lfloor {#1} \right\rfloor}
\renewcommand{\v}{\mathbf}
\renewcommand{\bar}{\overline}
\renewcommand{\hat}{\widehat}
\renewcommand{\tilde}{\widetilde}

\renewcommand{\epsilon}{\varepsilon}

\usepackage{listings, textcomp, color, verbatim}
\definecolor{deepgreen}{rgb}{0,0.4,0}

\title{\CuTe
	   Layout Representation and Algebra}
\author{
    Cris Cecka\\
    NVIDIA Research\\
    \url{ccecka@nvidia.com}
}
\date{}  

\begin{document}

\maketitle

\begin{abstract}
Modern architectures for high-performance computing and deep learning increasingly incorporate specialized tensor instructions, including tensor cores for matrix multiplication and hardware-optimized copy operations for multi-dimensional data. These instructions prescribe fixed, often complex data layouts that must be correctly propagated through the entire execution pipeline to ensure both correctness and optimal performance. We present \CuTe, a novel mathematical specification for representing and manipulating tensors. \CuTe introduces two key innovations: (1) a hierarchical layout representation that directly extends traditional flat-shape and flat-stride tensor representations, enabling the representation of complex mappings required by modern hardware instructions, and (2) a rich algebra of layout operations -- including concatenation, coalescence, composition, complementation, division, tiling, and inversion -- that enables sophisticated layout manipulation, derivation, verification, and static analysis. \CuTe layouts provide a framework for managing both data layouts and thread arrangements in GPU kernels, while the layout algebra enables powerful compile-time reasoning about layout properties and the expression of generic tensor transformations.

In this work, we demonstrate that \CuTe's abstractions significantly aid software development compared to traditional approaches, promote compile-time verification of architecturally prescribed layouts, facilitate the implementation of algorithmic primitives that generalize to a wide range of applications, and enable the concise expression of tiling and partitioning patterns required by modern specialized tensor instructions.

\CuTe has been successfully deployed in production systems, forming the foundation of NVIDIA's CUTLASS library and a number of related efforts including CuTe DSL.
We provide both mathematical foundations and a reference implementation in Python (PyCuTe)~\cite{NVlabs:PyCuTe}, establishing \CuTe as a principled approach to layout-aware programming for the tensor-centric computing era.
\end{abstract}

\tableofcontents

\section*{Acknowledgments}

The author thanks Vijay Thakkar for being an early adopter of \CuTe, for his pivotal role in its popularization and integration into CUTLASS, and for substantial contributions to its foundational design and deployment. He also thanks Andrew Kerr, Pradeep Ramani, and Manish Gupta for their insights into using \CuTe within CUTLASS and for driving its early adoption in the development of high-performance kernels. He thanks Muhammad Osama and Duane Merrill for early experimental work with \CuTe prototypes and low-level performance optimizations in initial \CuTe examples. Finally, the author thanks Michael Garland, Bastian Hagedorn, Jay Shah, and Amanda Liu for insightful feedback and for repeatedly challenging the \CuTe specifications.

\section{Introduction and Motivation}

Modern GPUs are increasingly optimized for tensor-centric computations, driven by the demands of deep learning and scientific computing. NVIDIA's Volta architecture~\cite{NVIDIA:Volta} introduced Tensor Cores, enabling efficient small-matrix multiplications directly in hardware. This capability expanded in Turing~\cite{NVIDIA:Turing} and Ampere~\cite{NVIDIA:Ampere} with specialized instructions for structured matrix movement within the GPU memory hierarchy. The Hopper~\cite{NVIDIA:Hopper} and Blackwell~\cite{NVIDIA:Blackwell} architectures further advance this paradigm, introducing copy instructions for efficiently transferring rank-5 tensors between global and shared memory and further expanding tensor core capabilities. Fully harnessing these tensor-oriented hardware features is critical for peak GPU performance, motivating high-performance programming models that can represent and manipulate these tensors efficiently.

The existing and emerging hardware depends critically on how multidimensional data is stored and accessed in multiple hierarchical memory spaces and in multiple hierarchical levels of parallelism. Data storage layouts have always affected performance by dictating how and when memory accesses occur, but as hardware instructions become larger and prescribe fixed layouts for inputs and outputs the effect on correctness also becomes vital. These layouts must be propagated through the entire execution pipeline to ensure correct invocation of hardware instructions and ensure optimized memory access patterns.

In this work, we present the foundational concepts of \CuTe, a specification for \underline{CU}DA \underline{Te}nsors, or \underline{C}ompute \underline{U}nified \underline{Te}nsors, designed to provide building blocks for writing peak-performance linear algebra libraries. At its core, \CuTe introduces two key innovations:
\begin{compactitem}
\item A {\em novel representation for tensor layouts}: \CuTe shapes, layouts, and tensors are inherently hierarchical, constructed from smaller nested instances. This hierarchy provides the means to represent complex mappings required by modern tensor instructions, but remains a strict extension of existing flat-shape and flat-stride representations found in libraries like {\tt BLAS}, {\tt torch.tensor}, {\tt numpy.ndarray}, and MATLAB.
\item A {\em novel algebra of operations defined over layouts}: \CuTe layouts support a rich set of operations, including concatenation, coalescence, composition, complementation, division, tiling, and inversion, which all result in new \CuTe layouts. These operations enable sophisticated partitioning, manipulation, verification, and derivation of tensor layouts demanded by modern tensor instructions.
\end{compactitem}

\CuTe's layout representation provides an intuitive framework for managing threads and data in writing generic algorithms. \CuTe's layout algebra provides an expressive approach for manipulating layouts and generating new layouts in the development of high-performance linear algebra kernels. These approaches enable:
\begin{compactitem}
\item Support for complex layouts and partitioning: \CuTe facilitates the representation of intricate layouts required by application-specific data patterns and complex partitioning patterns required by specialized tensor instructions.
\item Separation of concerns: Data layouts are declared independently of algorithmic logic, promoting clarity and modularity.
\item Static analysis and optimization: Sophisticated algebraic techniques empower the inspection, reordering, and partitioning of tensor arguments according to architectural constraints.
\end{compactitem}

\subsection{Related Work}

\CuTe is motivated by the need to support development of efficient tensor contractions, which are at the core of many scientific computing and machine learning applications. Conventional approaches for computing general tensor contractions rely on matricization, which involves logically or explicitly restructuring tensor data to perform computations using a sequence of calls to a Basic Linear Algebra Subprograms (BLAS) library.


BLAS provides efficient and portable implementations of core linear algebra operations, with highly optimized versions available for a wide range of architectures~\cite{blackford2002updated}. Among BLAS primitives, the GEneral Matrix Multiply (GEMM) routine is easily the most optimized and widely used operation in all of scientific computing and machine learning.

The BLAS-like Library Instantiation Software (BLIS) framework~\cite{VanZee:BLIS} extends GEMM by supporting non-unit strides in both row and column modes simultaneously, which addresses some challenges in handling irregular memory layouts without resorting to explicit memory copies. The strided-batched GEMM extension to BLAS further generalizes the primitive and allows its application to even more tensor contractions~\cite{Shi:BLASContractions}. Abstracting matrix layouts even further, a key insight motivating \CuTe is the use of multi-indices in tensor notation to enable the transformation of arbitrary tensor contractions into a canonical batched-GEMM primitive.

Many existing libraries rely on tensors and nearly all of them are based on the flat-shape and flat-stride representation. In Python, these include {\tt numpy.ndarray} and {\tt torch.tensor}

\noindent
\begin{minipage}[t]{0.48\textwidth}
\begin{python}
>>> import numpy
>>> a = numpy.ndarray([3,7,5])
>>> a.dtype
dtype('float64')
>>> a.shape
(3, 7, 5)
>>> a.strides
(280, 40, 8)
\end{python}
\end{minipage}
\hfill
\begin{minipage}[t]{0.48\textwidth}
\begin{python}
>>> import torch
>>> a = torch.empty(3,7,5)
>>> a.dtype
torch.float32
>>> a.shape
torch.Size([3, 7, 5])
>>> a.stride()
(35, 5, 1)
\end{python}
\end{minipage}

\noindent and in C++, {\tt std::mdspan}

\noindent
\begin{cpp}
std::mdspan a = std::mdspan(data, 3, 7, 5);
a.extent(0);  // 3
a.extent(1);  // 7
a.extent(2);  // 5
a.stride(0);  // 35
a.stride(1);  // 5
a.stride(2);  // 1
\end{cpp}

\noindent \CuTe supports these representations and strictly expands on them with generalizations to hierarchical shapes and strides to represent more complex layouts, non-integral strides, and non-integral layout codomains.

Independent generalizations of dense tensor representations include HeLayers~\cite{Aharoni:HeLayers}, ThunderKittens~\cite{Spector:ThunderKittens}, and the Linear Layouts~\cite{Tillet:LinearLayouts} approach used in OpenAI's Triton compiler~\cite{Tillet:TritonAI}. Thunderkittens implements a wide variety of bespoke types for register memory, shared memory, row/column-major tiles, row/column-major tiles of row/column-major subtiles, and prescribed access patterns for warps and threads. These types are written with the architectural layout requirements and memory hierarchy in mind, but the representational span is limited to the existing types and patterns. Linear Layouts are based on $\mathbb{F}_2$ linear algebra and provide a more general representation of tensor layouts as well as an avenue for layout analysis and generation. Linear Layouts' strict reliance on $\mathbb{F}_2$ makes some of the operators difficult for humans to inspect and limits the work to power-of-two shapes and strides, which is unacceptable to many applications.

Other approaches to layout analysis which inspired portions of this work also rely on $\mathbb{F}_2$ linear algebra, such as the work of Edelman et al.~\cite{Edelman:IndexTransforms}, Cormen et al.~\cite{Cormen:FastPermuting}, and Bouverot et al.~\cite{Bouverot:AffineIndex}. These works provide a foundation for analysis and algorithm generation, but are limited due to the application contexts in which they were written.

As the C++ implementation of \CuTe within CUTLASS v3~\cite{NVIDIA:cutlass_v3} is open-source with some documentation, it has already been referenced and used in independent works analyzing the \CuTe layouts and \CuTe algebraic operations. Bhaskaracharya et al.~\cite{Grover:ISLCuTe} analyzes \CuTe and Linear Layouts~\cite{Tillet:LinearLayouts} in the context of integer set relations (ISL), but omit stride abstractions that allow the representation of Linear Layouts as \CuTe layouts. LEGO~\cite{Tavakkoli:LEGO} uses a restricted form of \CuTe layouts and a primitive form of \CuTe composition to generate complex indexing within a code generator. Carlisle et al.~\cite{Carlisle:CategoryCuTe} analyzes \CuTe layouts and some operations on them in the context of category theory.
In this paper, we intend to provide a more definitive and formal treatment of \CuTe concepts and their applications as well as a reference implementation of \CuTe in Python.
A pure-Python reference implementation, PyCuTe~\cite{NVlabs:PyCuTe}, is available at \url{https://github.com/NVlabs/CuTe} and mirrors the definitions and post-conditions of this paper; it is distinct from CuTe DSL~\cite{cute_dsl_nvidia}, which targets JIT compilation of CUDA kernels, and from the C++ CuTe headers in CUTLASS.

The design and concepts of \CuTe have already demonstrated utility in several applications. For instance, \CuTe has been used within the Graphene tensor compiler~\cite{Hagedorn:Graphene}, where it plays a critical role in representing tensor operations. Additionally, the C++ implementation of \CuTe has been used in implementations of the Stream-K algorithm~\cite{Osama:StreamK} and is core to the development of NVIDIA's CUTLASS v3 library~\cite{kerr2017cutlass}. \CuTe has also been a core component in state-of-the-art implementations for large language models, including FlashAttention and each of its evolving generations~\cite{Dao:FlashAttention, Dao:FlashAttention2, Dao:FlashAttention3}, highlighting its relevance to cutting-edge research and applications in deep learning. Additionally, \CuTe is the basis for a number of related compiler projects including CuTe DSL~\cite{cute_dsl_nvidia}, a Python-based DSL for dynamic compilation of CUDA software for linear algebra applications.

\subsection{Canonical Loops and Loop Transformations}
\label{sec:loops}

The explicit calculation of loop indices is a common challenge in the development of high-performance linear algebra kernels. These calculations are difficult for programmers to get right and even more challenging to maintain. Rather than coupling information about data access with algorithmic logic, we prefer to write algorithmic logic clearly in terms of matrix/vector coordinates and abstract the data access patterns to the data layouts.

To illustrate, we first define the class of loop nests that can be addressed by the techniques developed in this work. Specifically, we consider a {\em standard loop form} to be a loop with a single index, starting at zero, bounded by a constant, and incremented by 1 each iteration.

For instance, consider the following loop:
\begin{cpp}
for (int m = 2; m <= 50; m += 3)
  A[m] = e(m);
\end{cpp}
This loop sets {\tt A[2]}, {\tt A[5]}, {\tt A[8]}, $\ldots$ to the result of a pure expression {\tt e(m)}. It can be transformed into a canonical loop as follows:
\begin{cpp}
for (int i = 0; i < 17; ++i)
  (A + 2)[3*i] = g(i);
\end{cpp}
Here, the pointer is offset by a loop-invariant constant, the loop stride is normalized to 1, the lower bound is transformed to zero, the upper bound is tight and non-inclusive, and the pure expression is transformed {\tt g(i) = e(3*i+2)}. It is now intuitive to interpret the above example as iterating through a logically 17-element vector, where the logical coordinate is strided by 3 to index the data at base address {\tt A + 2}. This program can be represented with the following data:
\begin{verbatim}
Accessor: A + 2
   Shape: 17
  Stride: 3
\end{verbatim}
Nested loops can be treated similarly. Consider the following two-dimensional loop nest:
\begin{cpp}
for (int n = 3; n < 43; n += 2)
  for (int m = 4; m <= 22; m += 5)
    A[p*m + q*n] = e(m,n);
\end{cpp}
which can be transformed into canonical loop form:
\begin{cpp}
for (int j = 0; j < 20; ++j)
  for (int i = 0; i < 4; ++i)
    (A + 4*p + 3*q)[5*p*i + 2*q*j] = g(i,j);
\end{cpp}
With the canonical loop nest, it is natural to interpret the transformed loop as iterating through a logically $4 {\times} 20$ matrix, where the logical row coordinate $i$ is strided by $5p$, and the logical column coordinate $j$ is strided by $2q$ to index the data at base address {\tt A + 4*p + 3*q}. This can be represented with the following data
\begin{verbatim}
Accessor: A + 4p + 3q
   Shape: ( 4, 20)
  Stride: (5p, 2q)
\end{verbatim}
A key observation is that the $4 {\times} 20$ matrix can also be interpreted as an $80$-element vector with non-uniform, semi-affine striding, expressed with an equivalent canonical loop form:
\begin{cpp}
for (int k = 0; k < 80; ++k)
  (A + 4*p + 3*q)[5*p*(k%4) + 2*q*(k/4)] = f(k);
\end{cpp}
where {\tt \%} is modulo and {\tt /} is integer floor-division. This transformation is the colexicographical bijection, \verb|(i,j) = (k%4,k/4)|, between 2D coordinates \verb|(i,j)| and 1D coordinates \verb|k|.
This bijection is equivalent to, and can be derived directly from, the shape represented previously. Thus, the shape representation can accept both 2D coordinates and 1D coordinates, providing a flexible and rank-agnostic framework for indexing data.

Furthermore, the canonical loop form also provides guidance for provably correct loop transformations that often appear in optimizing compilers for tensor computations. Consider the most general canonical loop nest:
\begin{cpp}
for (int in = 0; in < Nn; ++in)
  ...
    for (int i1 = 0; i1 < N1; ++i1)
      for (int i0 = 0; i0 < N0; ++i0)
        A[d0*i0 + d1*i1 + ... + dn*in] = e(i0,i1,...,in);
\end{cpp}
where $(N_0, N_1, \ldots, N_n)$ is the ``shape" of the computation and $(d_0, d_1, \ldots, d_n)$ are the ``strides" of the access pattern,
\begin{verbatim}
Accessor: A
   Shape: (N0,N1,...,Nn)
  Stride: (d0,d1,...,dn)
\end{verbatim}
Because there is a one-to-one correspondence between the $\text{Shape}:\text{Stride}$ information and the loop nest itself, rather than asking how to perform transformations on the loops -- splitting, transposition, concatenation, permutation, truncation, vectorization, etc -- we can instead ask ``What are valid ways to transform the $\text{Shape}:\text{Stride}$ representation and what operators provide those transformations?" Indeed, if $\v{L} = \text{Shape}:\text{Stride}$ represents the data access and the loop nest, what functions $P$ exist such that
\begin{align*}
\v{L}' = P(\v{L}) = \v{L} \circ P
\end{align*}
is a meaningful transformation of $\v{L} = \text{Shape}:\text{Stride}$ to a new loop nest $\v{L}' = \text{Shape}':\text{Stride}'$ with a potentially new shape and stride. These transformations, $P$, essentially {\em rewrite} the loop nest and, if defined properly, may themselves be composable, invertible, and provide functional-programming-like control of imperative loops. With considerations of the bijection between the 1D coordinates and the ND coordinates discussed above, this paper demonstrates that a very effective representation of these transformations is $P = \v{P} = \text{Shape}^*:\text{Stride}^*$, the same objects that we use to represent the data accesses and loop nests themselves.

\subsection{Tensors and Folding}

To further motivate shapes that can be indexed by ND coordinates in addition to 1D coordinates, we generalize observations made in~\cite{Shi:BLASContractions} mapping tensor contractions to canonical BLAS-like primitives.

In this work, tensors are denoted by bold letters, indices by lowercase letters, and the bounds of those indices by their corresponding uppercase letters. The rank of a tensor refers to the number of dimensions, also known as modes, it possesses. For example:
\begin{compactitem}
\item A scalar, $\alpha$, is a rank-0 tensor.
\item A vector, $\mathbf{a}_i$, is a rank-1 tensor with $0 \leq i < I$.
\item A matrix, $\mathbf{A}_{mn}$ is a rank-2 tensor with $0 \leq m < M$ and $0 \leq n < N$.
\item A three-way array, $\mathbf{A}_{mnp}$, is a rank-3 tensor with $0 \leq m < M$, $0 \leq n < N$, and $0 \leq p < P$.
\end{compactitem}
Summation is implied over repeated indices that appear only on a single side of an equation (Einstein notation), so an instance of a tensor contraction is
\begin{align}
\v{C}_{stqp} = \v{A}_{stupr} \, \v{B}_{qtru},
\label{eqn:contraction}
\end{align}
which represents the contraction of a rank-5 tensor with a rank-4 tensor to produce a rank-4 tensor. Contractions of this form are expressed compactly in the {\tt numpy.einsum} and {\tt torch.einsum} interfaces, for instance.

The above tensor contraction can be rewritten as
\begin{align*}
\v{C}_{(sp)(q)(t)} = \v{A}_{(sp)(ur)(t)} \, \v{B}_{(q)(ur)(t)},
\end{align*}
where the modes of the original tensor contraction have been grouped into four types:
\begin{compactitem}
\item \textbf{Row modes, $\hat{m}$}: Appear in $\v{A}$ and $\v{C}$, and not in $\v{B}$.
\item \textbf{Column modes, $\hat{n}$}: Appear in $\v{B}$ and $\v{C}$, and not in $\v{A}$.
\item \textbf{Reduction modes, $\hat{k}$}: Appear in $\v{A}$ and $\v{B}$, and not in $\v{C}$.
\item \textbf{Batch modes, $\hat{\ell}$}: Appear in $\v{A}$, $\v{B}$, and $\v{C}$.
\end{compactitem}
This is referred to as tensor {\em folding}. Folding a tensor need not require any explicit copy, but can instead simply be a change of view of the data.

As an explicit example of tensor folding, consider the $2 {\times} 2 {\times} 2$ tensor of 8 elements shown in the first row of Figure~\ref{fig:folding}. The flat representation holds a shape and a stride for each mode of the tensor to index into the physical data. The flat representation is identical to the representation that is adopted by {\tt std::mdspan} in C++, {\tt torch.tensor} in PyTorch, {\tt numpy.ndarray} in NumPy, among many other similar libraries. The $2 {\times} 2 {\times} 2$ tensor can be folded into a $4 {\times} 2$ matrix by folding the third mode into the first, as shown in the second row. The result admits the flat representation with a shape of $(4,2)$ and a stride of $(2,1)$. In principle, the $2 {\times} 2 {\times} 2$ tensor can also be folded into a $2 {\times} 4$ matrix by folding the third mode into the second mode. However, the result no longer admits a flat representation -- there is no integer that can represent the stride of the second mode.

\begin{figure}[ht]
\centering
\scalebox{0.6}{
\begin{tikzpicture}[x={(0cm,-1cm)},y={(1cm,0cm)},every node/.style={minimum size=1cm, outer sep=0pt}]
\node[fill={rgb,255:red,175;green,175;blue,255}] at (0,0) {\Large a\strut};
\node[fill={rgb,255:red,175;green,175;blue,255}] at (0,1) {\Large b\strut};
\node[fill={rgb,255:red,175;green,255;blue,175}] at (0,2) {\Large c\strut};
\node[fill={rgb,255:red,175;green,255;blue,175}] at (0,3) {\Large d\strut};
\node[fill={rgb,255:red,255;green,255;blue,175}] at (0,4) {\Large e\strut};
\node[fill={rgb,255:red,255;green,255;blue,175}] at (0,5) {\Large f\strut};
\node[fill={rgb,255:red,255;green,175;blue,175}] at (0,6) {\Large g\strut};
\node[fill={rgb,255:red,255;green,175;blue,175}] at (0,7) {\Large h\strut};
\draw[color=black,thick,shift={(-0.5,-0.5)}] (0,0) grid (1,8);
\node at (-1,0) {\Large{\texttt{0}}};
\node at (-1,1) {\Large{\texttt{1}}};
\node at (-1,2) {\Large{\texttt{2}}};
\node at (-1,3) {\Large{\texttt{3}}};
\node at (-1,4) {\Large{\texttt{4}}};
\node at (-1,5) {\Large{\texttt{5}}};
\node at (-1,6) {\Large{\texttt{6}}};
\node at (-1,7) {\Large{\texttt{7}}};
\node at (1,3.5) {\Large Physical data};
\end{tikzpicture}
}\\
\vspace{1em}
\begin{tabular}{|>{\centering\arraybackslash}m{3cm}|>{\centering\arraybackslash}m{3cm}|>{\centering\arraybackslash}m{3.5cm}|>{\centering\arraybackslash}m{3.5cm}|}
View Description & View Diagram & Flat Representation & \CuTe Representation \\
\hline\hline
\makecell{Tensor, $2 {\times} 2 {\times} 2$ \\ Rank-3} &
\scalebox{0.4}{
\begin{adjustbox}{margin=0.5cm}
\begin{tikzpicture}[x={(0cm,-1cm)},y={(1cm,0cm)},every node/.style={minimum size=1cm, outer sep=0pt},baseline=(current bounding box.center)]
\node[draw=black, fill={rgb,255:red,255;green,255;blue,175}] at (0,0) {\Large e\strut};
\node[draw=black, fill={rgb,255:red,255;green,255;blue,175}] at (0,1) {\Large f\strut};
\node[draw=black, fill={rgb,255:red,255;green,175;blue,175}] at (1,0) {\Large g\strut};
\node[draw=black, fill={rgb,255:red,255;green,175;blue,175}] at (1,1) {\Large h\strut};
\node[draw=black, fill={rgb,255:red,175;green,175;blue,255}] at (0.75,-0.75) {\Large a\strut};
\node[draw=black, fill={rgb,255:red,175;green,175;blue,255}] at (0.75,0.25) {\Large b\strut};
\node[draw=black, fill={rgb,255:red,175;green,255;blue,175}] at (1.75,-0.75) {\Large c\strut};
\node[draw=black, fill={rgb,255:red,175;green,255;blue,175}] at (1.75,0.25) {\Large d\strut};
\end{tikzpicture}
\end{adjustbox}
} &
\makecell{Shape: $(2,2,2)$\\ Stride: $(2,1,4)$} &
\makecell{Shape: $(2,2,2)$\\ Stride: $(2,1,4)$} \\
\hline
\makecell{Matrix, $4 {\times} 2$\\ Fold mode 2 into 0} &
\scalebox{0.4}{
\begin{adjustbox}{margin=0.5cm}
\begin{tikzpicture}[x={(0cm,-1cm)},y={(1cm,0cm)},every node/.style={minimum size=1cm, outer sep=0pt},baseline=(current bounding box.center)]
\node[draw=black, fill={rgb,255:red,175;green,175;blue,255}] at (0,0) {\Large a\strut};
\node[draw=black, fill={rgb,255:red,175;green,175;blue,255}] at (0,1) {\Large b\strut};
\node[draw=black, fill={rgb,255:red,175;green,255;blue,175}] at (1,0) {\Large c\strut};
\node[draw=black, fill={rgb,255:red,175;green,255;blue,175}] at (1,1) {\Large d\strut};
\node[draw=black, fill={rgb,255:red,255;green,255;blue,175}] at (2,0) {\Large e\strut};
\node[draw=black, fill={rgb,255:red,255;green,255;blue,175}] at (2,1) {\Large f\strut};
\node[draw=black, fill={rgb,255:red,255;green,175;blue,175}] at (3,0) {\Large g\strut};
\node[draw=black, fill={rgb,255:red,255;green,175;blue,175}] at (3,1) {\Large h\strut};
\end{tikzpicture}
\end{adjustbox}
} &
\makecell{Shape: $(4,2)$\\ Stride: $(2,1)$} &
\makecell{Shape: $((2,2),2)$\\ Stride: $((2,4),1)$} \\
\hline
\makecell{Matrix, $2 {\times} 4$\\ Fold mode 2 into 1} &
\scalebox{0.4}{
\begin{adjustbox}{margin=0.5cm}
\begin{tikzpicture}[x={(0cm,-1cm)},y={(1cm,0cm)},every node/.style={minimum size=1cm, outer sep=0pt},baseline=(current bounding box.center)]
\node[draw=black, fill={rgb,255:red,175;green,175;blue,255}] at (0,0) {\Large a\strut};
\node[draw=black, fill={rgb,255:red,175;green,175;blue,255}] at (0,1) {\Large b\strut};
\node[draw=black, fill={rgb,255:red,175;green,255;blue,175}] at (1,0) {\Large c\strut};
\node[draw=black, fill={rgb,255:red,175;green,255;blue,175}] at (1,1) {\Large d\strut};
\node[draw=black, fill={rgb,255:red,255;green,255;blue,175}] at (0,2) {\Large e\strut};
\node[draw=black, fill={rgb,255:red,255;green,255;blue,175}] at (0,3) {\Large f\strut};
\node[draw=black, fill={rgb,255:red,255;green,175;blue,175}] at (1,2) {\Large g\strut};
\node[draw=black, fill={rgb,255:red,255;green,175;blue,175}] at (1,3) {\Large h\strut};
\end{tikzpicture}
\end{adjustbox}
} &
\makecell{Shape: $(2,4)$\\ Stride: $(2,\text{\color{red}\ding{55}})$} &
\makecell{Shape: $(2,(2,2))$\\ Stride: $(2,(1,4))$} \\
\hline
\end{tabular}
\caption{A physical array of 8-elements viewed as a rank-3 tensor and folded into two matrices.}
\label{fig:folding}
\end{figure}

The \CuTe representation of the folded matrices is shown in the last column and emphasizes that the tensor folding really is just grouping modes together. In the case of the $4 {\times} 2$ matrix, the flat representation is called the {\em coalesced} version of the \CuTe representation, while in the case of the $2 {\times} 4$ matrix no such flat representation exists.

This generalized form of tensor folding allows all tensor contractions to be written in a single canonical contraction form,
\begin{align}
\mathbf{C}_{\hat{m}\hat{n}\hat{\ell}} = \mathbf{A}_{\hat{m}\hat{k}\hat{\ell}} \, \mathbf{B}_{\hat{n}\hat{k}\hat{\ell}}.
\end{align}
where each mode may be a single mode or a group of modes, which we call a {\em multi-mode}. Appealing to the canonical loops in Section~\ref{sec:loops}, regardless of whether the shape of $\hat{m}$ is $M$ or $(M_0,M_1)$, we can loop over it with 1D coordinates, $m$. Thus, any tensor contraction can be folded into a canonical {\tt batched-GEMM} and evaluated with a trivial reference implementation composed of four nested loops:
\begin{cpp}
for (int l = 0; l < L; ++l)
  for (int m = 0; m < M; ++m)
    for (int n = 0; n < N; ++n)
      for (int k = 0; k < K; ++k)
        C(m,n,l) += A(m,k,l) * B(n,k,l);
\end{cpp}
This simple implementation of {\tt batched-GEMM} can be used to evaluate a wide range of compatible tensor contractions, including any matrix-multiplication ({\tt GEMM}), tensor contraction ({\tt GETT}), and convolution ({\tt CONV}), with intelligent construction of folded layouts. See Section~\ref{sec:gemm} for details on generic {\tt GEMM} and its applications.

Optimizations can then focus on loop reordering, tiling, vectorization, and other common optimizations by transforming the order and rank of the loop nests. These transformations, it turns out, can very often be represented by layouts as permutations on the coordinates spaces of the algorithm. These transform layouts functionally compose with the data layouts to generate new loop nests that are guaranteed to be consistent with the original problem. See Section~\ref{sec:composition} for details on layout composition and application to generic partitioning.

\section{Layout Representation}

\CuTe layouts are versatile objects that are capable of representing a wide range of data and thread arrangements and have great utility in abstracting physical addresses and separating iteration order from storage order. In this section, we define the \CuTe representation of shapes, layouts, and tensors and construct their interaction with coordinates. \CuTe layouts enable generic algorithms that, with a single implementation, can be applied to any complex layout of data that may be folded into that algorithms' canonical form. Examples of such algorithms and their breadth of applications are provided in Section~\ref{sec:algorithms}.

\subsection{Tuples and HTuples}

The {\tt Tuple} and {\tt HTuple} concepts serve as foundational data structures throughout this work.

\begin{definition}
A $\texttt{Tuple}(\mathcal{T})$ is a finite, ordered list of elements selected from a set $\mathcal{T}$. For an $X = (X_0,X_1,\ldots,X_{n-1}) \in \texttt{Tuple}(\mathcal{X})$, we define the operations:
\begin{compactitem}
\item {\bf Rank:} rank($X$). The tuple length $n$.
\item {\bf Access:} $X_i$. The \texttt{i}th element of the \texttt{Tuple} $X$ for $0 \leq i < \text{rank}(X)$.
\end{compactitem}
\end{definition}

\noindent Where a $\texttt{Tuple}(\mathcal{T})$ is a flat collection of elements, an $\texttt{HTuple}(\mathcal{T})$ is a ``hierarchical tuple of $\mathcal{T}$s".

\begin{definition}
An $\texttt{HTuple}(\mathcal{T})$ is either an element of set $\mathcal{T}$ or a $\texttt{Tuple}(\texttt{HTuple}(\mathcal{T}))$. For an $X \in \texttt{HTuple}(\mathcal{X})$ we define the operations:
\begin{compactitem}
\item {\bf Rank:} rank($X$). If $X \in \texttt{Tuple}$, then the tuple length, else 1.
\item {\bf Access:} $X_i$. The \texttt{i}th element of an \texttt{HTuple} $X$ for $0 \leq i < \text{rank}(X)$.
\item {\bf Depth:} depth($X$). If $X \in \texttt{Tuple}$, then $1 + \text{max}(\text{depth}(X_0), \text{depth}(X_1), \ldots)$, else 0.
\end{compactitem}
\end{definition}

For instance,
\begin{align*}
31 \quad\quad (16,32) \quad\quad (3,-8,7) \quad\quad (2,(4,1),-1) \quad\quad ((4,6),(3,(2,2),8))
\end{align*}
are all instances of $\texttt{HTuple}(\Z)$.


When reasoning with \texttt{HTuple}s, it is useful to define a notion of {\em congruence} and {\em weak congruence}.
\begin{definition}
\emph{Congruence}, $\sim$, is an equivalence relation on \texttt{HTuple}s. For $P \in \texttt{HTuple}(\mathcal{P})$ and $S \in \texttt{HTuple}(\mathcal{S})$,
\begin{align*}
P \sim S \quad \text{ iff } \quad
\begin{cases}
P \in \mathcal{P} \text{ and } S \in \mathcal{S}, \text{ or} \\
P,S \in \texttt{Tuple} \text{ and } \text{rank}(P) = \text{rank}(S) \text{ and } \forall_i \ P_i \sim S_i
\end{cases}
\end{align*}
and we say that $P$ and $S$ are {\em congruent} and have the same {\em profile}.
\end{definition}
For instance,
\begin{align*}
(4,8) \sim (5,7) \quad \text{and} \quad (4,(2,4)) \sim (7,(3,2)) \quad \text{and} \quad (\v{v}, ((\v{p}, 3))) \sim (0, ((0, 0)))
\end{align*}
but $(4,8)$ and $(4,(2,4))$ are not congruent and $(4,(2,4))$ and $(0, ((0, 0)))$ are not congruent.

Similarly, weak congruence tests that the profile of one {\tt HTuple} is at least as {\em refined} as another.
\begin{definition}
\emph{Weak Congruence}, $\lesssim$, is a partial order on \texttt{HTuple}s. For $P \in \texttt{HTuple}(\mathcal{P})$ and $S \in \texttt{HTuple}(\mathcal{S})$,
\begin{align*}
P \lesssim S \quad \text{ iff } \quad
\begin{cases}
P \in \mathcal{P}, \text{ or} \\
P,S \in \texttt{Tuple} \text{ and } \text{rank}(P) = \text{rank}(S) \text{ and } \forall_i \ P_i \lesssim S_i
\end{cases}
\end{align*}
and we say that $P$ and $S$ are {\em weakly congruent}, $P$ {\em coarsens} the profile of $S$, and $S$ {\em refines} the profile of $P$.
\end{definition}
For instance,
\begin{align*}
30 \lesssim (a,b) \lesssim (\v{v},(0,\alpha)) \quad \text{and} \quad 30 \lesssim (a,b,c) \lesssim ((0,0),0,0)
\end{align*}
but $(a,b)$ and $(a,b,c)$ are not weakly congruent and $(\v{v},(0,\alpha))$ and $((0,0),0)$ are not weakly congruent.

\subsection{Shape}
\label{sec:shape}

Multidimensional arrays are often characterized by their \emph{shape}, a sequence of positive integers describing the extent of each mode. The 2D shape of an $M{\times}N$ matrix is represented as $(M,N)$ and is naturally indexed by coordinates $(m,n)$ with $0 \leq m < M$ and $0 \leq n < N$. A natural extension of this is to represent a shape as a hierarchical tuple of positive integers.

\begin{definition}
A {\em shape} is an $\texttt{HTuple}(\Z^+)$, where $\Z^+ = \{1,2,3,\ldots\}$ is the set of positive integers. The rank of a shape $S$ is the rank of the \texttt{HTuple}. The size of a shape $S$ is the product of its elements, denoted $\abs{S} = \prod_k \abs{S_k}$.
\end{definition}

What makes hierarchical shapes particularly useful is that they can be indexed by multiple coordinate systems. Consider the set of integers up to $N$,
\begin{align*}
\Z_N = \{ 0, 1, 2, \ldots, N-1 \}.
\end{align*}
\CuTe makes the observation that a 2D shape $(M,N)$ can also be interpreted to describe 1D $MN$ elements indexed by an integral coordinate $i$ with $0 \leq i < MN$ provided a bijection
\begin{align*}
S : \Z_{MN} \longleftrightarrow \Z_M \times \Z_N
\end{align*}
maps between the 1D integral coordinates $i \in \Z_{MN}$ and the 2D natural coordinates $(m,n) \in \Z_M \times \Z_N$.

Similarly, the 2D shape $(M,NP)$ can be interpreted as a hierarchical shape $(M,(N,P))$ indexed by natural coordinates $(m,(n,p))$ with $0 \leq m < M$, $0 \leq n < N$, $0 \leq p < P$. A similar bijection can be made
\begin{align*}
S : \Z_M \times \Z_{NP} \longleftrightarrow \Z_M \times (\Z_N \times \Z_P)
\end{align*}
to map between the 2D coordinates $(m,q) \in \Z_M \times \Z_{NP}$ and the natural coordinates $(m,(n,p)) \in \Z_M \times (\Z_N \times \Z_P)$.

A direct consequence of hierarchical shapes and coordinates is that tensor algorithms can be written for the shapes that are most natural to them (Section~\ref{sec:algorithms}) -- 1D shapes for vectors in {\tt COPY}, 2D shapes for matrices in {\tt GEMM}, 3D shapes for tensors in {\tt batched-GEMM}, etc -- while still accepting hierarchically shaped tensors that are folded to be weakly congruent with the algorithm's specification (Section~\ref{sec:compatibility}). Tensors of data, whose shape is often represented as a flat sequence of integers, can be arbitrarily folded into shapes accepted by generic tensor algorithms. Furthermore, because each mode of a tensor is associated with a stride (Section~\ref{sec:stride}) to index data, this folding of modes allows the representation of much more complex layouts of data beyond simple contiguous arrays in {\tt COPY} or row-major and col-major matrices in {\tt BLAS GEMM} (Section~\ref{sec:layout}).

In the following sections, we define this notion of compatibility and these relations between coordinate sets within a shape.

\subsubsection{Coordinate Sets and Compatibility}
\label{sec:compatibility}

As previously suggested, hierarchical shapes provide for indexing by multiple coordinate systems. Here, we define the coordinate set for a specific shape and a notion of compatibility between shapes to share coordinate sets between shapes.

\begin{definition}
A \emph{coordinate set} is a set $\Z_N = \{0, 1, 2, \ldots, N-1\}$ of non-negative integers or a Cartesian product of coordinate sets, $\Z_N \times \Z_M = \Z_{(N,M)}$.
\end{definition}

For instance, the following are examples of coordinate sets:
\begin{align*}
\Z_6 &= \{ 0, 1, 2, 3, 4, 5 \} \\
\Z_3 \times \Z_4 = \Z_{(3,4)} &= \{(0,0), (1,0), (2,0), (0,1), (1,1), (2,1), (0,2), (1,2), (2,2), (0,3), (1,3), (2,3) \} \\
(\Z_2 \times \Z_1) \times \Z_3 = \Z_{((2,1),3)} &= \{((0,0),0), ((1,0),0), ((0,0),1), ((1,0),1), ((0,0),2), ((1,0),2) \}
\end{align*}
A coordinate set $\Z_S$ is precisely the set of natural coordinates for a shape $S$. Other coordinate sets for a shape $S$ are any coordinate set for a shape that is {\em compatible} with $S$ and {\em coarsens} $S$.

\begin{definition}
\emph{Compatibility}, $\preceq$, is a partial order on the set of shapes. For shapes $P$ and $S$,
\begin{align*}
P \preceq S \quad \text{ iff } \quad
\begin{cases}
P \in \Z^+ \text{ and } P = \abs{S}, \text{ or} \\
P,S \in \texttt{Tuple} \text{ and } \text{rank}(P) = \text{rank}(S) \text{ and } \forall_i \ P_i \preceq S_i
\end{cases}
\end{align*}
and we say that $P$ and $S$ are {\em compatible}, $P$ {\em coarsens} $S$, and $S$ {\em refines} $P$.
\end{definition}

Compatibility requires that the two shapes be the same size, so the integral values of the \texttt{HTuple}s matter. For example,
\begin{align*}
30 \preceq (2,15) \preceq (2,(3,5)) \quad \text{and} \quad
30 \preceq (6,5) \preceq ((3,2),5)
\end{align*}
but $(2,(3,5))$ and $((3,2),5)$ are not compatible despite having the same size. They do, however, share a common compatible shape of $30$.

With the definition of a coordinate set and shape compatibility, we can define the set of all compatible coordinates for any given shape.

\begin{definition}
A shape $S$ defines a {\em set of compatible coordinate} sets, $\Z(S)$, as the coordinate sets of all shapes that coarsen $S$.
\begin{align}
\Z(S) = \{\Z_{S'} \, \mid \, S' \preceq S\}.
\end{align}
\end{definition}

Every shape has an integral coordinate set,
\begin{align*}
\{ 0, 1, 2, \ldots, \abs{S}-1 \} = \Z_{\abs{S}} \in \Z(S),
\end{align*}
and every rank-$r$ shape has a rank-$r$ coordinate set,
\begin{align*}
\{ (a_0, \ldots, a_{r-1}) \ \mid \ a_i \in \Z_{\abs{S_i}} \} = \Z_{(\abs{S_0}, \abs{S_1}, \ldots, \abs{S_{r-1}})} \in \Z(S).
\end{align*}

Note that if shape $P$ coarsens shape $S$, then $\Z(P) \subseteq \Z(S)$. This means that any coordinate within shape $P$ is also a coordinate within shape $S$.






\subsubsection{Coordinates}

In this section, we define classes of coordinates, define a bijection between the compatible coordinate sets of a shape, and provide examples of these coordinate mappings.

\begin{definition}
An \emph{in-bounds coordinate}, or simply {\em coordinate}, into a shape $S$ is an element of one of its coordinate sets, $c \in \Z_{S'} \in \Z(S)$. Note that a coordinate is always an $\texttt{HTuple}(\N)$. When intention is clear, we will simply write $c \in \Z(S)$.
\end{definition}

\begin{definition}
An \emph{integral coordinate} into a shape $S$ is a coordinate $\bar{c} \in \Z_{\abs{S}} \in \Z(S)$. Note that an integral coordinate is always an integer, $\bar{c} \in \N$.
\end{definition}

\begin{definition}
A \emph{natural coordinate} into a shape $S$ is a coordinate $\tilde{c} \in \Z_S \in \Z(S)$. Note that a natural coordinate is always an $\texttt{HTuple}(\N)$ that is congruent to the shape, $\tilde{c} \sim S$.
\end{definition}

To transform between in-bound coordinates, we construct an enumeration over the coordinate sets of a shape $S$ to define {\em coordinate lists}. In this work, we choose the colexicographical ordering, $<$, of coordinates defined by:
\begin{align*}
(a_0, \ldots, a_n) < (b_0, \ldots, b_n)
\quad \text{iff} \quad
\begin{cases}
    a_n < b_n, \text{ or} \\
    a_n = b_n \ \ \text{and} \ \ (a_0,...,a_{n-1}) < (b_0,...,b_{n-1})
\end{cases}
\end{align*}
and applied recursively as needed. The colexicographical enumeration defines a bijection on coordinate lists. The function
\begin{align}
\texttt{idx2crd} \colon \ \Z_{\abs{S}} &\to \Z_{(\abs{S_0}, \abs{S_1}, \ldots, \abs{S_{r-1}})}, \nonumber \\
i &\mapsto \Bigl(i \bmod \abs{S_0}, \floor{\frac{i}{\abs{S_0}}} \bmod \abs{S_1}, \ldots, \floor{\frac{i}{\prod_{k=0}^{r-3} \abs{S_k}}} \bmod \abs{S_{r-2}}, \floor{\frac{i}{\prod_{k=0}^{r-2} \abs{S_k}}} \Bigr)
\label{eqn:idx2crd}
\end{align}
maps the $i$th coordinate of $\Z_{\abs{S}}$ (the $i$th integral coordinate of shape $S$) to the $i$th coordinate of $\Z_{(\abs{S_0}, \abs{S_1}, \ldots, \abs{S_{r-1}})}$ (the $i$th natural coordinate of shape $(\abs{S_0}, \abs{S_1}, \ldots, \abs{S_{r-1}})$). Other bijections such as the reverse and/or reflected lexicographical or colexicographical orderings can be used as well.

The inverse of \texttt{idx2crd} is given by
\begin{align}
\texttt{crd2idx} \colon \ \Z_{(\abs{S_0}, \abs{S_1}, \ldots, \abs{S_{r-1}})} &\to \Z_{\abs{S}}, \nonumber \\
(c_0, c_1, \ldots, c_{r-1}) &\mapsto c_0 + c_1 \cdot \abs{S_0} + \ldots + c_{r-1} \cdot \prod_{k=0}^{r-2} \abs{S_k}
\label{eqn:crd2idx}
\end{align}
which maps the $i$th coordinate of $\Z_{(\abs{S_0}, \abs{S_1}, \ldots, \abs{S_{r-1}})}$ to the $i$th coordinate of $\Z_{\abs{S}}$.

If two shapes are compatible, $P \preceq S$, then coordinates in $\Z_P$ can be mapped to $\Z_S$ via repeated application of \texttt{idx2crd}, and coordinates in $\Z_S$ can be mapped to $\Z_P$ via repeated application of \texttt{crd2idx}. Following the co-lexicographical ordering of coordinates, Figure~\ref{fig:coordlists} tabulates the mappings from integral coordinates to natural coordinates for each shape.

\begin{figure}[ht]
\centering
\begin{subfigure}[t]{0.2\textwidth}
    \centering
    \begin{tabular}{|c|c|}
    $\Z_4$ \\
    \hline
    0 \\
    1 \\
    2 \\
    3 \\
    \hline
    \end{tabular}
    \caption*{$S = 4$ \\ $\Z(S) = \{\Z_4\}$}
\end{subfigure}
\begin{subfigure}[t]{0.3\textwidth}
    \centering
    \begin{tabular}{|c|c|}
    $\Z_6$   & $\Z_{(2,3)}$ \\
    \hline
    0        & $(0,0)$ \\
    1        & $(1,0)$ \\
    2        & $(0,1)$ \\
    3        & $(1,1)$ \\
    4        & $(0,2)$ \\
    5        & $(1,2)$ \\
    \hline
    \end{tabular}
    \caption*{$S = (2,3)$ \\ $\Z(S) = \{\Z_6, \Z_{(2,3)}\}$}
\end{subfigure}
\begin{subfigure}[t]{0.4\textwidth}
    \centering
    \begin{tabular}{|c|c|c|}
    $\Z_{12}$ & $\Z_{(6,2)}$ & $\Z_{((2,3),2)}$ \\
    \hline
    0         & $(0,0)$         & $((0,0),0)$ \\
    1         & $(1,0)$         & $((1,0),0)$ \\
    2         & $(2,0)$         & $((0,1),0)$ \\
    3         & $(3,0)$         & $((1,1),0)$ \\
    4         & $(4,0)$         & $((0,2),0)$ \\
    5         & $(5,0)$         & $((1,2),0)$ \\
    6         & $(0,1)$         & $((0,0),1)$ \\
    7         & $(1,1)$         & $((1,0),1)$ \\
    8         & $(2,1)$         & $((0,1),1)$ \\
    9         & $(3,1)$         & $((1,1),1)$ \\
    10        & $(4,1)$         & $((0,2),1)$ \\
    11        & $(5,1)$         & $((1,2),1)$ \\
    \hline
    \end{tabular}
    \caption*{$S = ((2,3),2)$ \\ $\Z(S) = \{\Z_{12}, \Z_{(6,2)}, \Z_{((2,3),2)}\}$}
\end{subfigure}
\caption{Examples of the sets of coordinate lists for shapes $S$.}
\label{fig:coordlists}
\end{figure}

\paragraph{Out-of-bounds Coordinates} In addition to the coordinate sets we have already defined, it is useful to define coordinates of specific profiles that may not be in the coordinate sets of a shape.

\begin{definition}
An \emph{admissible coordinate} into a shape $S$ is any coordinate $c \in \texttt{HTuple}(\Z)$ that is weakly congruent to the shape, $c \lesssim S$.
\end{definition}

\begin{definition}
An \emph{out-of-bounds coordinate} into a shape $S$ is any admissible coordinate $c \in {\tt HTuple}(\Z)$ that is not in-bounds, $c \notin \Z(S)$.
\end{definition}

\begin{definition}
A \emph{congruent coordinate} into a shape $S$ is any coordinate $c \in \texttt{HTuple}(\Z)$ that is congruent to the shape, $c \sim S$. This is denoted as $\Z^S = \{c \in \texttt{HTuple}(\Z) \ \mid \ c \sim S\}$.
\end{definition}

That is, $\Z_S$ is the finite set of coordinates bounded by shape $S$, and $\Z^S$ is the infinite set of all coordinates congruent to $S$. Because the values within the shape $S$ for $\Z^S$ don't matter, we will sometimes use $(\ast,\ast)$ as a placeholder profile. For instance,
\begin{align*}
\Z^{(\ast,\ast)} = \{ (a,b) \ \mid \ a,b \in \Z \} \quad \text{and} \quad \Z^{(\ast,(\ast,\ast))} = \{ (a,(b,c)) \ \mid \ a,b,c \in \Z \}
\end{align*}

Note that {\tt idx2crd} is well-defined for all integers, equivalently coordinates in $\Z^{\abs{S}}$, rather than simply the integers in $\Z_{\abs{S}}$. When it is evaluated on an integer $i \geq \abs{S}$, it will always return a coordinate $(c_0,c_1,\ldots,c_{r-1})$ that is out-of-bounds with respect to shape $(\abs{S_0},\abs{S_1},\ldots,\abs{S_{r-1}})$. In contrast, {\tt crd2idx} cannot guarantee an out-of-bounds result for an out-of-bounds coordinate input. Therefore, {\tt crd2idx} and {\tt idx2crd} are only inverses of each other when evaluated on in-bounds coordinates.

\subsection{Stride}
\label{sec:stride}

The previous section described shapes, their hierarchies, and coordinates for those shapes. To construct \textbf{layouts} of data, threads, or other objects, we define a mapping from coordinates within a shape to offsets.

\begin{definition}
A \emph{stride} $D$ for a shape $S$ is an $\texttt{HTuple}(\mathcal{D})$ that is congruent with the shape, $S \sim D$. This stride defines a mapping from a natural coordinate $\tilde{c} \in \Z_S$ to the codomain $\D$, given by
\begin{align}
\texttt{inner\_product} \colon \ &\Z \cdot \D \to \D, \nonumber \\
&c \cdot d \mapsto cd \nonumber\\
\texttt{inner\_product} \colon \ &\texttt{HTuple($\Z$)} \cdot \texttt{HTuple($\D$)} \to \D, \nonumber \\
&c \cdot d \mapsto \sum_i \texttt{inner\_product}(c_i, d_i)
\label{eqn:innerprod}
\end{align}
\end{definition}

In most cases, strides are also \texttt{HTuple($\Z$)}s, meaning $\D = \Z$. The resulting integer produced by \texttt{inner\_product} is typically interpreted as an offset within a data array. However, the concept of a stride element generalizes to any element of an integer-semimodule, which provides significant flexibility in the span of functions that layouts can represent.

\subsubsection{Integer-Semimodules}

\begin{definition}
An \emph{integer-semimodule} is a set $M$ equipped with an associative addition, $M + M \to M$, and a scalar multiplication, $\Z \cdot M \to M$. For $a,b \in \Z$ and $m,n,p \in M$, the addition and scalar multiplication satisfy
\begin{compactitem}
\item Multiplicative Identity: $1 \cdot m = m$,
\item Additive Associativity: $m + (n + p) = (m + n) + p$,
\item Multiplicative Associativity: $a \cdot (b \cdot m) = (ab) \cdot m$.
\end{compactitem}
Additive identities and inverses are not required, so $(M,+)$ is a semigroup. We write an integer-semimodule as $(M,+,\cdot)$ to denote the set $M$, the additive operation $+$, and the scalar multiplication $\cdot$.
\end{definition}

The integers, $\Z$, are an integer-semimodule. The rationals, $\Q$, are an integer-semimodule. The field $\mathbb{F}_2 = (\{0,1\},\text{XOR},\text{AND})$ of arithmetic operations modulo 2 is an integer-semimodule. Any Cartesian product or {\tt HTuple} of integer-semimodules is an integer-semimodule given elementwise addition and scalar multiplication.

A uniquely useful integer-semimodule is $(\Z^S,+,\cdot)$ with $\Z^S$ the set of all {\tt HTuple}($\Z$) congruent to $S$. For instance, the basis elements of rank-2 arithmetic tuples form an integer-semimodule:
\begin{align*}
e_0 = (1,0), \quad e_1 = (0,1), \quad \Z^{(\ast,\ast)} &= \{a \cdot e_0 + b \cdot e_1 \ \mid \ a, b \in \Z\}
\end{align*}
with integer scaling and addition defined element-wise:
\begin{align*}
a \cdot e_0 + b \cdot e_1 = (a, b)
\end{align*}

Thus, $e_0$, $e_1$, and any linear combination can be used as strides within a layout. By selecting stride elements from $\Z^S$, layouts can generate natural coordinates of a shape $S$ through the \texttt{inner\_product} operation.

\subsection{Layout}
\label{sec:layout}

\CuTe uses a shape $S$ and a stride $D$ to define a {\em layout function}, or just {\em layout}. The shape $S$ defines the domain(s) of the layout function, while the stride $D$ defines the codomain of the layout function.

An alternate interpretation of the shape presented previously is that it is a map from the set of all coordinate lists to the natural coordinates. This map is bijective so that each coordinate is mapped to a unique and equivalent natural coordinate within the shape. Similarly, a stride is a map from the natural coordinates of a shape to some codomain.
\begin{align*}
S \ &\colon \ Z \leftrightarrow \Z_S, \ \forall Z \in \Z(S) \\
D \ &\colon \ \Z_S \to \D
\end{align*}
The $\leftrightarrow$ are the functions {\tt idx2crd} \eqref{eqn:idx2crd} and {\tt crd2idx} \eqref{eqn:crd2idx} that map between natural coordinates of compatible shapes, while the $\to$ mapping is the \texttt{inner\_product} \eqref{eqn:innerprod} between a natural coordinate and the stride.

The composition of the shape and the stride defines the layout function, which is a map from the set of all coordinate lists to the codomain.

\begin{definition}
A \emph{layout} $\v{L} = D \circ S$ is the functional composition of a shape $S$ and a stride $D$, where $S \sim D$, and defines a mapping $Z \to \D$ for each $Z \in \Z(S)$.
\end{definition}

\subsubsection{Notations and Operations}

We write layouts $\v{L}$ in different notations depending on context. For instance:
\begin{align*}
\begin{array}{ccc} (4, & (3, & 2)) \\ (2, & (8, & 1)) \end{array} = \begin{array}{c} S \\ D \end{array} \quad\quad \text{or} \quad\quad (4, (3, 2)) : (2, (8, 1)) = S : D  \quad\quad \text{or} \quad\quad (2, (8, 1)) \circ (4, (3, 2)) = D \circ S
\end{align*}
where the last style emphasizes that the shape and stride themselves can be interpreted as functions that compose to define the layout function.

Since the domain(s) of a layout are determined by its shape, layout properties align closely with shape properties. For layouts $\v{L} = S : D$ and $\v{U} = X : Y$, we define:
\begin{compactitem}
\item $\text{rank}(\v{L}) = \text{rank}(S)$: The rank of a layout is the rank of its shape.
\item $\text{depth}(\v{L}) = \text{depth}(S)$: The depth of a layout is the depth of its shape.
\item $\abs{\v{L}} = \abs{S}$: The size of a layout is the size of its shape.
\item $\v{L}_i = S_i : D_i$: The $i$th sublayout is constructed from the $i$th element of its shape and stride.
\item $\Z(\v{L}) = \Z(S)$: The coordinate sets of a layout are the coordinate sets of its shape.
\item $\v{L} \sim \v{U} \Leftrightarrow S \sim X$: The congruence of two layouts is the congruence of their shapes.
\item $\v{L} \preceq \v{U} \Leftrightarrow S \preceq X$: The compatibility of two layouts is the compatibility of their shapes.
\end{compactitem}

As an example of layout evaluation, consider the layout $\v{L} = ((2,2),(4,2)):((1,8),(2,16))$ from Figure~\ref{fig:layouts:mixed} and the integral coordinate $22 \in Z_{32} \in Z(\v{L})$, then
\begin{align*}
\v{L}(22) = \v{L}(2,5) = \v{L}((0,1),(1,1)) = 26
\end{align*}
where we have shown the integral coordinate, the equivalent 2D coordinate, the equivalent natural coordinate, and the computed offset.

The in-bounds domain of a layout is the set of all coordinates $c \in \Z(\v{L})$. Layouts can also be evaluated on out-of-bounds coordinates, analogous to arrays that can be evaluated on out-of-bounds indices with undefined behavior. So while the domain of a layout is the finite set $\Z(\v{L})$, the {\em extended domain} is the infinite set $\Z^\v{L}$ of all $\texttt{HTuple}(\Z)$ that are weakly congruent with $S$.

We also make the distinction between the codomain and the image of a layout. The codomain of a layout is $\D$, which is often infinite (for instance, $\Z$ or $\Z^D$). The image of a layout is finite and is the range of values the layout produces for all coordinates in the domain.
\begin{align*}
\text{image}(\v{L}) = \v{L}(\Z(\v{L})) = \v{L}(\Z_{\abs{\v{L}}}) \subseteq \text{codomain}(\v{L})
\end{align*}

\subsubsection{Layout Examples}

Having defined shapes, strides, and layouts, we now present examples that demonstrate \CuTe layouts are a strict generalization of common flat N-dimensional layouts.

\begin{figure}[ht]
\centering
\begin{subfigure}[t]{0.3\textwidth}
    \centering
    \begin{tikzpicture}[thick,scale=1.10, every node/.style={scale=1.10}]
        \draw[step=0.5cm,color=black] (-2,-1) grid (2,1);
        \matrix[matrix of nodes,nodes={inner sep=0pt,text width=.5cm,align=center,minimum height=.5cm}]{
            0 & 4 & 8  & 12 & 16 & 20 & 24 & 28 \\
            1 & 5 & 9  & 13 & 17 & 21 & 25 & 29 \\
            2 & 6 & 10 & 14 & 18 & 22 & 26 & 30 \\
            3 & 7 & 11 & 15 & 19 & 23 & 27 & 31 \\};
        \draw[line width=2pt,rounded corners,opacity=0.2]
        (-1.75,+0.75) -- (-1.75,-0.75) --
        (-1.25,+0.75) -- (-1.25,-0.75) --
        (-0.75,+0.75) -- (-0.75,-0.75) --
        (-0.25,+0.75) -- (-0.25,-0.75) --
        (+0.25,+0.75) -- (+0.25,-0.75) --
        (+0.75,+0.75) -- (+0.75,-0.75) --
        (+1.25,+0.75) -- (+1.25,-0.75) --
        (+1.75,+0.75) -- (+1.75,-0.75);
    \end{tikzpicture}
    \subcaption{\tabular[t]{@{}l@{}}Col-Major \\ $(4,8):(1,4)$\endtabular}
\end{subfigure}
\begin{subfigure}[t]{0.3\textwidth}
    \centering
    \begin{tikzpicture}[thick,scale=1.10, every node/.style={scale=1.10}]
        \draw[step=0.5cm,color=black] (-2,-1) grid (2,1);
        \matrix[matrix of nodes,nodes={inner sep=0pt,text width=.5cm,align=center,minimum height=.5cm}]{
            0  & 1  & 2  & 3  & 4  & 5  & 6  & 7  \\
            8  & 9  & 10 & 11 & 12 & 13 & 14 & 15 \\
            16 & 17 & 18 & 19 & 20 & 21 & 22 & 23 \\
            24 & 25 & 26 & 27 & 28 & 29 & 30 & 31 \\};
        \draw[line width=2pt,rounded corners,opacity=0.2]
        (-1.75,+0.75) -- (+1.75,+0.75) --
        (-1.75,+0.25) -- (+1.75,+0.25) --
        (-1.75,-0.25) -- (+1.75,-0.25) --
        (-1.75,-0.75) -- (+1.75,-0.75);
    \end{tikzpicture}
    \subcaption{\tabular[t]{@{}l@{}}Row-Major \\ $(4,8):(8,1)$\endtabular}
\end{subfigure}
\begin{subfigure}[t]{0.3\textwidth}
    \centering
    \begin{tikzpicture}[thick,scale=1.10, every node/.style={scale=1.10}]
        \draw[step=0.5cm,color=black] (-2,-1) grid (2,1);
        \matrix[matrix of nodes,nodes={inner sep=0pt,text width=.5cm,align=center,minimum height=.5cm}]{
            0 & 5 & 10 & 15 & 20 & 25 & 30 & 35 \\
            1 & 6 & 11 & 16 & 21 & 26 & 31 & 36 \\
            2 & 7 & 12 & 17 & 22 & 27 & 32 & 37 \\
            3 & 8 & 13 & 18 & 23 & 28 & 33 & 38 \\};
        \draw[line width=2pt,rounded corners,opacity=0.2]
        (-1.75,+0.75) -- (-1.75,-0.75) --
        (-1.25,+0.75) -- (-1.25,-0.75) --
        (-0.75,+0.75) -- (-0.75,-0.75) --
        (-0.25,+0.75) -- (-0.25,-0.75) --
        (+0.25,+0.75) -- (+0.25,-0.75) --
        (+0.75,+0.75) -- (+0.75,-0.75) --
        (+1.25,+0.75) -- (+1.25,-0.75) --
        (+1.75,+0.75) -- (+1.75,-0.75);
    \end{tikzpicture}
    \subcaption{\tabular[t]{@{}l@{}}Col-Major Padded \\ $(4,8):(1,5)$\endtabular}
\end{subfigure}
\begin{subfigure}[t]{0.3\textwidth}
    \centering
    \begin{tikzpicture}[thick,scale=1.10, every node/.style={scale=1.10}]
        \draw[step=0.5cm,color=black] (-2,-1) grid (2,1);
        \matrix[matrix of nodes,nodes={inner sep=0pt,text width=.5cm,align=center,minimum height=.5cm}]{
            0  & 1  & 2  & 3  & 16 & 17 & 18 & 19 \\
            4  & 5  & 6  & 7  & 20 & 21 & 22 & 23 \\
            8  & 9  & 10 & 11 & 24 & 25 & 26 & 27 \\
            12 & 13 & 14 & 15 & 28 & 29 & 30 & 31 \\};
        \draw[line width=2pt,rounded corners,opacity=0.2]
        (-1.75,+0.75) -- (-0.25,+0.75) --
        (-1.75,+0.25) -- (-0.25,+0.25) --
        (-1.75,-0.25) -- (-0.25,-0.25) --
        (-1.75,-0.75) -- (-0.25,-0.75) --
        (+0.25,+0.75) -- (+1.75,+0.75) --
        (+0.25,+0.25) -- (+1.75,+0.25) --
        (+0.25,-0.25) -- (+1.75,-0.25) --
        (+0.25,-0.75) -- (+1.75,-0.75);
    \end{tikzpicture}
    \subcaption{\tabular[t]{@{}l@{}}Col-Major Interleave \\ $(4,(4,2)):(4,(1,16))$\endtabular}
\end{subfigure}
\begin{subfigure}[t]{0.3\textwidth}
    \centering
    \begin{tikzpicture}[thick,scale=1.10, every node/.style={scale=1.10}]
        \draw[step=0.5cm,color=black] (-2,-1) grid (2,1);
        \matrix[matrix of nodes,nodes={inner sep=0pt,text width=.5cm,align=center,minimum height=.5cm}]{
            0  & 2  & 4  & 6  & 16 & 18 & 20 & 21 \\
            1  & 3  & 5  & 7  & 17 & 19 & 22 & 23 \\
            8  & 10 & 12 & 14 & 24 & 26 & 28 & 30 \\
            9  & 11 & 13 & 15 & 25 & 27 & 29 & 31 \\};
        \draw[line width=2pt,rounded corners,opacity=0.2]
        (-1.75,+0.75) -- (-1.75,+0.25) --
        (-1.25,+0.75) -- (-1.25,+0.25) --
        (-0.75,+0.75) -- (-0.75,+0.25) --
        (-0.25,+0.75) -- (-0.25,+0.25) --
        (-1.75,-0.25) -- (-1.75,-0.75) --
        (-1.25,-0.25) -- (-1.25,-0.75) --
        (-0.75,-0.25) -- (-0.75,-0.75) --
        (-0.25,-0.25) -- (-0.25,-0.75) --
        (+0.25,+0.75) -- (+0.25,+0.25) --
        (+0.75,+0.75) -- (+0.75,+0.25) --
        (+1.25,+0.75) -- (+1.25,+0.25) --
        (+1.75,+0.75) -- (+1.75,+0.25) --
        (+0.25,-0.25) -- (+0.25,-0.75) --
        (+0.75,-0.25) -- (+0.75,-0.75) --
        (+1.25,-0.25) -- (+1.25,-0.75) --
        (+1.75,-0.25) -- (+1.75,-0.75);
    \end{tikzpicture}
    \subcaption{\tabular[t]{@{}l@{}}Mixed \\ $((2,2),(4,2)):((1,8),(2,16))$\endtabular}
  \label{fig:layouts:mixed}
\end{subfigure}
\begin{subfigure}[t]{0.3\textwidth}
    \centering
    \begin{tikzpicture}[thick,scale=1.10, every node/.style={scale=1.10}]
        \draw[step=0.5cm,color=black] (-2,-1) grid (2,1);
        \matrix[matrix of nodes,nodes={inner sep=0pt,text width=.5cm,align=center,minimum height=.5cm}]{
            0  & 0  & 4  & 4  & 8  & 8  & 12  & 12  \\
            0  & 0  & 4  & 4  & 8  & 8  & 12  & 12  \\
            2  & 2  & 6  & 6  & 10 & 10 & 14  & 14  \\
            2  & 2  & 6  & 6  & 10 & 10 & 14  & 14  \\};
        \draw[line width=2pt,rounded corners,opacity=0.2]
        (-1.75,+0.75) --
        (-1.75,-0.25) --
        (-0.75,+0.75) --
        (-0.75,-0.25) --
        (+0.25,+0.75) --
        (+0.25,-0.25) --
        (+1.25,+0.75) --
        (+1.25,-0.25);
    \end{tikzpicture}
    \subcaption{\tabular[t]{@{}l@{}}Blocked Broadcast \\ $((2,2),(2,4)):((0,2),(0,4))$\endtabular}
  \label{fig:layouts:broadcast}
\end{subfigure}
\caption{Examples of layouts compatible with shape $(4,8)$ plotted as $4 {\times} 8$ matrices. These layouts use integer strides and hierarchical shapes to define functions useful for data tensors. The light line follows the general order of the layout offsets.}
\label{fig:layouts}
\end{figure}

Figure~\ref{fig:layouts} illustrates examples of data layouts commonly encountered in dense linear algebra libraries, such as CUTLASS. Each layout is represented as a mapping from logical coordinates $(m,n) \in \Z_{(4,8)}$ to an offset $k \in \Z$. These offsets can, for instance, be used to index elements within a data array. The common row-major, column-major, and padded layouts are trivially represented by \CuTe layouts, while the interleaved and mixed layouts demonstrate that the representation set of layouts is strictly expanded by using nested shapes and strides. In particular, it is clear that grids-of-tiles-of-data are immediately representable with \CuTe layouts.

\begin{figure}[ht]
\centering
\begin{subfigure}[t]{0.45\textwidth}
    \centering
    \begin{tikzpicture}[thick,scale=0.9, every node/.style={scale=0.9}]
      \foreach \row in {0,...,3} {
        \foreach \col in {0,...,7} {
          \draw (0.9*\col,-0.5*\row) rectangle ++(0.9,-0.5) node[pos=.5] {(\row,\col)};
        }
      }
    \end{tikzpicture}
    \subcaption{\tabular[t]{@{}l@{}}Identity Coordinate \\ $(4,8):(e_0,e_1)$\endtabular}
    \label{fig:layouts:coord0}
\end{subfigure}
\begin{subfigure}[t]{0.45\textwidth}
    \centering
    \begin{tikzpicture}[thick,scale=0.9, every node/.style={scale=0.9}]
      \foreach \row in {0,...,3} {
        \foreach \cola in {0,...,3} {
          \foreach \colb in {0,...,1} {
            \pgfmathtruncatemacro{\secondcoord}{\row + 6*\colb}
            \draw (0.9*\cola+3.6*\colb,-0.5*\row) rectangle ++(0.9,-0.5) node[pos=.5] {(\cola,\secondcoord)};
          }
        }
      }
    \end{tikzpicture}
    \subcaption{\tabular[t]{@{}l@{}}Transposed Block Coordinate \\ $(4,(4,2)):(e_1,(e_0,6e_1))$\endtabular}
    \label{fig:layouts:coord1}
\end{subfigure}
\begin{subfigure}[t]{0.45\textwidth}
    \centering
    \begin{tikzpicture}[thick,scale=1, every node/.style={scale=1}]
        \draw[step=0.5cm,color=black] (-3,-1) grid (3,1);
        \matrix[matrix of nodes,nodes={inner sep=0pt,text width=.5cm,align=center,minimum height=.5cm}]{
            0 & 5 & 10 & 15 & 16 & 21 & 26 & 31 & 32 & 37 & 42 & 47 \\
            1 & 4 & 11 & 14 & 17 & 20 & 27 & 30 & 33 & 36 & 43 & 46 \\
            2 & 7 &  8 & 13 & 18 & 23 & 24 & 29 & 34 & 39 & 40 & 45 \\
            3 & 6 &  9 & 12 & 19 & 22 & 25 & 28 & 35 & 38 & 41 & 44 \\};
    \end{tikzpicture}
    \subcaption{\tabular[t]{@{}l@{}}Binary Swizzle \\ $(4,(4,3)):(f_1,(f_5,f_{16}))$ \\ $f_i \in F_2 = (\Z,XOR,\cdot)$\endtabular}
    \label{fig:layouts:swizzle}
\end{subfigure}
\caption{Examples of layouts plotted with two-dimensional coordinates. These layouts use strides from integer-semimodules to define functions that are useful in tracking coordinates and ``swizzling" offsets in non-affine patterns.}
\label{fig:layouts2}
\end{figure}

Figure~\ref{fig:layouts2} illustrates examples of layouts constructed with integer-semimodule strides rather than integer strides. Figures~\ref{fig:layouts:coord0} and~\ref{fig:layouts:coord1} demonstrate layouts that generate coordinates. In future sections, we will discover that these coordinate layouts transform symmetrically to their data layout counterparts and often serve as utilities for detecting and predicating out-of-bounds accesses into data tensors. Additionally, coordinate tensors have proven very useful for instructions like TMA in Hopper and Blackwell that consume coordinates as instruction arguments rather than addresses. Figure~\ref{fig:layouts:swizzle} uses an integer-semimodule where group addition is replaced with binary XOR. This can be used to generate so-called swizzle patterns of data which are useful to prevent bank conflicts in read and write access patterns to shared memory. These examples highlight the uniformity of \CuTe's layout concept and the generality of functions it can be used to represent.

\subsubsection{Completeness}

Every function $f$ with $f(0) = 0$ and finite domain $\Z_N$ can be represented as the functional composition of the finite sequence of \CuTe layouts. This means that \CuTe layouts are a generating set under functional composition. Such a function $f$ can be represented as sequence of compositions
\begin{align*}
f &\equiv (2,2,2,\ldots):(f(1),f(2),f(3),\ldots) \ \circ \ (3,1):(1,4) \ \circ \ (4,1):(1,6) \ \circ \ \cdots \ \circ \ (N-1,1):(1,2(N-2)).
\end{align*}
For all $i \in \Z_N / \{0\}$, the rightmost $N-3$ layouts map $i \to 2^{i-1}$ and the leftmost layout maps $2^{i-1} \to f(i)$. Note that we've used the extended domain, $\Z$, rather than the in-bounds domain, $\Z_N$, to evaluate the intermediate layouts above.

\subsubsection{Semi-Linearity}

The shape-stride definition, $\v{L} = D \circ S$, along with generalized integer-module strides yield a particularly useful linear-algebraic view of layout functions,
\begin{align}
\v{L}(c) = (D \circ S)(c) = d \cdot S(c) = d \cdot \tilde{c}.
\end{align}
The shape function is a convenient semi-affine bijection into the natural coordinates, $\tilde{c} \in \Z^S$, and the stride function is a linear function of the natural coordinates. Indeed, for two natural coordinates $\tilde{c}_0, \tilde{c}_1 \in \Z^S$, the layout function is linear,
\begin{align*}
\v{L}(\alpha \tilde{c}_0 + \beta \tilde{c}_1) = d \cdot (\alpha \tilde{c}_0 + \beta \tilde{c}_1) = \alpha (d \cdot \tilde{c}_0) + \beta (d \cdot \tilde{c}_1) = \alpha \v{L}(\tilde{c}_0) + \beta \v{L}(\tilde{c}_1),
\end{align*}
because the shape function is the identity and the stride function is linear. Note that for arbitrary coordinates $c_0, c_1 \in \Z(S)$, the layout is not linear because the shape function is not linear,
\begin{align*}
\tilde{c} = S(c_0 + c_1) \neq S(c_0) + S(c_1) = \tilde{c}_0 + \tilde{c}_1.
\end{align*}
As a result, the layout function is linear in the natural coordinates, $\tilde{c} \in \Z^S$, but nonlinear in the arbitrary coordinates, $c \in \Z(S)$.

In the natural coordinates, $d \cdot \tilde{c}$ can be interpreted as a generalized matrix-vector product,
\begin{align*}
\v{L}(c) = d \cdot \tilde{c} = \v{D} \ \tilde{c}
\end{align*}
where $\v{D}$ is a matrix with elements selected from an integer-semimodule $\D$. In the most common case with integer strides, $\D = \Z$, this is a matrix-vector product with $\v{D} \in \Z^{1 \times n}$. When strides are selected from the coordinate integer-semimodule, $\D = (\Z^S, +, \cdot)$, this is a matrix-vector product with $\v{D} \in \Z^{m \times n}$. When strides are binary sequences, $\D = F_2^m = (\Z_{2^m}, XOR, \cdot)$, this is a matrix-vector product with $\v{D} \in \mathbb{F}_2^{m \times n}$.

\begin{table}[ht]
\begin{center}
\begin{tabular}{|>{$}c<{$}|>{$}c<{$}|c|}
\v{L} & \text{Linear Form: } r = \v{D} \ \tilde{c} & Comment \\ \hline \hline
((2,2),(4,\ 2)):((1,8),(2,16)) & r = \begin{bmatrix}1 & 8 & 2 & 16\end{bmatrix} \begin{bmatrix} \tilde{c}_0 \\ \tilde{c}_1 \\ \tilde{c}_2 \\ \tilde{c}_3 \end{bmatrix} & \makecell{Integer strides are columns \\ of $1 {\times} n$ $\Z$-matrix} \\ \hline
(4,(4,2)):(e_1,(e_0,6e_1)) & \begin{bmatrix}r_0 \\ r_1\end{bmatrix} = \begin{bmatrix}0 & 1 & 0 \\ 1 & 0 & 6 \end{bmatrix} \begin{bmatrix} \tilde{c}_0 \\ \tilde{c}_1 \\ \tilde{c}_2 \end{bmatrix} &
\makecell{Coordinate strides are columns \\ of $m {\times} n$ $\Z$-matrix} \\ \hline
\makecell{(4,4) : (f_1,f_5) \\ \equiv ((2,2),(2,2)):((f_1,f_2),(f_5,f_{10}))} &
\footnotesize
\begin{bmatrix}r_0 \\ r_1 \\ r_2 \\ r_3\end{bmatrix} =
\begin{bmatrix}
1 & 0 & 1 & 0 \\
0 & 1 & 0 & 1 \\
0 & 0 & 1 & 0 \\
0 & 0 & 0 & 1 \\
\end{bmatrix}
\begin{bmatrix} \tilde{c}_0 \\ \tilde{c}_1 \\ \tilde{c}_2 \\ \tilde{c}_3 \end{bmatrix} &
\makecell{Binary strides are columns \\ of $m {\times} n$ $\mathbb{F}_2$-matrix} \\ \hline
\end{tabular}
\end{center}
\caption{Layouts and their associated linear forms.}
\label{tab:linear_forms}
\end{table}

For instance, Table~\ref{tab:linear_forms} shows the linear forms of an integer layout, a coordinate layout, and a binary layout. Well-studied transformations of the $\mathbb{F}_2$ linear forms include the Bit Permute Complement (BCP) and Bit Matrix Multiply Complement (BMMC) transforms~\cite{Edelman:IndexTransforms,Cormen:FastPermuting,Bouverot:AffineIndex}, which have recently found significant application to SIMT GPU programming~\cite{Tillet:LinearLayouts}. In those works, the transforms under consideration are
\begin{align}
f(\v{v}) = \v{A} \v{v} + \v{b}
\label{eqn:linear}
\end{align}
where $\v{A}$ is an $m {\times} n$ matrix of binary elements, $\v{v}$ is a binary vector of length $n$, $\v{b}$ is a binary vector of length $m$, and all arithmetic is performed modulo 2 (in the finite field with two elements, $\mathbb{F}_2$).

Subsequent sections of this paper define algebraic operations on \CuTe layouts that are generalized from linear algebra. Group composition on \CuTe layouts can be interpreted as a generalization of matrix-multiplication. Right-inverses and left-inverses of \CuTe layouts can be interpreted as Moore-Penrose pseudo-inverses in linear algebra. Indeed, similar linear algebraic expressions can be found in BCP and BMMC analysis~\cite{Edelman:IndexTransforms,Cormen:FastPermuting,Bouverot:AffineIndex,Tillet:LinearLayouts}, which all use factorizations, inverses, and matrix products in analysis of algorithms. The \CuTe layout algebra can be interpreted as a generalization of the linear algebraic BCP and BMMC operations beyond the $\mathbb{F}_2$ field. In particular, the \CuTe layout algebra was principally motivated by the general integer strides found in exotic data layouts for tensors. In this work, we aim to demonstrate that more general functional definitions of these operations are often computable for \CuTe layouts and useful for a broad range of analyses and applications.

\subsection{Tensor}
\label{sec:tensor}

Finally, we define tensors which are the central object of \CuTe by binding a layout to an accessor. The accessor is effectively a random-access, pointer-like object for the layout.

\begin{definition}
An {\em accessor} is an object that supports offset and dereference operations:
\begin{align*}
e + d &\to e', &&\text{{\bf offset} accessor $e$ by $d \in \D$ to produce another accessor $e'$;} \\
*e &\to v, &&\text{{\bf dereference} accessor $e$ to produce value $v$;} \\
e[d] &\to *(e+d), &&\text{{\bf subscript} operator as a common convenience.}
\end{align*}
\end{definition}
When $\D = \Z$, common implementations of an accessor include raw pointers (e.g., \texttt{T*}), arrays (e.g. \texttt{T[N]}), and random-access iterators (e.g. \texttt{thrust::counting\_iterator} and \texttt{thrust::transform\_iterator}, etc). When $\D$ is a more exotic integer-semimodule, the accessor must be compatible with the semigroup's addition operation.

\begin{definition}
A {\em tensor} is defined by the composition of an accessor, $e$, with a layout, $\v{L}$, expressed as $T = e \circ \v{L}$. A tensor evaluates the layout by mapping a coordinate $c \in \Z(\v{L})$ to the codomain $\D$, offsets the accessor accordingly, and dereferences the result to obtain the tensor's value. Formally,
\begin{align}
T(c) = (e \circ \v{L})(c) = *(e + \v{L}(c)) = e[\v{L}(c)],
\label{eqn:tensor}
\end{align}
yields the value of the tensor at coordinate $c$.
\end{definition}

Most tensors are data layouts that use a memory address as an accessor. For instance, a memory address $p$ can be used as a pointer accessor $\{p\}$ with normal offset and dereference operations to construct a data tensor,
\begin{align*}
\{p\} + b &\to \{p+b\} \\
*\{p\} &\to *p \\
T &= \{p\} \circ \v{L}.
\end{align*}
In addition, all layouts illustrated in Figure~\ref{fig:layouts} can be transformed into implicit tensors by composing them with a counting iterator $\{a\}$ which dereferences to a stored offset $a \in \Z$,
\begin{align*}
\{a\} + b &\to \{a+b\} \\
*\{a\} &\to a \\
T &= \{a\} \circ \v{L}.
\end{align*}
Similarly, the coordinate layouts illustrated in Figure~\ref{fig:layouts:coord0} and~\ref{fig:layouts:coord1} can be transformed into implicit tensors by composing them with an accessor $\{(a,b)\}$ that offsets with coordinates and dereferences to a stored coordinate $(a,b) \in \Z^{(\star,\star)}$,
\begin{align*}
\{(a,b)\} + (c,d) &\to \{(a+c,b+d)\} \\
*\{(a,b)\} &\to (a,b) \\
T &= \{(a,b)\} \circ \v{L}.
\end{align*}



\subsubsection{Slicing}
\label{sec:slicing}

A tensor may be either {\em fully evaluated} or {\em partially evaluated} through \emph{slicing}. As a \CuTe Tensor can be thought of as a Layout with an offset, arbitrary slicing can be performed along any mode(s) of a natural coordinate.

\begin{itemize}
\item \textbf{Full evaluation}: Applying Eq.~\eqref{eqn:tensor} with a complete coordinate $c$ results in a value.
\item \textbf{Partial evaluation (Slicing)}: When slicing with an incomplete coordinate $c = c' + c^*$, where $c^*$ represents the unspecified portion of $c$, the result is a new tensor. The operation is expressed as:
\begin{align}
T(c) = (e \circ \v{L})(c'+ c^*) = (e + \v{L}(c')) \circ \v{L}(c^*) = e' \circ \v{L}(c^*) = T'(c^*),
\end{align}
\end{itemize}
where $\v{L}(c')$ can be fully evaluated and accumulated into $e$ and $\v{L}(c^*)$ is the sublayout of $\v{L}$ that remains unevaluated. Slicing creates a sub-tensor that can be further evaluated or manipulated.

\begin{figure}[p]
\centering
\scalebox{0.6}{
\begin{tikzpicture}[x={(0cm,-1cm)},y={(1cm,0cm)},every node/.style={minimum size=1cm, outer sep=0pt}]
\node[fill=white] at (0,0) {\Large 0};
\node[fill=white] at (0,1) {\Large 2};
\node[fill=white] at (0,2) {\Large 15};
\node[fill=white] at (0,3) {\Large 17};
\node[fill=white] at (0,4) {\Large 30};
\node[fill=white] at (0,5) {\Large 32};
\node[fill=white] at (0,6) {\Large 100};
\node[fill=white] at (0,7) {\Large 102};
\node[fill=white] at (0,8) {\Large 115};
\node[fill=white] at (0,9) {\Large 117};
\node[fill=white] at (0,10) {\Large 130};
\node[fill=white] at (0,11) {\Large 132};
\node[fill=white] at (1,0) {\Large 4};
\node[fill=white] at (1,1) {\Large 6};
\node[fill=white] at (1,2) {\Large 19};
\node[fill=white] at (1,3) {\Large 21};
\node[fill=white] at (1,4) {\Large 34};
\node[fill=white] at (1,5) {\Large 36};
\node[fill=white] at (1,6) {\Large 104};
\node[fill=white] at (1,7) {\Large 106};
\node[fill=white] at (1,8) {\Large 119};
\node[fill=white] at (1,9) {\Large 121};
\node[fill=white] at (1,10) {\Large 134};
\node[fill=white] at (1,11) {\Large 136};
\node[fill=white] at (2,0) {\Large 8};
\node[fill=white] at (2,1) {\Large 10};
\node[fill=white] at (2,2) {\Large 23};
\node[fill=white] at (2,3) {\Large 25};
\node[fill=white] at (2,4) {\Large 38};
\node[fill=white] at (2,5) {\Large 40};
\node[fill=white] at (2,6) {\Large 108};
\node[fill=white] at (2,7) {\Large 110};
\node[fill=white] at (2,8) {\Large 123};
\node[fill=white] at (2,9) {\Large 125};
\node[fill=white] at (2,10) {\Large 138};
\node[fill=white] at (2,11) {\Large 140};
\node[fill=white] at (3,0) {\Large 1};
\node[fill=white] at (3,1) {\Large 3};
\node[fill=white] at (3,2) {\Large 16};
\node[fill=white] at (3,3) {\Large 18};
\node[fill=white] at (3,4) {\Large 31};
\node[fill=white] at (3,5) {\Large 33};
\node[fill=white] at (3,6) {\Large 101};
\node[fill=white] at (3,7) {\Large 103};
\node[fill=white] at (3,8) {\Large 116};
\node[fill=white] at (3,9) {\Large 118};
\node[fill=white] at (3,10) {\Large 131};
\node[fill=white] at (3,11) {\Large 133};
\node[fill=white] at (4,0) {\Large 5};
\node[fill=white] at (4,1) {\Large 7};
\node[fill=white] at (4,2) {\Large 20};
\node[fill=white] at (4,3) {\Large 22};
\node[fill=white] at (4,4) {\Large 35};
\node[fill=white] at (4,5) {\Large 37};
\node[fill=white] at (4,6) {\Large 105};
\node[fill=white] at (4,7) {\Large 107};
\node[fill=white] at (4,8) {\Large 120};
\node[fill=white] at (4,9) {\Large 122};
\node[fill=white] at (4,10) {\Large 135};
\node[fill=white] at (4,11) {\Large 137};
\node[fill=white] at (5,0) {\Large 9};
\node[fill=white] at (5,1) {\Large 11};
\node[fill=white] at (5,2) {\Large 24};
\node[fill=white] at (5,3) {\Large 26};
\node[fill=white] at (5,4) {\Large 39};
\node[fill=white] at (5,5) {\Large 41};
\node[fill=white] at (5,6) {\Large 109};
\node[fill=white] at (5,7) {\Large 111};
\node[fill=white] at (5,8) {\Large 124};
\node[fill=white] at (5,9) {\Large 126};
\node[fill=white] at (5,10) {\Large 139};
\node[fill=white] at (5,11) {\Large 141};
\draw[color=black,thick,shift={(-0.5,-0.5)}] (0,0) grid (6,12);
\node at (0,-1) {\Large{\texttt{0}}};
\node at (1,-1) {\Large{\texttt{1}}};
\node at (2,-1) {\Large{\texttt{2}}};
\node at (3,-1) {\Large{\texttt{3}}};
\node at (4,-1) {\Large{\texttt{4}}};
\node at (5,-1) {\Large{\texttt{5}}};
\node at (-1,0) {\Large{\texttt{0}}};
\node at (-1,1) {\Large{\texttt{1}}};
\node at (-1,2) {\Large{\texttt{2}}};
\node at (-1,3) {\Large{\texttt{3}}};
\node at (-1,4) {\Large{\texttt{4}}};
\node at (-1,5) {\Large{\texttt{5}}};
\node at (-1,6) {\Large{\texttt{6}}};
\node at (-1,7) {\Large{\texttt{7}}};
\node at (-1,8) {\Large{\texttt{8}}};
\node at (-1,9) {\Large{\texttt{9}}};
\node at (-1,10) {\Large{\texttt{10}}};
\node at (-1,11) {\Large{\texttt{11}}};
\node at (6.25,5.5) {\huge Tensor $\v{A} = \{0\} \circ ((3,2),((2,3),2)):((4,1),((2,15),100))$};
\end{tikzpicture}
}\\
\vspace{1em}
\begin{tabular}{|>{\centering\arraybackslash}m{3cm}|>{\centering\arraybackslash}m{5cm}|>{\centering\arraybackslash}m{5cm}|}
Slice & Sliced Tensor & Slice Diagram \\
\hline\hline
$\v{A}(2,\_)$ & $\{8\} \circ ((2,3),2):((2,15),100)$ &
\scalebox{0.4}{
\begin{adjustbox}{margin=0.5cm}
\scalebox{0.8}{
\begin{tikzpicture}[x={(0cm,-1cm)},y={(1cm,0cm)},every node/.style={minimum size=1cm, outer sep=0pt}]
\node[fill=white] at (0,0) {\Large 0};
\node[fill=white] at (0,1) {\Large 2};
\node[fill=white] at (0,2) {\Large 15};
\node[fill=white] at (0,3) {\Large 17};
\node[fill=white] at (0,4) {\Large 30};
\node[fill=white] at (0,5) {\Large 32};
\node[fill=white] at (0,6) {\Large 100};
\node[fill=white] at (0,7) {\Large 102};
\node[fill=white] at (0,8) {\Large 115};
\node[fill=white] at (0,9) {\Large 117};
\node[fill=white] at (0,10) {\Large 130};
\node[fill=white] at (0,11) {\Large 132};
\node[fill=white] at (1,0) {\Large 4};
\node[fill=white] at (1,1) {\Large 6};
\node[fill=white] at (1,2) {\Large 19};
\node[fill=white] at (1,3) {\Large 21};
\node[fill=white] at (1,4) {\Large 34};
\node[fill=white] at (1,5) {\Large 36};
\node[fill=white] at (1,6) {\Large 104};
\node[fill=white] at (1,7) {\Large 106};
\node[fill=white] at (1,8) {\Large 119};
\node[fill=white] at (1,9) {\Large 121};
\node[fill=white] at (1,10) {\Large 134};
\node[fill=white] at (1,11) {\Large 136};
\node[fill=black!30] at (2,0) {\Large 8};
\node[fill=black!30] at (2,1) {\Large 10};
\node[fill=black!30] at (2,2) {\Large 23};
\node[fill=black!30] at (2,3) {\Large 25};
\node[fill=black!30] at (2,4) {\Large 38};
\node[fill=black!30] at (2,5) {\Large 40};
\node[fill=black!30] at (2,6) {\Large 108};
\node[fill=black!30] at (2,7) {\Large 110};
\node[fill=black!30] at (2,8) {\Large 123};
\node[fill=black!30] at (2,9) {\Large 125};
\node[fill=black!30] at (2,10) {\Large 138};
\node[fill=black!30] at (2,11) {\Large 140};
\node[fill=white] at (3,0) {\Large 1};
\node[fill=white] at (3,1) {\Large 3};
\node[fill=white] at (3,2) {\Large 16};
\node[fill=white] at (3,3) {\Large 18};
\node[fill=white] at (3,4) {\Large 31};
\node[fill=white] at (3,5) {\Large 33};
\node[fill=white] at (3,6) {\Large 101};
\node[fill=white] at (3,7) {\Large 103};
\node[fill=white] at (3,8) {\Large 116};
\node[fill=white] at (3,9) {\Large 118};
\node[fill=white] at (3,10) {\Large 131};
\node[fill=white] at (3,11) {\Large 133};
\node[fill=white] at (4,0) {\Large 5};
\node[fill=white] at (4,1) {\Large 7};
\node[fill=white] at (4,2) {\Large 20};
\node[fill=white] at (4,3) {\Large 22};
\node[fill=white] at (4,4) {\Large 35};
\node[fill=white] at (4,5) {\Large 37};
\node[fill=white] at (4,6) {\Large 105};
\node[fill=white] at (4,7) {\Large 107};
\node[fill=white] at (4,8) {\Large 120};
\node[fill=white] at (4,9) {\Large 122};
\node[fill=white] at (4,10) {\Large 135};
\node[fill=white] at (4,11) {\Large 137};
\node[fill=white] at (5,0) {\Large 9};
\node[fill=white] at (5,1) {\Large 11};
\node[fill=white] at (5,2) {\Large 24};
\node[fill=white] at (5,3) {\Large 26};
\node[fill=white] at (5,4) {\Large 39};
\node[fill=white] at (5,5) {\Large 41};
\node[fill=white] at (5,6) {\Large 109};
\node[fill=white] at (5,7) {\Large 111};
\node[fill=white] at (5,8) {\Large 124};
\node[fill=white] at (5,9) {\Large 126};
\node[fill=white] at (5,10) {\Large 139};
\node[fill=white] at (5,11) {\Large 141};
\draw[color=black,thick,shift={(-0.5,-0.5)}] (0,0) grid (6,12);
\end{tikzpicture}
}
\end{adjustbox}
} \\
\hline
$\v{A}(\_,5)$ & $\{32\} \circ (3,2):(4,1)$ &
\scalebox{0.4}{
\begin{adjustbox}{margin=0.5cm}
\scalebox{0.8}{
\begin{tikzpicture}[x={(0cm,-1cm)},y={(1cm,0cm)},every node/.style={minimum size=1cm, outer sep=0pt}]
\node[fill=white] at (0,0) {\Large 0};
\node[fill=white] at (0,1) {\Large 2};
\node[fill=white] at (0,2) {\Large 15};
\node[fill=white] at (0,3) {\Large 17};
\node[fill=white] at (0,4) {\Large 30};
\node[fill=black!30] at (0,5) {\Large 32};
\node[fill=white] at (0,6) {\Large 100};
\node[fill=white] at (0,7) {\Large 102};
\node[fill=white] at (0,8) {\Large 115};
\node[fill=white] at (0,9) {\Large 117};
\node[fill=white] at (0,10) {\Large 130};
\node[fill=white] at (0,11) {\Large 132};
\node[fill=white] at (1,0) {\Large 4};
\node[fill=white] at (1,1) {\Large 6};
\node[fill=white] at (1,2) {\Large 19};
\node[fill=white] at (1,3) {\Large 21};
\node[fill=white] at (1,4) {\Large 34};
\node[fill=black!30] at (1,5) {\Large 36};
\node[fill=white] at (1,6) {\Large 104};
\node[fill=white] at (1,7) {\Large 106};
\node[fill=white] at (1,8) {\Large 119};
\node[fill=white] at (1,9) {\Large 121};
\node[fill=white] at (1,10) {\Large 134};
\node[fill=white] at (1,11) {\Large 136};
\node[fill=white] at (2,0) {\Large 8};
\node[fill=white] at (2,1) {\Large 10};
\node[fill=white] at (2,2) {\Large 23};
\node[fill=white] at (2,3) {\Large 25};
\node[fill=white] at (2,4) {\Large 38};
\node[fill=black!30] at (2,5) {\Large 40};
\node[fill=white] at (2,6) {\Large 108};
\node[fill=white] at (2,7) {\Large 110};
\node[fill=white] at (2,8) {\Large 123};
\node[fill=white] at (2,9) {\Large 125};
\node[fill=white] at (2,10) {\Large 138};
\node[fill=white] at (2,11) {\Large 140};
\node[fill=white] at (3,0) {\Large 1};
\node[fill=white] at (3,1) {\Large 3};
\node[fill=white] at (3,2) {\Large 16};
\node[fill=white] at (3,3) {\Large 18};
\node[fill=white] at (3,4) {\Large 31};
\node[fill=black!30] at (3,5) {\Large 33};
\node[fill=white] at (3,6) {\Large 101};
\node[fill=white] at (3,7) {\Large 103};
\node[fill=white] at (3,8) {\Large 116};
\node[fill=white] at (3,9) {\Large 118};
\node[fill=white] at (3,10) {\Large 131};
\node[fill=white] at (3,11) {\Large 133};
\node[fill=white] at (4,0) {\Large 5};
\node[fill=white] at (4,1) {\Large 7};
\node[fill=white] at (4,2) {\Large 20};
\node[fill=white] at (4,3) {\Large 22};
\node[fill=white] at (4,4) {\Large 35};
\node[fill=black!30] at (4,5) {\Large 37};
\node[fill=white] at (4,6) {\Large 105};
\node[fill=white] at (4,7) {\Large 107};
\node[fill=white] at (4,8) {\Large 120};
\node[fill=white] at (4,9) {\Large 122};
\node[fill=white] at (4,10) {\Large 135};
\node[fill=white] at (4,11) {\Large 137};
\node[fill=white] at (5,0) {\Large 9};
\node[fill=white] at (5,1) {\Large 11};
\node[fill=white] at (5,2) {\Large 24};
\node[fill=white] at (5,3) {\Large 26};
\node[fill=white] at (5,4) {\Large 39};
\node[fill=black!30] at (5,5) {\Large 41};
\node[fill=white] at (5,6) {\Large 109};
\node[fill=white] at (5,7) {\Large 111};
\node[fill=white] at (5,8) {\Large 124};
\node[fill=white] at (5,9) {\Large 126};
\node[fill=white] at (5,10) {\Large 139};
\node[fill=white] at (5,11) {\Large 141};
\draw[color=black,thick,shift={(-0.5,-0.5)}] (0,0) grid (6,12);
\end{tikzpicture}
}
\end{adjustbox}
} \\
\hline
$\v{A}(2,((0,\_),\_))$ & $\{8\} \circ (3,2):(15,100)$ &
\scalebox{0.4}{
\begin{adjustbox}{margin=0.5cm}
\scalebox{0.8}{
\begin{tikzpicture}[x={(0cm,-1cm)},y={(1cm,0cm)},every node/.style={minimum size=1cm, outer sep=0pt}]
\node[fill=white] at (0,0) {\Large 0};
\node[fill=white] at (0,1) {\Large 2};
\node[fill=white] at (0,2) {\Large 15};
\node[fill=white] at (0,3) {\Large 17};
\node[fill=white] at (0,4) {\Large 30};
\node[fill=white] at (0,5) {\Large 32};
\node[fill=white] at (0,6) {\Large 100};
\node[fill=white] at (0,7) {\Large 102};
\node[fill=white] at (0,8) {\Large 115};
\node[fill=white] at (0,9) {\Large 117};
\node[fill=white] at (0,10) {\Large 130};
\node[fill=white] at (0,11) {\Large 132};
\node[fill=white] at (1,0) {\Large 4};
\node[fill=white] at (1,1) {\Large 6};
\node[fill=white] at (1,2) {\Large 19};
\node[fill=white] at (1,3) {\Large 21};
\node[fill=white] at (1,4) {\Large 34};
\node[fill=white] at (1,5) {\Large 36};
\node[fill=white] at (1,6) {\Large 104};
\node[fill=white] at (1,7) {\Large 106};
\node[fill=white] at (1,8) {\Large 119};
\node[fill=white] at (1,9) {\Large 121};
\node[fill=white] at (1,10) {\Large 134};
\node[fill=white] at (1,11) {\Large 136};
\node[fill=black!30] at (2,0) {\Large 8};
\node[fill=white] at (2,1) {\Large 10};
\node[fill=black!30] at (2,2) {\Large 23};
\node[fill=white] at (2,3) {\Large 25};
\node[fill=black!30] at (2,4) {\Large 38};
\node[fill=white] at (2,5) {\Large 40};
\node[fill=black!30] at (2,6) {\Large 108};
\node[fill=white] at (2,7) {\Large 110};
\node[fill=black!30] at (2,8) {\Large 123};
\node[fill=white] at (2,9) {\Large 125};
\node[fill=black!30] at (2,10) {\Large 138};
\node[fill=white] at (2,11) {\Large 140};
\node[fill=white] at (3,0) {\Large 1};
\node[fill=white] at (3,1) {\Large 3};
\node[fill=white] at (3,2) {\Large 16};
\node[fill=white] at (3,3) {\Large 18};
\node[fill=white] at (3,4) {\Large 31};
\node[fill=white] at (3,5) {\Large 33};
\node[fill=white] at (3,6) {\Large 101};
\node[fill=white] at (3,7) {\Large 103};
\node[fill=white] at (3,8) {\Large 116};
\node[fill=white] at (3,9) {\Large 118};
\node[fill=white] at (3,10) {\Large 131};
\node[fill=white] at (3,11) {\Large 133};
\node[fill=white] at (4,0) {\Large 5};
\node[fill=white] at (4,1) {\Large 7};
\node[fill=white] at (4,2) {\Large 20};
\node[fill=white] at (4,3) {\Large 22};
\node[fill=white] at (4,4) {\Large 35};
\node[fill=white] at (4,5) {\Large 37};
\node[fill=white] at (4,6) {\Large 105};
\node[fill=white] at (4,7) {\Large 107};
\node[fill=white] at (4,8) {\Large 120};
\node[fill=white] at (4,9) {\Large 122};
\node[fill=white] at (4,10) {\Large 135};
\node[fill=white] at (4,11) {\Large 137};
\node[fill=white] at (5,0) {\Large 9};
\node[fill=white] at (5,1) {\Large 11};
\node[fill=white] at (5,2) {\Large 24};
\node[fill=white] at (5,3) {\Large 26};
\node[fill=white] at (5,4) {\Large 39};
\node[fill=white] at (5,5) {\Large 41};
\node[fill=white] at (5,6) {\Large 109};
\node[fill=white] at (5,7) {\Large 111};
\node[fill=white] at (5,8) {\Large 124};
\node[fill=white] at (5,9) {\Large 126};
\node[fill=white] at (5,10) {\Large 139};
\node[fill=white] at (5,11) {\Large 141};
\draw[color=black,thick,shift={(-0.5,-0.5)}] (0,0) grid (6,12);
\end{tikzpicture}
}
\end{adjustbox}
} \\
\hline
$\v{A}((\_,1),(\_,0))$ & $\{1\} \circ (3,(2,3)):(4,(2,15))$ &
\scalebox{0.4}{
\begin{adjustbox}{margin=0.5cm}
\scalebox{0.8}{
\begin{tikzpicture}[x={(0cm,-1cm)},y={(1cm,0cm)},every node/.style={minimum size=1cm, outer sep=0pt}]
\node[fill=white] at (0,0) {\Large 0};
\node[fill=white] at (0,1) {\Large 2};
\node[fill=white] at (0,2) {\Large 15};
\node[fill=white] at (0,3) {\Large 17};
\node[fill=white] at (0,4) {\Large 30};
\node[fill=white] at (0,5) {\Large 32};
\node[fill=white] at (0,6) {\Large 100};
\node[fill=white] at (0,7) {\Large 102};
\node[fill=white] at (0,8) {\Large 115};
\node[fill=white] at (0,9) {\Large 117};
\node[fill=white] at (0,10) {\Large 130};
\node[fill=white] at (0,11) {\Large 132};
\node[fill=white] at (1,0) {\Large 4};
\node[fill=white] at (1,1) {\Large 6};
\node[fill=white] at (1,2) {\Large 19};
\node[fill=white] at (1,3) {\Large 21};
\node[fill=white] at (1,4) {\Large 34};
\node[fill=white] at (1,5) {\Large 36};
\node[fill=white] at (1,6) {\Large 104};
\node[fill=white] at (1,7) {\Large 106};
\node[fill=white] at (1,8) {\Large 119};
\node[fill=white] at (1,9) {\Large 121};
\node[fill=white] at (1,10) {\Large 134};
\node[fill=white] at (1,11) {\Large 136};
\node[fill=white] at (2,0) {\Large 8};
\node[fill=white] at (2,1) {\Large 10};
\node[fill=white] at (2,2) {\Large 23};
\node[fill=white] at (2,3) {\Large 25};
\node[fill=white] at (2,4) {\Large 38};
\node[fill=white] at (2,5) {\Large 40};
\node[fill=white] at (2,6) {\Large 108};
\node[fill=white] at (2,7) {\Large 110};
\node[fill=white] at (2,8) {\Large 123};
\node[fill=white] at (2,9) {\Large 125};
\node[fill=white] at (2,10) {\Large 138};
\node[fill=white] at (2,11) {\Large 140};
\node[fill=black!30] at (3,0) {\Large 1};
\node[fill=black!30] at (3,1) {\Large 3};
\node[fill=black!30] at (3,2) {\Large 16};
\node[fill=black!30] at (3,3) {\Large 18};
\node[fill=black!30] at (3,4) {\Large 31};
\node[fill=black!30] at (3,5) {\Large 33};
\node[fill=white] at (3,6) {\Large 101};
\node[fill=white] at (3,7) {\Large 103};
\node[fill=white] at (3,8) {\Large 116};
\node[fill=white] at (3,9) {\Large 118};
\node[fill=white] at (3,10) {\Large 131};
\node[fill=white] at (3,11) {\Large 133};
\node[fill=black!30] at (4,0) {\Large 5};
\node[fill=black!30] at (4,1) {\Large 7};
\node[fill=black!30] at (4,2) {\Large 20};
\node[fill=black!30] at (4,3) {\Large 22};
\node[fill=black!30] at (4,4) {\Large 35};
\node[fill=black!30] at (4,5) {\Large 37};
\node[fill=white] at (4,6) {\Large 105};
\node[fill=white] at (4,7) {\Large 107};
\node[fill=white] at (4,8) {\Large 120};
\node[fill=white] at (4,9) {\Large 122};
\node[fill=white] at (4,10) {\Large 135};
\node[fill=white] at (4,11) {\Large 137};
\node[fill=black!30] at (5,0) {\Large 9};
\node[fill=black!30] at (5,1) {\Large 11};
\node[fill=black!30] at (5,2) {\Large 24};
\node[fill=black!30] at (5,3) {\Large 26};
\node[fill=black!30] at (5,4) {\Large 39};
\node[fill=black!30] at (5,5) {\Large 41};
\node[fill=white] at (5,6) {\Large 109};
\node[fill=white] at (5,7) {\Large 111};
\node[fill=white] at (5,8) {\Large 124};
\node[fill=white] at (5,9) {\Large 126};
\node[fill=white] at (5,10) {\Large 139};
\node[fill=white] at (5,11) {\Large 141};
\draw[color=black,thick,shift={(-0.5,-0.5)}] (0,0) grid (6,12);
\end{tikzpicture}
}
\end{adjustbox}
} \\
\hline
$\v{A}((\_,0),((0,\_),1))$ & $\{100\} \circ (3,3):(4,15)$ &
\scalebox{0.4}{
\begin{adjustbox}{margin=0.5cm}
\scalebox{0.8}{
\begin{tikzpicture}[x={(0cm,-1cm)},y={(1cm,0cm)},every node/.style={minimum size=1cm, outer sep=0pt}]
\node[fill=white] at (0,0) {\Large 0};
\node[fill=white] at (0,1) {\Large 2};
\node[fill=white] at (0,2) {\Large 15};
\node[fill=white] at (0,3) {\Large 17};
\node[fill=white] at (0,4) {\Large 30};
\node[fill=white] at (0,5) {\Large 32};
\node[fill=black!30] at (0,6) {\Large 100};
\node[fill=white] at (0,7) {\Large 102};
\node[fill=black!30] at (0,8) {\Large 115};
\node[fill=white] at (0,9) {\Large 117};
\node[fill=black!30] at (0,10) {\Large 130};
\node[fill=white] at (0,11) {\Large 132};
\node[fill=white] at (1,0) {\Large 4};
\node[fill=white] at (1,1) {\Large 6};
\node[fill=white] at (1,2) {\Large 19};
\node[fill=white] at (1,3) {\Large 21};
\node[fill=white] at (1,4) {\Large 34};
\node[fill=white] at (1,5) {\Large 36};
\node[fill=black!30] at (1,6) {\Large 104};
\node[fill=white] at (1,7) {\Large 106};
\node[fill=black!30] at (1,8) {\Large 119};
\node[fill=white] at (1,9) {\Large 121};
\node[fill=black!30] at (1,10) {\Large 134};
\node[fill=white] at (1,11) {\Large 136};
\node[fill=white] at (2,0) {\Large 8};
\node[fill=white] at (2,1) {\Large 10};
\node[fill=white] at (2,2) {\Large 23};
\node[fill=white] at (2,3) {\Large 25};
\node[fill=white] at (2,4) {\Large 38};
\node[fill=white] at (2,5) {\Large 40};
\node[fill=black!30] at (2,6) {\Large 108};
\node[fill=white] at (2,7) {\Large 110};
\node[fill=black!30] at (2,8) {\Large 123};
\node[fill=white] at (2,9) {\Large 125};
\node[fill=black!30] at (2,10) {\Large 138};
\node[fill=white] at (2,11) {\Large 140};
\node[fill=white] at (3,0) {\Large 1};
\node[fill=white] at (3,1) {\Large 3};
\node[fill=white] at (3,2) {\Large 16};
\node[fill=white] at (3,3) {\Large 18};
\node[fill=white] at (3,4) {\Large 31};
\node[fill=white] at (3,5) {\Large 33};
\node[fill=white] at (3,6) {\Large 101};
\node[fill=white] at (3,7) {\Large 103};
\node[fill=white] at (3,8) {\Large 116};
\node[fill=white] at (3,9) {\Large 118};
\node[fill=white] at (3,10) {\Large 131};
\node[fill=white] at (3,11) {\Large 133};
\node[fill=white] at (4,0) {\Large 5};
\node[fill=white] at (4,1) {\Large 7};
\node[fill=white] at (4,2) {\Large 20};
\node[fill=white] at (4,3) {\Large 22};
\node[fill=white] at (4,4) {\Large 35};
\node[fill=white] at (4,5) {\Large 37};
\node[fill=white] at (4,6) {\Large 105};
\node[fill=white] at (4,7) {\Large 107};
\node[fill=white] at (4,8) {\Large 120};
\node[fill=white] at (4,9) {\Large 122};
\node[fill=white] at (4,10) {\Large 135};
\node[fill=white] at (4,11) {\Large 137};
\node[fill=white] at (5,0) {\Large 9};
\node[fill=white] at (5,1) {\Large 11};
\node[fill=white] at (5,2) {\Large 24};
\node[fill=white] at (5,3) {\Large 26};
\node[fill=white] at (5,4) {\Large 39};
\node[fill=white] at (5,5) {\Large 41};
\node[fill=white] at (5,6) {\Large 109};
\node[fill=white] at (5,7) {\Large 111};
\node[fill=white] at (5,8) {\Large 124};
\node[fill=white] at (5,9) {\Large 126};
\node[fill=white] at (5,10) {\Large 139};
\node[fill=white] at (5,11) {\Large 141};
\draw[color=black,thick,shift={(-0.5,-0.5)}] (0,0) grid (6,12);
\end{tikzpicture}
}
\end{adjustbox}
} \\
\hline
$\v{A}((1,\_),((\_,0),\_))$ & $\{4\} \circ (2,(2,2)):(1,(2,100))$ &
\scalebox{0.4}{
\begin{adjustbox}{margin=0.5cm}
\scalebox{0.8}{
\begin{tikzpicture}[x={(0cm,-1cm)},y={(1cm,0cm)},every node/.style={minimum size=1cm, outer sep=0pt}]
\node[fill=white] at (0,0) {\Large 0};
\node[fill=white] at (0,1) {\Large 2};
\node[fill=white] at (0,2) {\Large 15};
\node[fill=white] at (0,3) {\Large 17};
\node[fill=white] at (0,4) {\Large 30};
\node[fill=white] at (0,5) {\Large 32};
\node[fill=white] at (0,6) {\Large 100};
\node[fill=white] at (0,7) {\Large 102};
\node[fill=white] at (0,8) {\Large 115};
\node[fill=white] at (0,9) {\Large 117};
\node[fill=white] at (0,10) {\Large 130};
\node[fill=white] at (0,11) {\Large 132};
\node[fill=black!30] at (1,0) {\Large 4};
\node[fill=black!30] at (1,1) {\Large 6};
\node[fill=white] at (1,2) {\Large 19};
\node[fill=white] at (1,3) {\Large 21};
\node[fill=white] at (1,4) {\Large 34};
\node[fill=white] at (1,5) {\Large 36};
\node[fill=black!30] at (1,6) {\Large 104};
\node[fill=black!30] at (1,7) {\Large 106};
\node[fill=white] at (1,8) {\Large 119};
\node[fill=white] at (1,9) {\Large 121};
\node[fill=white] at (1,10) {\Large 134};
\node[fill=white] at (1,11) {\Large 136};
\node[fill=white] at (2,0) {\Large 8};
\node[fill=white] at (2,1) {\Large 10};
\node[fill=white] at (2,2) {\Large 23};
\node[fill=white] at (2,3) {\Large 25};
\node[fill=white] at (2,4) {\Large 38};
\node[fill=white] at (2,5) {\Large 40};
\node[fill=white] at (2,6) {\Large 108};
\node[fill=white] at (2,7) {\Large 110};
\node[fill=white] at (2,8) {\Large 123};
\node[fill=white] at (2,9) {\Large 125};
\node[fill=white] at (2,10) {\Large 138};
\node[fill=white] at (2,11) {\Large 140};
\node[fill=white] at (3,0) {\Large 1};
\node[fill=white] at (3,1) {\Large 3};
\node[fill=white] at (3,2) {\Large 16};
\node[fill=white] at (3,3) {\Large 18};
\node[fill=white] at (3,4) {\Large 31};
\node[fill=white] at (3,5) {\Large 33};
\node[fill=white] at (3,6) {\Large 101};
\node[fill=white] at (3,7) {\Large 103};
\node[fill=white] at (3,8) {\Large 116};
\node[fill=white] at (3,9) {\Large 118};
\node[fill=white] at (3,10) {\Large 131};
\node[fill=white] at (3,11) {\Large 133};
\node[fill=black!30] at (4,0) {\Large 5};
\node[fill=black!30] at (4,1) {\Large 7};
\node[fill=white] at (4,2) {\Large 20};
\node[fill=white] at (4,3) {\Large 22};
\node[fill=white] at (4,4) {\Large 35};
\node[fill=white] at (4,5) {\Large 37};
\node[fill=black!30] at (4,6) {\Large 105};
\node[fill=black!30] at (4,7) {\Large 107};
\node[fill=white] at (4,8) {\Large 120};
\node[fill=white] at (4,9) {\Large 122};
\node[fill=white] at (4,10) {\Large 135};
\node[fill=white] at (4,11) {\Large 137};
\node[fill=white] at (5,0) {\Large 9};
\node[fill=white] at (5,1) {\Large 11};
\node[fill=white] at (5,2) {\Large 24};
\node[fill=white] at (5,3) {\Large 26};
\node[fill=white] at (5,4) {\Large 39};
\node[fill=white] at (5,5) {\Large 41};
\node[fill=white] at (5,6) {\Large 109};
\node[fill=white] at (5,7) {\Large 111};
\node[fill=white] at (5,8) {\Large 124};
\node[fill=white] at (5,9) {\Large 126};
\node[fill=white] at (5,10) {\Large 139};
\node[fill=white] at (5,11) {\Large 141};
\draw[color=black,thick,shift={(-0.5,-0.5)}] (0,0) grid (6,12);
\end{tikzpicture}
}
\end{adjustbox}
} \\
\hline
\end{tabular}
\caption{A $6 {\times} 12$ tensor sliced along logical subboundaries to extract subtensors.}
\label{fig:slicing}
\end{figure}

Figure~\ref{fig:slicing} illustrates examples of slicing into a $6 {\times} 12$ matrix to extract rows, columns, and submatrices. These use a counting-iterator accessor denoted by $\{a\}$ to indicate an offset of $a \in \Z$. Slicing involves accumulating into the tensor offset and determining the unevaluated portion of the layout.

In many tensor libraries like {\tt numpy.ndarray}, {\tt torch.tensor}, and MATLAB, slicing is supported with notation similar to above: write {\tt my\_matrix[2,:]} to extract the second row of a matrix and {\tt my\_matrix[:,4]} to extract the fourth column of a matrix. These libraries also support ranged slicing, such as {\tt my\_matrix[2:4,1:3]} to extract the submatrix from the second to the fourth row and the first to the third column. \CuTe does not support ranged slicing as it finds ranged slicing to be problematic for several reasons:
\begin{compactitem}
\item Ranged slicing can't express all of the slices demonstrated in Figure~\ref{fig:slicing}. The last slice example cannot be expressed with ranged slicing on only the rows and columns of a matrix.
\item Ranged slicing promotes patterns like
\begin{python}
# Extract a TILE_SIZE of data for each thr_id
thr_data = my_data[thr_id*TILE_SIZE:(thr_id+1)*TILE_SIZE]
\end{python}
to retrieve a ``tile'' of data local to each thread. This pattern conflates the {\tt TILE\_SIZE}, which is very often a static constant that a program wants to optimize over, with a {\tt thr\_id}, which is a fundamentally dynamic index local to each thread. Instead, \CuTe prefers a two-stage permute-and-slice approach like
\begin{python}
# Permute and reshape my_data tensor into shape (TILE_SIZE,NumTiles)
# This is a wrapper around the more general composition(my_data, Transform_Layout)
tiled_data = logical_divide(my_data, TILE_SIZE)  # (TILE_SIZE, NumTiles)
assert size[1](tiled_data) == NumThrs            # Verify NumTiles == NumThrs
# Slice tiled tensor to retrieve TILE_SIZE local to thr_id
thr_data = thr_tile_data[None, thr_id]           # TILE_SIZE
\end{python}
which separates the distinct parameters {\tt TILE\_SIZE} and {\tt thr\_id}, is easier for a compiler to reason about and propagate static information for (e.g. {\tt thr\_data} has size {\tt TILE\_SIZE}), and is equally as flexible. See Section~\ref{sec:layouttv} for details and examples on generic partitioning and Section~\ref{sec:logical_divide} for details and examples on logical divide.
\item Ranged slicing can express slices that are impossible to represent with a \CuTe layout. For instance, with $\v{A}$ as in Figure~\ref{fig:slicing}, the slice {\tt $\v{A}$(0,0:2:12)} could result in $\{0\} \circ (3,2):(15,100)$ while the slice {\tt $\v{A}$(0:2:6,0)} cannot be represented due to the incompatibility of the requested slice with the layout of $\v{A}$. This is principally due to an incompatibility of the requested tile and the layout of the tensor and, therefore, we prefer this error to be detected and exposed at the permute-and-reshape stage ({\tt composition}) rather than the slicing stage. See Section~\ref{sec:composition_impl} for details and examples on admissibility conditions for layout group composition.
\end{compactitem}

\subsection{Applications}
\label{sec:algorithms}

\CuTe provides a compact representation for a set of layouts that is strictly larger than can be represented with traditional flat shapes and strides. In contrast, libraries like CUTLASS v2~\cite{kerr2017cutlass} and Thunderkittens~\cite{Spector:ThunderKittens} implement each layout individually and manually. This approach is labor-intensive, error-prone, and requires significant development time. To illustrate the complexity, the CUTLASS v2 code base contains nearly 300 separately implemented layouts spread across 87 files, collectively amounting to approximately 55,000 lines of code. Furthermore, many algorithms in CUTLASS v2 are designed to operate with only a limited subset of these layouts, exacerbating maintenance and scalability challenges. By comparison, \CuTe's core layout representation, along with the associated algebra for manipulating layouts, requires only 3,000 lines of code and is capable of representing all 300 layouts found in CUTLASS v2 and more. Algorithms implemented in \CuTe can enforce constraints on the rank or shape of their input, but remain compatible with any layout that satisfies these preconditions. This decoupling of algorithm logic from specific data or thread layouts results in more flexible and composable code.

The benefits above were recognized and \CuTe now forms the basis of CUTLASS v3, CUTLASS v4, and CuTe DSL, which are all built on top of \CuTe's core layout representation and algebra. The complex data layouts and partitioning patterns required by modern tensor instructions are represented and manipulated with \CuTe's single, consistent, and composable representation that has been robust across multiple generations of NVIDIA's instruction set architecture.

In this section, we detail the use of only the layout and tensor representation to provide powerful generic implementations of two of the most fundamental algorithms: {\tt COPY} and {\tt GEMM}. These algorithms are implemented with \CuTe tensors and are used to illustrate logical implementations being applicable to a wide range of applications. These algorithms, though often optimal on their own, serve as excellent generic reference implementations for optimized versions. The layout algebra in Section~\ref{sec:layoutalgebra} provides methods for inspecting and manipulating layouts to perform these optimizations.

\subsubsection{COPY}
\label{sec:copy}

A generic {\tt COPY} algorithm written with \CuTe tensors is

\noindent
\begin{minipage}[t]{0.55\textwidth}
\begin{cpp}
// @pre size(src) == size(dst)
template <class TS, class SLayout,
          class TD, class DLayout>
void
copy(Tensor<TS,SLayout> const& src,  // N
     Tensor<TD,DLayout>      & dst)  // N
{
  for (int i = 0; i < size(dst); ++i) {
    dst(i) = src(i);
  }
}
\end{cpp}
\end{minipage}
\hfill
\begin{minipage}[t]{0.40\textwidth}
\begin{python}
# @pre size(src) == size(dst)
def copy(src : Tensor,   # N
         dst : Tensor):  # N
  for i in range(size(dst)):
    dst[i] = src[i]
\end{python}
\end{minipage}

\noindent where the precondition specifies that the size of the tensors is equal. Equivalently written in the tensor argument comments, both tensors are compatible with a shape $N$.

This simple implementation of \texttt{COPY} accommodates a wide range of applications by varying the layouts of the source and destination tensors. Table~\ref{tab:copy} provides examples of common applications and their associated source and destination layouts.

\begin{table}[ht]
\centering
\begin{tabular}{|c|c|c|}
\textbf{Application} & \textbf{Source Layout} & \textbf{Destination Layout} \\ \hline\hline
1D Arrays & $8:1$ & $8:1$ \\ \hline
ND Arrays & $(8,2,3):(1,16,32)$ & $(8,2,3):(1,16,32)$ \\ \hline
Gather & $(2,3,2):(42,1,128)$ & $12:1$\\ \hline
Scatter & $12:1$ & $(2,3,2):(42,1,128)$\\ \hline
Broadcast & $7:0$ & $7:1$\\ \hline
Constant & $7:0$ & $7:0$\\ \hline
Transpose & $(8,3):(1,8)$ & $(8,3):(3,1)$\\ \hline
Tensor Transpose & $(8,(3,5)):(1,(57,8))$ & $(8,15):(1,8)$\\ \hline
\end{tabular}
\caption{Applications of {\tt COPY} with example source and destination layouts.}
\label{tab:copy}
\end{table}

Any tensor of any rank can be copied to any tensor of any other rank. In that sense, {\tt COPY} is a rank-1 algorithm regardless of the arguments' ranks. This is a version of rank-agnostic programming.

When the \texttt{idx2crd} function for the source and destination tensors is computationally inexpensive to evaluate (e.g., when the tensor shapes are statically known), the above implementation never actually generates dynamic coordinate transformations. In such cases, the loop can be unrolled, \texttt{idx2crd} can be statically applied to the loop index \texttt{i}, and the \texttt{inner\_product} computation incurs minimal overhead in computing offsets. This is a version of static analysis and optimization since the layout shape and/or strides are often known at compile-time and available to the compiler. If \texttt{idx2crd} does incur a runtime cost, the provided implementation still serves as a robust reference to validate optimized versions that may further inspect domains and transform layouts using operations detailed in Section~\ref{sec:layoutalgebra}.

\subsubsection{GEMM}
\label{sec:gemm}

A generic {\tt GEMM} algorithm written with \CuTe tensors is

\noindent
\begin{minipage}[t]{0.54\textwidth}
\begin{lstlisting}[
  language=C++,                      % choose the language of the code
	basicstyle=\ttfamily\small,              % the size of the fonts that are used for the code
	keywordstyle=\color{purple}\bf,        % keyword style
	%identifierstyle=\color{blue},     % identifier style
	%emphstyle=\color{green}\bf,        % emphasis style
	commentstyle=\color{green!35!black}\slshape, % comment style
	%stringstyle=\color{gray},          % string literal style
	%aboveskip=\baselineskip,           % skip space when starting code environment
	xleftmargin=1pt, xrightmargin=1pt, % code margins
	frame=single,                        % adds a frame around the code
	%numbers=left,                      % where to put the line-numbers
	%numberstyle=\tiny,                 % the size of the fonts that are used for the line-numbers
	%numbersep=4pt,                     % how far the line-numbers are from the code
	%captionpos=b,
  ]
// @pre M: size<0>(A) == size<0>(C)
// @pre N: size<0>(B) == size<1>(C)
// @pre K: size<1>(A) == size<1>(B)
template <class TA, class ALayout,
          class TB, class BLayout,
          class TC, class CLayout>
void
gemm(Tensor<TA,ALayout> const& A,  // (M,K)
     Tensor<TB,BLayout> const& B,  // (N,K)
     Tensor<TC,CLayout>      & C)  // (M,N)
{
  for (int k = 0; k < size<1>(B); ++k) {
    for (int n = 0; n < size<0>(B); ++n) {
      for (int m = 0; m < size<0>(A); ++m) {
        C(m,n) += A(m,k) * B(n,k);
  } } }
}
\end{lstlisting}
\end{minipage}
\hfill
\begin{minipage}[t]{0.43\textwidth}
\begin{python}
# @pre M: size[0](A) == size[0](C)
# @pre N: size[0](B) == size[1](C)
# @pre K: size[1](A) == size[1](B)
def gemm(A : Tensor,   # (M,K)
         B : Tensor,   # (N,K)
         C : Tensor):  # (M,N)
  for k in range(size[1](B)):
    for n in range(size[0](B)):
      for m in range(size[0](A)):
        C[m,n] += A[m,k] * B[n,k]
\end{python}
\end{minipage}
where the precondition specifies the logical constraints of the {\tt GEMM} algorithm. In the comments of each tensor parameter, we write the shape that each tensor must be compatible with.

This simple implementation of {\tt GEMM} (and a {\tt batched-GEMM} extension) can encompass a variety of applications by varying the layouts of the tensors. In Table~\ref{tab:gemm}, some common applications and example layouts are shown. These include all of the N-T variants of the BLAS GEMM and the generically strided (\texttt{dm*}, \texttt{dn*}, \texttt{dk*}) variants of the BLIS GEMM. This also functions as a fully generic tensor-tensor contraction ({\tt GETT}) where the tensors are folded into the appropriate matrix shape by grouping logical row modes, column modes, reduction modes, and batched modes~\cite{Shi:BLASContractions}. Creating a layout (as a functional composition of \CuTe layouts) that implements the {\tt im2col} transformation~\cite{cuDNN} also allows {\tt GEMM} to implement {\tt CONV}, which is core to many modern machine learning applications.

\begin{table}[ht]
\centering
{\small
\begin{tabular}{|c|c|c|c|}
\textbf{Application} & $\v{A}$-\textbf{Layout} & $\v{B}$-\textbf{Layout} & $\v{C}$-\textbf{Layout} \\ \hline\hline
NT GEMM & $(M,K):(1,{\tt lda})$ & $(N,K):(1,{\tt ldb})$ & $(M,N):(1,{\tt ldc})$ \\ \hline
TN GEMM & $(M,K):({\tt lda},1)$ & $(N,K):({\tt ldb},1)$ & $(M,N):(1,{\tt ldc})$ \\ \hline
NTT GEMM & $(N,K):(1,{\tt ldb})$ & $(M,K):(1,{\tt lda})$ & $(N,M):(1,{\tt ldc})$ \\ \hline
BLIS GEMM & $(M,K):({\tt dma},{\tt dka})$ & $(N,K):({\tt dnb},{\tt dkb})$ & $(M,N):({\tt dmc},{\tt dnc})$ \\ \hline
GETT & $((M_1,M_2),K):((1,W),X)$ & $(N,K):(K,1)$ & $((M_1,M_2),N):((1,Y),Z)$ \\ \hline
GETT & $(M,(K_1,K_2)):(1,(W,X))$ & $(N,(K_1,K_2)):(Y,(1,Z))$ & $(M,N):(1,M)$ \\ \hline
CONV & $(K,(C,T,R,S)):D_A$ & $((N,Z,P,Q),(C,T,R,S)):D_B$ & $(K,(N,Z,P,Q)):D_C$ \\ \hline
\end{tabular}
\caption{Applications of {\tt GEMM} with example layouts.}
\label{tab:gemm}
}
\end{table}

By abstracting the fused-multiply-add operation and providing sufficiently powerful tiling utilities, this algorithm can be adapted and applied recursively at each level of an architectural hierarchy.

\section{Layout Algebra}
\label{sec:layoutalgebra}

While \CuTe layouts are only a subset of the space of all possible functions, they are capable of representing a strictly larger set of layout functions than the traditional flat-shape and flat-stride representations found in libraries like {\tt BLAS}, {\tt torch.tensor}, and {\tt numpy.ndarray}.

Beyond their representational power, a key utility of \CuTe layouts lies in their ability to be manipulated and combined to create new layouts. This is achieved through a core set of algebraic operations defined over layouts, which can be further used to construct higher-level operations.

In this section, we define {\em layout homomorphisms} -- operations that take \CuTe layout(s) and produce a \CuTe layout that satisfies some functional properties.

\subsection{Concatenate}
\label{sec:concatenation}

A layout can be expressed as the \emph{concatenation} of its sublayouts,
\begin{align*}
\v{L} &= S : D\\
&= (S_0, S_1, \ldots, S_n) : (D_0, D_1, \ldots, D_n) \\
&= (S_0 : D_0, S_1 : D_1, \ldots, S_n : D_n) \\
&= (\v{L}_0, \v{L}_1, \ldots, \v{L}_n)
\end{align*}
such that
\begin{align}
\forall c = (c_0,c_1,\ldots,c_n) \in \Z(\v{L}), \ \ \v{L}(c) = \v{L}_0(c_0) + \v{L}_1(c_1) + \cdots + \v{L}_n(c_n).
\label{eqn:concat}
\end{align}
For concatenation admissibility, the functional property~\eqref{eqn:concat} implies that the codomain of all sublayouts must be contained in the same integer-semimodule. For instance, any two layouts with integer strides may be concatenated, but the layouts $4:2$ and $3:e_0$ cannot be concatenated.

Noting that every sublayout of a layout is also a layout, it is useful to observe that any algebraic operation that manipulates a layout can also be applied to any individual sublayout. We often call these ``by-mode operations," and every operation that is defined in this section (coalesce, composition, complement, logical divide, etc) can also be applied by-mode. This approach is expressed with the combinator
\begin{align}
\v{A} \star  \langle \v{B}, \v{C} \rangle = (\v{A}_0, \v{A}_1) \star  \langle \v{B}, \v{C} \rangle = (\v{A}_0 \star  \v{B}, \v{A}_1 \star  \v{C}),
\label{eqn:layout_combinator}
\end{align}
where the $\star$ is some operation on two layouts and the $\langle \rangle$ notation represents a tuple of layouts, distinguishing it from a concatenation of layouts.

\subsection{Coalesce}
\label{sec:coalesce}

Given a layout $\v{A}$, a \emph{coalesced} layout $\v{R}$ is a layout that satisfies:
\begin{align}
\text{Consistent integral domain:} && \abs{\v{R}} &= \abs{\v{A}}, \\
\text{Flattened or integral shape:} && \text{depth}(\v{R}) &\leq 1, \\
\text{Consistent integral evaluation:} && \forall \bar{c} \in \Z_{\abs{\v{A}}}, \ \ \v{R}(\bar{c}) &= \v{A}(\bar{c}).
\end{align}

The \texttt{coalesce} operation ``simplifies" the layout $\v{A}$ by treating it as a function over integers and potentially collapsing its shape to a shallower representation. While this process may remove rank and hierarchical information, modify coordinate sets, and merge multiple modes of $\v{A}$, it guarantees that the layout remains functionally equivalent as a mapping over its integral coordinates.

In practice, when referencing a coalesced layout, we typically mean the coalesced layout that achieves minimal rank.

As an example, the coalesced layout of $(2,(1,6)):(1,(6,2))$ is $12:1$. Alternatively, this layout can be coalesced by-mode with Eq.~\eqref{eqn:layout_combinator}, where we apply \texttt{coalesce} to each mode independently:
\begin{align*}
\texttt{coalesce}((2,(1,6)):(1,(6,2)), \ \langle \ast,\ast \rangle) &= \texttt{coalesce}((2:1,(1,6):(6,2)), \ \langle \ast,\ast \rangle) \\
&= (\texttt{coalesce}(2:1, \ \ast), \ \texttt{coalesce}((1,6):(6,2), \ \ast)) \\
&= (\texttt{coalesce}(2:1), \ \texttt{coalesce}((1,6):(6,2))) \\
&= (2:1,6:2) \\
&= (2,6):(1,2).
\end{align*}
Similarly, the rank-2 layout $((4,3),5):((15,1),3)$ coalesces to $(4,15):(15,1)$ and the by-mode coalesced layout is $((4,3),5):((15,1),3)$ because the row and column layouts remain unchanged when individually coalesced. The rank-2 layout $(4,(3,5)):(15,(1,3))$ also coalesces to $(4,15):(15,1)$ and the by-mode coalesced layout is $(4,15):(15,1)$ because the second mode can be coalesced individually to $15:1$.

\subsection{Composition}
\label{sec:composition}

Given layouts $\v{A}$ and $\v{B}$, the \emph{group composition} layout $\v{R} = \v{A} \circ \v{B}$ satisfies:
\begin{align}
\label{eqn:compose_compat}
\text{Domain compatibility:} && \v{B} &\preceq \v{R}, \\
\label{eqn:compose_result}
\text{Functional composition:} && \forall c \in \Z(\v{B}), \ \ \v{R}(c) &= \v{A}(\v{B}(c)).
\end{align}

In this formulation, $\v{B}$ determines the shape and coordinate sets of the resulting layout by defining the domain of $\v{R}$, while $\v{A}$ determines the codomain of $\v{R}$. The compatibility condition ensures that {\em all} coordinates of $\v{B}$ can also be used as coordinates of $\v{R}$. For admissibility of both group and functional composition, the codomain of $\v{B}$ must be compatible with the domain of $\v{A}$, which typically means that the codomain of $\v{B}$ is a set of coordinates that are congruent to one of the coordinate sets in $\Z(\v{A})$.

\subsubsection{Composition Properties}

\paragraph{Identity Layouts.}
For any shape $S$, an \emph{identity layout} $\v{I}_S$ satisfies
\begin{align*}
\forall c \in \Z_S, \ \ \v{I}_S(c) = c.
\end{align*}
Note that $\v{I}_S$ may actually take on any shape $P$ so long as $S \preceq P$. For example, the following are all identity layouts $\v{I}_{24}$ and satisfy $\v{L}(i) = i$ for all $i \in \Z_{24}$:
\begin{align*}
24 : 1, \quad (4,6) : (1,4), \quad (3,(4,2)) : (1,(3,12)).
\end{align*}
Similarly, the following are both identity layouts $\v{I}_{(4,6)}$ and satisfy $\v{L}(i,j) = (i,j)$ for all $(i,j) \in \Z_{(4,6)}$:
\begin{align*}
(4,6) : (e_0,e_1), \quad (4,(3,2)) : (e_0,(e_1,3e_1)).
\end{align*}
For a layout $\v{B}$ with codomain $\Z_D$, any $\v{I}_D$ serves as a \emph{group composition left identity},
\begin{align*}
\v{I}_D \circ \v{B} = \v{B}.
\end{align*}
For a layout $\v{A}$ with domain $\Z_S$, the layout $\v{I}_S$ with shape $S$ serves as a \emph{group composition right identity},
\begin{align*}
\v{A} \circ \v{I}_S = \v{A}.
\end{align*}

\paragraph{Associative Property.}
Given layouts $\v{A}$, $\v{B}$, and $\v{C}$, and the condition
\begin{align}
\text{image}(\v{C}) \subseteq \Z(\v{B}) \quad \text{and} \quad \text{image}(\v{B}) \subseteq \Z(\v{A}),
\label{eqn:layout_image}
\end{align}
then
\begin{align*}
\v{A} \circ (\v{B} \circ \v{C}) = (\v{A} \circ \v{B}) \circ \v{C}.
\end{align*}
Note that composition is still often possible when Eq.~\eqref{eqn:layout_image} is not satisfied, but associativity may be lost. For instance,
\begin{align*}
(5,3):(1,7) \ \circ \ [4:1 \ \circ \ 2:5] &&= (5,3):(1,7) \ \circ \ 2:5 &&= 2:7 \\
[(5,3):(1,7) \ \circ \ 4:1] \ \circ \ 2:5 &&= 4:1 \ \circ \ 2:5 &&= 2:5
\end{align*}
yield different results because the range of $2:5$ is not contained in the domain of $4:1$. However, both expressions do evaluate properly given the out-of-bounds behavior of {\tt idx2crd} and {\tt inner\_product}.

\subsubsection{Evaluation and Restrictions}
\label{sec:composition_impl}

The evaluation of group composition can be constructively derived from the layout evaluation operations in Eq.~\eqref{eqn:idx2crd} and Eq.~\eqref{eqn:innerprod}. In this section, we determine a set of admissibility conditions on layouts $\v{A}$ and $\v{B}$ that are required to successfully compute group composition.

First, we require functional compatibility between $\v{A}$ and $\v{B}$: the codomain of $\v{B}$ must be weakly congruent with the shape of $\v{A}$. Most commonly, $\v{B}$'s codomain is the integers and because all layouts $\v{A}$ accept integral coordinates the layouts are functionally compatible. More generally, when $\v{B}$ produces coordinates, those coordinates must be congruent to some coordinate in $\Z(\v{A})$.

A general approach would equate the evaluations of $(\v{A} \circ \v{B})(i)$ and $\v{R}(i)$. In practice, we have found it sufficient to analyze a base case with the simplest possible $\v{B} = s : d$ and reduce other evaluations to that case.

\paragraph{Base Case.}

Let $\v{B} = s : d$ with $s \in \Z^+$ and $d \in \N$. Let $\v{A} = S:D = (S_0, S_1, \ldots, S_R):(D_0, D_1, \ldots, D_R)$ be a layout with $S_r \in \Z^+$ and $D_r \in \mathcal{D}$. Define the exclusive prefix-product of $S = (S_0, S_1, \ldots, S_R)$ as
\begin{align*}
\bar{S}_r = \prod_{k=0}^{r-1} S_k.
\end{align*}
Then, the composition $\v{R} = \v{A} \circ \v{B}$ for each $i = 0, 1, \ldots, s-1$ is evaluated as follows:
\begin{align}
\v{R}(i) = (\v{A} \circ \v{B})(i) &= (D \circ S \circ d \circ s)(i) \nonumber \\
&= D(S(d(s(i)))) \nonumber\\
&= \texttt{inner\_product}(\texttt{idx2crd}_S(\texttt{inner\_product}(\texttt{idx2crd}_s(i), d)), D) \nonumber \\
&= \texttt{inner\_product}(\texttt{idx2crd}_S(\texttt{inner\_product}(i, d)), D) \nonumber \\
&= \texttt{inner\_product}(\texttt{idx2crd}_S(i \cdot d), D) \nonumber \\
&= \sum_{r=0}^{R-1} \left(\floor{\frac{i \cdot d}{\bar{S}_r}} \bmod S_r\right) \cdot D_r + \floor{\frac{i \cdot d}{\bar{S}_R}} \cdot D_R
\label{eqn:ABi} \\
&= \sum_{r=0}^{R-1} \left(\floor{\frac{i}{\bar{S}'_r}} \bmod S'_r\right) \cdot D'_r + \floor{\frac{i}{\bar{S}'_R}} \cdot D'_R. \nonumber
\end{align}
The result $\v{R}$, if it exists, is defined as the layout $\v{R} = S':D' = (S'_0, S'_1, \ldots, S'_R):(D'_0, D'_1, \ldots, D'_R)$. We aim to determine $S'$, $D'$, and the conditions under which they exist.

First, because the codomain of $\v{B}$ is the integers, we can assume that the layout $\v{A}$ has been coalesced. By definition, this cannot change the evaluation of the composition.

Next, observe that if $(s-1) \cdot d < \bar{S}_r$, then $\floor{i \cdot d / \bar{S}_q} = 0$ for all $i < s$ and $q > r$. That is, all of the modes $q > r$ of $\v{A}$ have no contribution to the result. Without loss of generality, we proceed under the assumption that $(s-1) \cdot d \geq \bar{S}_R$, which can always be achieved through truncation or extension of the shape of $\v{A}$. This assumption does not alter Eq.~\eqref{eqn:ABi} but simplifies the analysis by removing redundant cases.

We impose the {\bf stride divisibility condition}
\begin{align}
\bar{S}_r \mid d \ \text{ or } \ d \mid \bar{S}_r \ \text{ for each } r = 0, \ldots, R,
\label{eqn:divis_stride}
\end{align}
and define
\begin{align*}
\delta_r = \ceil{\frac{d}{\bar{S}_r}}, \quad \rho_r = \ceil{\frac{\bar{S}_r}{d}}.
\end{align*}
Then,
\begin{align*}
\sum_{r=0}^{R-1} \left(\floor{\frac{i \cdot d}{\bar{S}_r}} \bmod S_r\right) \cdot D_r + \floor{\frac{i \cdot d}{\bar{S}_R}} \cdot D_R &=
\sum_{r=0}^{R-1} \left(\floor{i \cdot \frac{\delta_r}{\rho_r}} \bmod S_r\right) \cdot D_r + \floor{i \cdot \frac{\delta_R}{\rho_R}} \cdot D_R.
\end{align*}
Using the identity $(a b) \bmod (n b) = (a \bmod n) \cdot b$, and noting that $\delta_r \mid S_r$, we have:
\begin{align*}
&= \sum_{r=0}^{R-1} \left(\floor{i \cdot \frac{1}{\rho_r}} \bmod \frac{S_r}{\delta_r}\right) \cdot (D_r \cdot \delta_r) + \floor{i \cdot \frac{1}{\rho_R}} \cdot (D_R \cdot \delta_R).
\end{align*}
Once we verify that
\begin{align*}
\rho_r = \ceil{\frac{\bar{S}_r}{d}} = \ceil{\frac{\prod_{k=0}^{r-1} S_k}{d}} = \ceil{\frac{S_0}{d}} \ceil{\frac{\prod_{k=1}^{r-1} S_k}{\ceil{d / S_0}}} = \cdots = \prod_{k=0}^{r-1} \ceil{\frac{S_k}{\ceil{d / \bar{S}_k}}} = \prod_{k=0}^{r-1} \ceil{\frac{S_k}{\delta_k}} = \bar{S}'_r,
\end{align*}
the resulting shape and strides can be identified as
\begin{align*}
S'_r &= \frac{S_r}{\delta_r}, \\
D'_r &= D_r \cdot \delta_r.
\end{align*}

To satisfy the compatibility condition~\eqref{eqn:compose_compat}, the shape must also satisfy $\abs{S'} = \bar{S}'_{R+1} = s$. To accomplish this, we also impose the {\bf shape divisibility condition}:
\begin{align}
\ceil{\frac{\bar{S}_r}{d}} \mid s \ \text{ for each } r = 0, \ldots, R,
\label{eqn:divis_shape}
\end{align}
and modify the shape to
\begin{align*}
S'_r &= \frac{S_r}{\delta_r}, \quad S'_R = \frac{s}{\rho_R}, \\
D'_r &= D_r \cdot \delta_r,
\end{align*}
where the term $s / \rho_R$ is imposed to truncate or extend the resulting shape to size $s$.

Thus, we have computed the shape and strides of the result, $\v{R} = S':D'$, provided that the stride divisibility condition~\eqref{eqn:divis_stride} and shape divisibility condition~\eqref{eqn:divis_shape} are satisfied by $\v{A}$ and $\v{B}$.

\paragraph{Reductive Case: Distributive.}

To express more general compositions in terms of the base case, we use a distributive property of the composition operation over concatenation of sublayouts of $\v{B}$,
\begin{align}
\v{A} \circ \v{B} &= \v{A} \circ (\v{B}_0, \v{B}_1, \ldots) = (\v{A} \circ \v{B}_0, \v{A} \circ \v{B}_1, \ldots).
\label{eqn:layout_distributivity}
\end{align}
This is satisfied when the shape $S$ of $\v{A}$ distributes over the sum of the sublayouts of $\v{B}$,
\begin{align}
(\v{A} \circ \v{B})(c)
&= D(S(\v{B}(c))) \notag \\
&= D(S(\v{B}_0(c_0) + \v{B}_1(c_1) + \cdots)) && \text{concatenation of $\v{B} = (\v{B}_0, \v{B}_1, \ldots)$} \notag \\
&\stackrel{?}{=} D(S(\v{B}_0(c_0)) + S(\v{B}_1(c_1)) + \cdots) && \text{conditioned on distributivity of $S$ over $\v{B}_i$} \label{eqn:shape_distributivity} \\
&= D(S(\v{B}_0(c_0))) + D(S(\v{B}_1(c_1))) + \cdots && \text{linearity of $D$} \notag \\
&= (\v{A} \circ \v{B}_0, \v{A} \circ \v{B}_1,\ldots)(c) \notag
\end{align}
Either of the following two conditions are sufficient to satisfy Eq~\eqref{eqn:shape_distributivity}:
\begin{compactitem}
\item The codomain of $\v{B}$ is congruent to the shape $S$. In this case, the application of $S$ is the identity transform, which distributes.
\item When $S$ is not the identity transform, then it must be the case that
\begin{align*}
\floor{\frac{\sum_i c_i \cdot d_i}{\bar{S}_r}} \bmod S_r &= \sum_i \floor{\frac{c_i \cdot d_i}{\bar{S}_r}} \bmod S_r \ \ \text{ for each } r = 0, \ldots, R-1, \text{ and } \\
\floor{\frac{\sum_i c_i \cdot d_i}{\bar{S}_R}} &= \sum_i \floor{\frac{c_i \cdot d_i}{\bar{S}_R}}
\end{align*}
which occurs when
\begin{compactenum}
\item All base sublayouts strides, $d_i$, satisfy the stride divisibility condition~\eqref{eqn:divis_stride}, and
\item All base sublayouts $s_i : d_i$ have mutually segregated images,
\begin{align*}
\forall i,j, \ \ s_i \cdot d_i \leq d_j \ \ \text{ or } \ \ s_j \cdot d_j \leq d_i
\end{align*}
\end{compactenum}
\end{compactitem}

\paragraph{Reductive Case: Coordinate.}

When $\v{B}$ produces coordinates, those coordinates must be congruent to some coordinate in $\Z(\v{A})$ and the strides of $\v{B}$ must each be a scaled basis element of a coordinate integer-semimodule: $d = a \cdot e_i \in \Z^{S'}$ with $S' \preceq \v{A}$. When these conditions are met, then we observe
\begin{align*}
\v{A} \circ \v{B} = \v{A} \circ (a \cdot e_i) \circ s = \v{A}_i \circ a \circ s = \v{A}_i \circ \v{B}',
\end{align*}
which returns to the base case where $\v{B}' = s : a$ is a rank-1, depth-0 layout with an integral codomain. For $\v{B}$'s with other integer-semimodule strides, special care must be taken to map $\v{B}$'s codomain to the domain of $\v{A}$, and additional restrictions on $\v{A}$ and $\v{B}$ may be required.

\subsubsection{Intuition and Divisibility}

Compositions with rank-1 left-hand layouts $\v{A}$ are trivial because $S_R$ does not appear in Eq.~\eqref{eqn:ABi}:
\begin{align*}
(S_0):(D_0) \ \circ \  s:d = s:D_0 \cdot d.
\end{align*}
In this case, there are no non-trivial divisibility checks because $R = 0$, $\bar{S}_0 = 1$, $\delta_0 = d$, and $\rho_0 = 1$. Note that this means group composition is still possible even when $\text{image}(\v{B}) \not\subseteq \Z(\v{A})$,
\begin{align*}
7:11 \ \circ \ 3:4 = 3:44,
\end{align*}
and $\v{B}$ does not need to be mutually disjoint in order to be distributive,
\begin{align*}
7:11 \ \circ \ (3,5):(6,3) = (3,5):(66,33).
\end{align*}

For compositions with layouts $\v{A}$ of greater rank, the intuitive strategy involves two steps:
\begin{compactitem}
\item Determine an intermediate layout that produces every $d$th element of $\v{A}$ by ``dividing out'' the first $d$ elements from $\v{A}$.
\item Fix the size of the intermediate strided layout to $s$ by ``keeping'' the first $s$ elements.
\end{compactitem}
For example,
\begin{align*}
(4,6,8,10):(2,3,5,7) \ \circ \  6:12
\end{align*}
is equivalent to
\begin{align*}
(4,6,8,10):(2,3,5,7) \ \circ \  6:12 = (1,2,8,10):(X,9,5,7) \ \circ \  6:1,
\end{align*}
where the first 12 elements of $\v{A}$ are ``divided out,'' and the strides are scaled accordingly. Then, the first 6 elements of the modified layout are ``kept,'' resulting in:
\begin{align*}
(1,2,8,10):(X,9,5,7) \ \circ \  6:1 = (2,3):(9,5).
\end{align*}
Alternatively, we can ``keep'' the first $6 \cdot 12$ elements and then ``divide out'' the first 12, as follows:
\begin{align*}
(4,6,8,10):(2,3,5,7) \ \circ \  6:12 = (4,6,3):(2,3,5) \ \circ \  6:12 = (2,3):(9,5).
\end{align*}
The divisibility conditions ensure that the progressive divisions in the ``divide out'' and ``keep'' steps always yield integers for the resulting shape.

\paragraph{Violations of Divisibility Conditions.}
Certain compositions violate the stride or shape divisibility conditions. For example,
\begin{align*}
(4,6,8):(2,3,5) \ \circ \  6:3
\end{align*}
violates the stride divisibility condition~\eqref{eqn:divis_stride} because $\bar{S}_1 = 4$ and $d = 3$ are not divisible. There is no layout that can represent every third element of the layout $(4,6,8):(2,3,5)$.

Similarly, the composition
\begin{align*}
(4,6,8):(2,3,5) \ \circ \  6:1
\end{align*}
violates the shape divisibility condition~\eqref{eqn:divis_shape}, since: $
\left\lceil \frac{\bar{S}_1}{d} \right\rceil = \left\lceil \frac{4}{1} \right\rceil = 4$
does not divide $s = 6$. There is no layout that can represent the first 6 elements of $(4,6,8):(2,3,5)$.

Ultimately, this means \CuTe layouts are not strictly closed under group composition. However, in practice, violations of divisibility conditions are often due to conceptual application errors, layout/hardware incompatibilities, programmer error, and other such issues. These errors can very often be caught at compile time and thereby contribute to program safety and development velocity.

\paragraph{Apparent Violations.}

Consider the composition
\begin{align*}
(4,2,8):(3,12,97) \ \circ \  3:3,
\end{align*}
which seemingly violates the stride divisibility condition because $\bar{S}_1 = 4$ and $d = 3$ are not divisible. This issue can be resolved by coalescing $\v{A}$ (as described in Sec.~\ref{sec:composition_impl}) and truncating $\v{A}$ (as addressed in the $(s-1) \cdot d < \bar{S}_r$ case). The equivalent composition produces the result
\begin{align*}
(4,2,8):(3,12,97) \ \circ \  3:3 = (8,8):(3,97) \ \circ \  3:3 = (8):(3) \ \circ \  3:3 = 3:9.
\end{align*}
However, the following compositions fail divisibility conditions:
\begin{align*}
(4,2,8):(3,12,97) \ \circ \  4:3 && (4,2,8):(3,15,97) \ \circ \  3:3.
\end{align*}
The example on the left cannot be sufficiently truncated, while the example on the right cannot be coalesced.

\subsubsection{Application: Partitioning Example}
\label{sec:layouttv}

Composition lies at the heart of the \CuTe layout algebra, enabling operations such as reshaping, restriding, permuting, partitioning, tiling, and extracting sublayouts. This section demonstrates the application of composition for partitioning a data tensor using an arbitrary thread-value pattern.

\begin{figure}[ht]
\centering
\scalebox{0.6}{
\begin{tikzpicture}[x={(0cm,-1cm)},y={(1cm,0cm)},every node/.style={minimum size=1cm, outer sep=0pt}]
\node[fill={rgb,255:red,175;green,175;blue,255}] at (0,0) {\shortstack{T0 \\ V0}};
\node[fill={rgb,255:red,175;green,175;blue,255}] at (0,1) {\shortstack{T0 \\ V1}};
\node[fill={rgb,255:red,175;green,255;blue,175}] at (0,2) {\shortstack{T1 \\ V0}};
\node[fill={rgb,255:red,175;green,255;blue,175}] at (0,3) {\shortstack{T1 \\ V1}};
\node[fill={rgb,255:red,255;green,255;blue,175}] at (0,4) {\shortstack{T2 \\ V0}};
\node[fill={rgb,255:red,255;green,255;blue,175}] at (0,5) {\shortstack{T2 \\ V1}};
\node[fill={rgb,255:red,255;green,175;blue,175}] at (0,6) {\shortstack{T3 \\ V0}};
\node[fill={rgb,255:red,255;green,175;blue,175}] at (0,7) {\shortstack{T3 \\ V1}};
\node[fill={rgb,255:red,210;green,210;blue,255}] at (1,0) {\shortstack{T4 \\ V0}};
\node[fill={rgb,255:red,210;green,210;blue,255}] at (1,1) {\shortstack{T4 \\ V1}};
\node[fill={rgb,255:red,210;green,255;blue,210}] at (1,2) {\shortstack{T5 \\ V0}};
\node[fill={rgb,255:red,210;green,255;blue,210}] at (1,3) {\shortstack{T5 \\ V1}};
\node[fill={rgb,255:red,255;green,255;blue,210}] at (1,4) {\shortstack{T6 \\ V0}};
\node[fill={rgb,255:red,255;green,255;blue,210}] at (1,5) {\shortstack{T6 \\ V1}};
\node[fill={rgb,255:red,255;green,210;blue,210}] at (1,6) {\shortstack{T7 \\ V0}};
\node[fill={rgb,255:red,255;green,210;blue,210}] at (1,7) {\shortstack{T7 \\ V1}};
\node[fill={rgb,255:red,175;green,175;blue,255}] at (2,0) {\shortstack{T8 \\ V0}};
\node[fill={rgb,255:red,175;green,175;blue,255}] at (2,1) {\shortstack{T8 \\ V1}};
\node[fill={rgb,255:red,175;green,255;blue,175}] at (2,2) {\shortstack{T9 \\ V0}};
\node[fill={rgb,255:red,175;green,255;blue,175}] at (2,3) {\shortstack{T9 \\ V1}};
\node[fill={rgb,255:red,255;green,255;blue,175}] at (2,4) {\shortstack{T10 \\ V0}};
\node[fill={rgb,255:red,255;green,255;blue,175}] at (2,5) {\shortstack{T10 \\ V1}};
\node[fill={rgb,255:red,255;green,175;blue,175}] at (2,6) {\shortstack{T11 \\ V0}};
\node[fill={rgb,255:red,255;green,175;blue,175}] at (2,7) {\shortstack{T11 \\ V1}};
\node[fill={rgb,255:red,210;green,210;blue,255}] at (3,0) {\shortstack{T12 \\ V0}};
\node[fill={rgb,255:red,210;green,210;blue,255}] at (3,1) {\shortstack{T12 \\ V1}};
\node[fill={rgb,255:red,210;green,255;blue,210}] at (3,2) {\shortstack{T13 \\ V0}};
\node[fill={rgb,255:red,210;green,255;blue,210}] at (3,3) {\shortstack{T13 \\ V1}};
\node[fill={rgb,255:red,255;green,255;blue,210}] at (3,4) {\shortstack{T14 \\ V0}};
\node[fill={rgb,255:red,255;green,255;blue,210}] at (3,5) {\shortstack{T14 \\ V1}};
\node[fill={rgb,255:red,255;green,210;blue,210}] at (3,6) {\shortstack{T15 \\ V0}};
\node[fill={rgb,255:red,255;green,210;blue,210}] at (3,7) {\shortstack{T15 \\ V1}};
\node[fill={rgb,255:red,175;green,175;blue,255}] at (4,0) {\shortstack{T16 \\ V0}};
\node[fill={rgb,255:red,175;green,175;blue,255}] at (4,1) {\shortstack{T16 \\ V1}};
\node[fill={rgb,255:red,175;green,255;blue,175}] at (4,2) {\shortstack{T17 \\ V0}};
\node[fill={rgb,255:red,175;green,255;blue,175}] at (4,3) {\shortstack{T17 \\ V1}};
\node[fill={rgb,255:red,255;green,255;blue,175}] at (4,4) {\shortstack{T18 \\ V0}};
\node[fill={rgb,255:red,255;green,255;blue,175}] at (4,5) {\shortstack{T18 \\ V1}};
\node[fill={rgb,255:red,255;green,175;blue,175}] at (4,6) {\shortstack{T19 \\ V0}};
\node[fill={rgb,255:red,255;green,175;blue,175}] at (4,7) {\shortstack{T19 \\ V1}};
\node[fill={rgb,255:red,210;green,210;blue,255}] at (5,0) {\shortstack{T20 \\ V0}};
\node[fill={rgb,255:red,210;green,210;blue,255}] at (5,1) {\shortstack{T20 \\ V1}};
\node[fill={rgb,255:red,210;green,255;blue,210}] at (5,2) {\shortstack{T21 \\ V0}};
\node[fill={rgb,255:red,210;green,255;blue,210}] at (5,3) {\shortstack{T21 \\ V1}};
\node[fill={rgb,255:red,255;green,255;blue,210}] at (5,4) {\shortstack{T22 \\ V0}};
\node[fill={rgb,255:red,255;green,255;blue,210}] at (5,5) {\shortstack{T22 \\ V1}};
\node[fill={rgb,255:red,255;green,210;blue,210}] at (5,6) {\shortstack{T23 \\ V0}};
\node[fill={rgb,255:red,255;green,210;blue,210}] at (5,7) {\shortstack{T23 \\ V1}};
\node[fill={rgb,255:red,175;green,175;blue,255}] at (6,0) {\shortstack{T24 \\ V0}};
\node[fill={rgb,255:red,175;green,175;blue,255}] at (6,1) {\shortstack{T24 \\ V1}};
\node[fill={rgb,255:red,175;green,255;blue,175}] at (6,2) {\shortstack{T25 \\ V0}};
\node[fill={rgb,255:red,175;green,255;blue,175}] at (6,3) {\shortstack{T25 \\ V1}};
\node[fill={rgb,255:red,255;green,255;blue,175}] at (6,4) {\shortstack{T26 \\ V0}};
\node[fill={rgb,255:red,255;green,255;blue,175}] at (6,5) {\shortstack{T26 \\ V1}};
\node[fill={rgb,255:red,255;green,175;blue,175}] at (6,6) {\shortstack{T27 \\ V0}};
\node[fill={rgb,255:red,255;green,175;blue,175}] at (6,7) {\shortstack{T27 \\ V1}};
\node[fill={rgb,255:red,210;green,210;blue,255}] at (7,0) {\shortstack{T28 \\ V0}};
\node[fill={rgb,255:red,210;green,210;blue,255}] at (7,1) {\shortstack{T28 \\ V1}};
\node[fill={rgb,255:red,210;green,255;blue,210}] at (7,2) {\shortstack{T29 \\ V0}};
\node[fill={rgb,255:red,210;green,255;blue,210}] at (7,3) {\shortstack{T29 \\ V1}};
\node[fill={rgb,255:red,255;green,255;blue,210}] at (7,4) {\shortstack{T30 \\ V0}};
\node[fill={rgb,255:red,255;green,255;blue,210}] at (7,5) {\shortstack{T30 \\ V1}};
\node[fill={rgb,255:red,255;green,210;blue,210}] at (7,6) {\shortstack{T31 \\ V0}};
\node[fill={rgb,255:red,255;green,210;blue,210}] at (7,7) {\shortstack{T31 \\ V1}};
\draw[color=black,thick,shift={(-0.5,-0.5)}] (0,0) grid (8,8);
\node at (0,-1) {\Large{\texttt{0}}};
\node at (1,-1) {\Large{\texttt{1}}};
\node at (2,-1) {\Large{\texttt{2}}};
\node at (3,-1) {\Large{\texttt{3}}};
\node at (4,-1) {\Large{\texttt{4}}};
\node at (5,-1) {\Large{\texttt{5}}};
\node at (6,-1) {\Large{\texttt{6}}};
\node at (7,-1) {\Large{\texttt{7}}};
\node at (-1,0) {\Large{\texttt{0}}};
\node at (-1,1) {\Large{\texttt{1}}};
\node at (-1,2) {\Large{\texttt{2}}};
\node at (-1,3) {\Large{\texttt{3}}};
\node at (-1,4) {\Large{\texttt{4}}};
\node at (-1,5) {\Large{\texttt{5}}};
\node at (-1,6) {\Large{\texttt{6}}};
\node at (-1,7) {\Large{\texttt{7}}};
\end{tikzpicture}
}
\caption{The thread-value partitioning of an $8 {\times} 8$ C-matrix required by the Ampere FP64 Tensor Core.}
\label{fig:TVLayout}
\end{figure}

Consider a layout of data that refines shape $(8,8)$. We aim to partition this data using the thread-value pattern shown in Figure~\ref{fig:TVLayout}, which is the logical partitioning pattern of a specific Ampere Tensor Core. The thread-value partitioning pattern for this instruction can be represented by the layout
\begin{align*}
\texttt{ThrValLayoutC:} \quad ((4,8),2):((16,1),8),
\end{align*}
which acts as a map from $(\texttt{thread\_idx}, \texttt{value\_idx})$ to the 1D coordinate within the $8 {\times} 8$ matrix. As illustrated, Figure~\ref{fig:TVLayout} actually plots the inverse for human interpretability as it displays the mapping from 2D coordinates within the $8 {\times} 8$ matrix to the $(\texttt{thread\_idx}, \texttt{value\_idx})$ pair. This {\tt ThrValLayoutC} is considered static metadata of the Ampere Tensor Core instruction -- it is intrinsic to the instruction and hardware's function and, prior to \CuTe, was only described in the instruction set architecture (ISA) with images similar to Figure~\ref{fig:TVLayout}.

Any $8 {\times} 8$ data layout can be permuted by composing it with the thread-value {\tt ThrValLayoutC}. Each composition produces a layout compatible with shape $(32,2)$ and defines the mapping between $(\texttt{thread\_idx}, \texttt{value\_idx})$ and data offsets. For example, Table~\ref{tab:TVLayoutExamples} shows examples of data layouts being transformed via composition by the above {\tt ThrValLayoutC}.

\begin{table}[ht]
\centering
\begin{tabular}{|c|c|c|c|}
Data Name & Data Layout $8 {\times} 8$ ($\v{A}$) & TV Layout $32 {\times} 2$ ($\v{B}$) & Result $32 {\times} 2$ ($\v{R}$) \\ \hline \hline
{\tt ColMajor} & $(8,8):(1,8)$ & $((4,8),2):((16,1),8)$ & $((4,8),2):((16,1),8)$ \\ \hline
{\tt RowMajor} & $(8,8):(8,1)$ & $((4,8),2):((16,1),8)$ & $((4,8),2):((2,8),1)$ \\ \hline
{\tt Padded} & $(8,8):(1,9)$ & $((4,8),2):((16,1),8)$ & $((4,8),2):((18,1),9)$ \\ \hline
{\tt ColInterleaved} & $((4,2),(2,4)):((2,16),(1,8))$ & $((4,8),2):((16,1),8)$ & $((4,(4,2)),2):((8,(2,16)),1)$ \\ \hline
{\tt Swizzled} & $(8,8):(f_1,f_9)$ & $((4,8),2):((16,1),8)$ & $((4,8),2):((f_{18},f_1),f_9)$ \\ \hline
{\tt Coordinate} & $(8,8):(e_0,e_1)$ & $((4,8),2):((16,1),8)$ & $((4,8),2):((2e_1,e_0),e_1)$ \\ \hline
\end{tabular}
\caption{Examples of data layouts being transformed via composition by a TV Layout.}
\label{tab:TVLayoutExamples}
\end{table}

The following code demonstrates a common pattern that uses thread-value partitioning:
\begin{python}
smem_data = Tensor(MyAccessor, MyLayout8x8)     # Tensor:        Coord -> Offset
tv_layout   = Layout(((4,8),2), ((16,1),8))     # TV Layout: (Thr,Val) -> Coord
smem_tv     = composition(smem_data, tv_layout) # Compose:   (Thr,Val) -> Offset
smem_v      = smem_tv[thr_id, None]             # Slice by thread to get subtensor
copy(smem_v, rmem_data)                         # Copy to register tensor/array
\end{python}
This pattern occurs extremely often in SIMD programming, where each processing element receives a symmetric partition of some parent data. In general, \CuTe recognizes that arbitrary partitioning can be defined as composition (permutation and/or reshaping) followed by slicing. As mentioned in Section~\ref{sec:slicing}, this compose-and-slice pattern separates out the partitioning pattern (the {\tt ThrValLayoutC}) from the realized slice of data (the slice with {\tt thr\_id}). Since the partitioning pattern is very often compile-time metadata related to instructions or optimization parameters and the slice is very often a runtime program identifier index, this pattern is much more capable of propagating static information and reducing runtime overhead.

\subsubsection{By-mode Composition and Tilers}
\label{sec:tiler}

Group composition can be applied by-mode, enabling operations on subdomains of layouts. For instance, it is often desirable to apply composition on the rows of a matrix and the columns of a matrix independently. We use the combinator in Eq~\eqref{eqn:layout_combinator} with the $\star$ operator identified as group composition $\circ$ to apply individual compositions to independent sublayouts of a layout.

We call the tuples of layouts in Eq~\eqref{eqn:layout_combinator} a {\em Tiler}.
\begin{definition}
A {\em Tiler} is an $\texttt{HTuple}(\texttt{Tile})$, where each \texttt{Tile} is
\begin{compactitem}
\item a \textit{Layout} $S:D$, or
\item a \textit{Integer} $S$, treated equivalently to the layout $S:1$.
\end{compactitem}
\end{definition}
This definition means that all layouts are tilers, all integers are tilers, and it provides for shapes like $(4,8)$ to be used as tilers. This common case is exemplified in Figure~\ref{fig:48tile}, where a shape serves as a convenient tiler for extracting and working with sublayouts.

\begin{figure}[ht]
\centering
\begin{subfigure}{0.22\textwidth}
\centering
\resizebox{\linewidth}{!}{
\begin{tikzpicture}[x={(0cm,-1cm)},y={(1cm,0cm)},every node/.style={minimum size=1cm, outer sep=0pt}]
\foreach \i in {0,1,2,3} {
\foreach \j in {0,1,2,3,4,5,6,7} {
\node[fill=gray] at (\i,\j) {};
}}
\draw[color=black,thick,shift={(-0.5,-0.5)}] (0,0) grid (8,16);
\end{tikzpicture}}
\caption{$\langle 4:1,8:1 \rangle \equiv \langle 4,8 \rangle$}
\label{fig:48tile}
\end{subfigure}\quad
\begin{subfigure}{0.22\textwidth}
\centering
\resizebox{\linewidth}{!}{
\begin{tikzpicture}[x={(0cm,-1cm)},y={(1cm,0cm)},every node/.style={minimum size=1cm, outer sep=0pt}]
\foreach \i in {0,1,4,5} {
\foreach \j in {0,1,2,3,4,5,6,7} {
\node[fill=gray] at (\i,\j) {};
}}
\draw[color=black,thick,shift={(-0.5,-0.5)}] (0,0) grid (8,16);
\end{tikzpicture}}
\caption{$\langle (2,2):(1,4),8:1 \rangle$}
\end{subfigure}\quad
\begin{subfigure}{0.22\textwidth}
\centering
\resizebox{\linewidth}{!}{
\begin{tikzpicture}[x={(0cm,-1cm)},y={(1cm,0cm)},every node/.style={minimum size=1cm, outer sep=0pt}]
\foreach \i in {0,1,4,5} {
\foreach \j in {0,2,4,6,8,10,12,14} {
\node[fill=gray] at (\i,\j) {};
}}
\draw[color=black,thick,shift={(-0.5,-0.5)}] (0,0) grid (8,16);
\end{tikzpicture}}
\caption{$\langle (2,2):(1,4),8:2 \rangle$}
\end{subfigure}\quad
\begin{subfigure}{0.22\textwidth}
\centering
\resizebox{\linewidth}{!}{
\begin{tikzpicture}[x={(0cm,-1cm)},y={(1cm,0cm)},every node/.style={minimum size=1cm, outer sep=0pt}]
\foreach \i in {0,2,4,6} {
\foreach \j in {0,2,4,6,8,10,12,14} {
\node[fill=gray] at (\i,\j) {};
}}
\draw[color=black,thick,shift={(-0.5,-0.5)}] (0,0) grid (8,16);
\end{tikzpicture}}
\caption{$\langle 4:2,8:2 \rangle$}
\end{subfigure}
\caption{Examples of tilers $\v{T}$ that extract a $4 {\times} 8$ sublayout from an arbitrary $8{\times}16$ layout $\v{L}$ via $\v{L} \circ \v{T}$.}
\label{fig:tilers}
\end{figure}

Figure~\ref{fig:tilers} illustrates how tilers can be applied via composition to any layout. The figure demonstrates taking an arbitrary layout compatible with shape $(8,16)$ and applying various tilers to extract elements from the columns and rows individually to produce a sublayout compatible with shape $(4,8)$. For instance, the tiler $\langle 4:1, 8:1 \rangle$ extracts the first 4 elements from each column and the first 8 elements from each row, producing a sublayout compatible with shape $(4,8)$.

Additionally, there is an equivalence between composition of tilers and composition of coordinate layouts. For a layout $\v{L} = (M,N) : (X,Y)$ and a tiler $\v{T} = \langle \v{T}_0, \v{T}_1 \rangle$, we note that
\begin{align*}
\v{L} \ \circ \ \v{T} \ \equiv \ \v{L} \ \circ \ (M,N):(e_0,e_1) \ \circ \ \v{T}
\end{align*}
meaning that $(M,N):(e_0,e_1) \ \circ \ \v{T}$ has the same action on $\v{L}$ that $\v{T}$ does. Here, $(M,N):(e_0,e_1)$ is an identity layout, mapping it's 2D natural coordinates to 2D coordinates. This formulation of tilers cements the notion that shapes can be considered maps from coordinates to natural coordinates as used in the definition of Layout in Section~\ref{sec:layout}. Specifically, all of the following objects should be considered equivalent when used in composition:
\begin{align*}
(4,8) \equiv \langle 4, 8 \rangle \equiv (4:1, 8:1) \equiv (4,8):(e_0,e_1)
\end{align*}

\subsection{Inverse}
\label{sec:inverse}

Layouts may be injective, surjective, or bijective and admit right-, left-, full-, and quasi-inverses. When layouts are interpreted as functions from coordinates to offsets, inverse layouts may be interpreted as functions from offsets to coordinates. Layout inverses are very useful in determining where within a layout certain offsets exist, extracting groups of specific offsets, or determining the common sublayout of two individual layouts.

In this section, we define the generalized left- and right-inverses of a layout and provide application examples for their use.

\subsubsection{Right-Inverse}

A {\em right-(pseudo)inverse} of a layout $\v{L} : \Z_{\abs{\v{L}}} \to \D$ is an injective layout $\v{L}^{\ddagger} : \D_{\v{L}^{\ddagger}} \to \Z_{\abs{\v{L}}}$ that satisfies
\begin{align}
\forall k \in \D_{\v{L}^{\ddagger}}, \ \ \v{L}^\ddagger(\v{L}(\v{L}^\ddagger(k))) &= \v{L}^\ddagger(k)
\label{eqn:right-inverse}
\end{align}
The above general definition is useful in determining properties of $\v{L}^{\ddagger}$ in the case that $\v{L}$ has a general integer-semimodule codomain $\D$. In the common case $\D = \Z$, the canonical right-inverse definition is recovered,
\begin{align*}
\forall k \in \Z_{\abs{\v{L}^{\ddagger}}}, \ \ \v{L}(\v{L}^{\ddagger}(k)) = k,
\end{align*}
but with more general codomains like $\D = \Z^{(\ast,\ast)}$, we can conclude from~\eqref{eqn:right-inverse} that $\D \lesssim \v{L}^\ddagger$, the right-inverse's domain coarsens the profile of the codomain of $\v{L}$ so it can accept offsets from $\v{L}$.

If a layout $\v{L}$ has a right-inverse layout $\v{L}^\ddagger$, then $\abs{\v{L}^\ddagger} \leq \abs{\v{L}}$. In practice, when referencing the right-inverse of a layout, we typically mean the right-inverse with the maximum size. For instance, the right-inverses of the largest size of the layouts in Figure~\ref{fig:layouts} and~\ref{fig:layouts2} are shown in Table~\ref{tab:rinv}.

\begin{table}[ht]
\begin{center}
\begin{tabular}{|>{$}c<{$}|>{$}c<{$}|c|}
\v{L} & \v{L}^\ddagger & \textbf{Comments} \\ \hline \hline
(4,8):(1,4) & 32:1 & \\ \hline
(4,8):(8,1) & (8,4):(4,1) \\ \hline
(3,7,5):(5,15,1) & (5,21):(21,1) & Non-power-of-two sizes/strides are fine. \\ \hline
(4,8):(1,5) & 4:1 & Smaller result for non-contiguous images. \\ \hline
(4,(4,2)):(4,(1,16)) & (4,4,2):(4,1,16) & Result domain congruent with codomain $\Z$. \\ \hline
((2,2),(4,2)):((1,8),(2,16)) & (2,4,2,2):(1,4,2,16) \\ \hline
((2,2),(2,4)):((0,1),(0,2)) & (2,4):(2,8) & Stride-0 modes do not contribute. \\ \hline
((2,2),(2,4)):((0,2),(0,4)) & 1:0 & Trivial right-inverse. \\ \hline
(4,8):(e_0,e_1) & (4,8):(1,4) & Result domain congruent with codomain $\Z^{(\ast,\ast)}$. \\ \hline
(4,(4,2)):(e_1,(e_0,6e_1)) & (4,4):(4,1) & Smaller result and domain compatible. \\ \hline
(4,(4,3)):(f_1,(f_5,f_{16})) & (4,4,3):(f_1,f_5,f_{16}) & Possible to define for  integer-module strides. \\ \hline
\end{tabular}
\end{center}
\caption{Examples of layout right-inverses.}
\label{tab:rinv}
\end{table}

When the layout $\v{L}$ is a bijection on $\Z_{\abs{\v{L}}}$, then the right-inverse is also the {\em inverse}, $\v{L}^{-1}$, of the layout $\v{L}$ and satisfies
\begin{align*}
\forall k \in \Z_{\abs{\v{L}}}, \ \ \v{L}(\v{L}^{-1}(k)) = k = \v{L}^{-1}(\v{L}(k))
\end{align*}
If a layout $\v{L}$ has a full-inverse $\v{L}^{-1}$ we call that layout {\em compact}.


\subsubsection{Application: Vectorization Example}

\begin{figure}[ht]
	\centering
	\begin{subfigure}[t]{0.35\textwidth}
		\centering
		\begin{tikzpicture}[thick,scale=1.10, every node/.style={scale=1.10}]
			\begin{scope}[xshift=-1.5cm]
			\matrix[matrix of nodes,nodes={draw,inner sep=0pt,text width=.5cm,align=center,minimum height=.5cm}]
			at (0,0) {
				0 & 4 & 8  & 12 \\
				1 & 5 & 9  & 13 \\
				2 & 6 & 10 & 14 \\
				3 & 7 & 11 & 15 \\};
			\fill [gray,fill opacity=0.4] (-1.05,0) rectangle (-0.50,1.05);
			\draw (0,-1) node[anchor=north,align=center] {\footnotesize $(4,4):(1,4)$ \\ \footnotesize Source};
			\end{scope}
			\begin{scope}[xshift=+1.5cm]
			\matrix[matrix of nodes,nodes={draw,inner sep=0pt,text width=.5cm,align=center,minimum height=.5cm}]
			at (0,0) {
				0 & 2 & 4  & 6 \\
				1 & 3 & 5  & 7 \\
				8 & 10 & 12 & 14 \\
				9 & 11 & 13 & 15 \\};
			\fill [gray,fill opacity=0.4] (-1.05,0) rectangle (-0.50,1.05);
			\draw (0,-1) node[anchor=north,align=center] {\footnotesize $((2,2),4):((1,8),2)$ \\ \footnotesize Destination};
		    \end{scope}
		\end{tikzpicture}
		\caption{A 2-element common subvector.} \label{fig:autovec1}
	\end{subfigure}
    \quad\quad\quad\quad\quad\quad
	\begin{subfigure}[t]{0.5\textwidth}
	\centering
	\begin{tikzpicture}[thick,scale=1.10, every node/.style={scale=1.10}]
		\begin{scope}[xshift=-2cm]
		\matrix[matrix of nodes,nodes={draw,inner sep=0pt,text width=.5cm,align=center,minimum height=.5cm}]
		at (0,0) {
			0  & 4  & 1  & 5  \\
			8  & 12 & 9  & 13 \\
			2  & 6  & 3  & 7  \\
			10 & 14 & 11 & 15 \\};
		\fill [gray,fill opacity=0.4] (-1.05,+0.50) rectangle (-0.50,1.05);
		\fill [gray,fill opacity=0.4] (-1.05,-0.50) rectangle (-0.50,0.00);
		\fill [gray,fill opacity=0.4] (    0,+0.50) rectangle (+0.50,1.05);
		\fill [gray,fill opacity=0.4] (    0,-0.50) rectangle (+0.50,0.00);
		\draw (0,-1) node[anchor=north,align=center] {\footnotesize$((2,2),(2,2)):((8,2),(4,1))$ \\ \footnotesize Source};
		\end{scope}
		\begin{scope}[xshift=+2cm]
		\matrix[matrix of nodes,nodes={draw,inner sep=0pt,text width=.5cm,align=center,minimum height=.5cm}]
		at (0,0) {
			0 & 8 & 1  & 9 \\
			4 & 12 & 5  & 13 \\
			2 & 10 & 3 & 11 \\
			6 & 14 & 7 & 15 \\};
		\fill [gray,fill opacity=0.4] (-1.05,+0.50) rectangle (-0.50,1.05);
		\fill [gray,fill opacity=0.4] (-1.05,-0.50) rectangle (-0.50,0.00);
		\fill [gray,fill opacity=0.4] (    0,+0.50) rectangle (+0.50,1.05);
		\fill [gray,fill opacity=0.4] (    0,-0.50) rectangle (+0.50,0.00);
		\draw (0,-1) node[anchor=north,align=center] {\footnotesize $((2,2),(2,2)):((4,2),(8,1))$ \\ \footnotesize Destination};
		\end{scope}
	\end{tikzpicture}
	\caption{A 4-element common subvector.} \label{fig:autovec2}
\end{subfigure}
\caption{Examples of common subvectors between layouts.}
\label{fig:autovec}
\end{figure}

Right inverses are extremely useful in inspecting data layouts and determining if and where contiguous elements exist. As an immediate example, the right-inverse of the layouts $(4,8):(1,4)$ and $(4,8):(8,1)$ are $32:1$ and $(8,4):(4,1)$ respectively. This means that both layouts, because their right-inverses have size-32, index into 32 contiguous physical elements.

As a more involved example, a common pattern for copy, called a vectorizing-copy in \CuTe, attempts to find the maximum number of elements that can be copied at once between two tensors. The right-inverse allows \CuTe to determine the {\em maximum common sublayout} between two layouts and, with additional information regarding hardware capabilities and physical alignment of pointers and strides, can algebraically determine the number and location of elements that can be safely vectorized to perform the copy.

Two examples of vectorization are shown in Figure~\ref{fig:autovec}. In the first image~\ref{fig:autovec1}, the coordinates of the source and destination layouts match for element offsets $k \in \{0,1\}$ with coordinates $\bar{c} \in \{0,1\}$. Provided the accessor, data types, and alignments allow for it, the copy width could be expanded to accommodate two elements per instruction. In the second image~\ref{fig:autovec2}, the coordinates of the source and destination layouts match for element offsets $k \in \{0,1,2,3\}$ with coordinates $\bar{c} = \{0,2,8,10\}$. The copy width could be expanded to accommodate four elements per instruction.

In general, for layout $\v{A} : \Z(\v{A}) \to \Z_\alpha$ and $\v{B} : \Z(\v{B}) \to \Z_\beta$ with $\abs{\v{A}} = \abs{\v{B}}$, we wish to find the largest integer $K$ such that the coordinates match
\begin{align*}
\forall \, k \in \Z_K, \ \ \v{A}^\ddagger(k) = \v{B}^\ddagger(k)
\end{align*}
and this can be computed efficiently via finding $K$ such that
\begin{align*}
\forall \, k \in \Z_K, \ \ k = \v{A}(\v{B}^\ddagger(k)) = (\v{A} \circ \v{B}^\ddagger)(k) \quad \text{or} \quad
\forall \, k \in \Z_K, \ \ k = \v{B}(\v{A}^\ddagger(k)) = (\v{B} \circ \v{A}^\ddagger)(k)
\end{align*}
which is simply the size of the identity portion (the stride-1 mode) of $\v{A} \circ  \v{B}^\ddagger = (\v{I}_K, \v{X})$ or $\v{B} \circ \v{A}^\ddagger = (\v{I}_K, \v{Y})$.

Moreover, the integral coordinates of these mutually contiguous elements is given by
\begin{align}
\v{A}^\ddagger \circ  \floor{\v{B} \circ \v{A}^\ddagger}_K = \v{A}^\ddagger \circ \v{I}_K = \floor{\v{A}^\ddagger}_K  \ \colon \ \Z_K \to \Z_{\abs{A}} \nonumber \\
\v{B}^\ddagger \circ  \floor{\v{A} \circ \v{B}^\ddagger}_K = \v{B}^\ddagger \circ \v{I}_K = \floor{\v{B}^\ddagger}_K  \ \colon \ \Z_K \to \Z_{\abs{B}}
\label{eqn:common_layout}
\end{align}
where $\floor{\cdot}_K$ is a truncated-at-size-$K$ operation. This layout yields the logical coordinates of the first $K$ physical elements in the data and could be used to extract out the common subvectors from each via logical divide or zipped divide (Section~\ref{sec:logical_divide}).

In practice, consider the reference function {\tt COPY} in Section~\ref{sec:copy}. The implementation may internally permute the iteration order without altering the observable behavior. Permuting the iteration order is equivalent to applying a consistent permutation to the layouts of the source and destination tensors. By selecting the permutation to match the layout defined in Eq.~\eqref{eqn:common_layout}, all contiguous elements can be permuted to the first mode, enabling special handling by vectorized instructions. Furthermore, this procedure can not only detect if and where the elements to be vectorized are located, but also the maximum number of elements that can be safely vectorized as well. This provides a framework for the {\bf AutoVectorization} optimization opportunity.

\subsubsection{Left-Inverse}

A {\em left-(pseudo)inverse} of a layout $\v{L} : \Z_{\abs{\v{L}}} \to \D$ is a layout $\v{L}^{\dagger} : \D_{\v{L}^\dagger} \to \Z_{\abs{\v{L}}}$ that satisfies
\begin{align}
\forall k \in \Z_{\abs{\v{L}}}, \ \ \v{L}(\v{L}^{\dagger}(\v{L}(k))) = \v{L}(k)
\end{align}
The left inverse may not be unique and may take any values for inputs that are not in the image of $\v{L}$. The above general definition is useful in determining properties of $\v{L}^{\dagger}$ in the case that $\v{L}$ has a general integer-semimodule codomain $\D$. In the common case where $\v{L}$ is injective, the canonical left-inverse definition is recovered,
\begin{align*}
\forall k \in \Z_{\abs{\v{L}}}, \ \ \v{L}^\dagger(\v{L}(k)) = k.
\end{align*}

The left-inverses of the layouts in Figure~\ref{fig:layouts} and~\ref{fig:layouts2} are shown in Table~\ref{tab:linv}.

\begin{table}[ht]
\begin{center}
\begin{tabular}{|>{$}c<{$}|>{$}c<{$}|c|}
\v{L} & \v{L}^\dagger & Comment \\ \hline \hline
(4,8):(1,4) & 32:1 \\ \hline
(4,8):(8,1) & (8,4):(4,1) & Same as right-inverse for contiguous images. \\ \hline
(3,7,5):(5,15,1) & (5,21):(21,1) & Non-power-of-two sizes/strides are fine. \\ \hline
(4,8):(1,5) & (5,8):(1,4) & Larger result for non-contiguous images. \\ \hline
(4,(4,2)):(4,(1,16)) & (4,4,2):(4,1,16) & Result domain congruent with codomain $\Z$. \\ \hline
((2,2),(4,2)):((1,8),(2,16)) & (2,4,2,2):(1,4,2,16) \\ \hline
((2,2),(2,4)):((0,2),(0,4)) & (2,2,4):(0,2,8) & Result is not unique -- any mode-0 stride. \\ \hline
((2,2),(2,4)):((0,1),(0,2)) & (2,4):(2,8) \\ \hline
(4,8):(e_0,e_1) & (4,8):(1,4) & Result domain congruent with codomain $\Z^{(\ast,\ast)}$. \\ \hline
(4,(4,2)):(e_1,(e_0,6e_1)) & (4,(6,2)):(4,(1,16)) & Larger result and domain compatible. \\ \hline
(4,(4,3)):(f_1,(f_5,f_{16})) & (4,4,3):(f_1,f_5,f_{16}) & Possible to define for integer-module strides. \\ \hline
\end{tabular}
\end{center}
\caption{Examples of layout left-inverses.}
\label{tab:linv}
\end{table}

\subsubsection{Application: Admissibility Example}

Left-inverses are useful for determining the existence and location of specific offsets produced by a data layout.

\begin{figure}[ht]
\centering
\begin{subfigure}[b]{0.45\textwidth}
    \centering
    \includegraphics[width=0.3\textwidth]{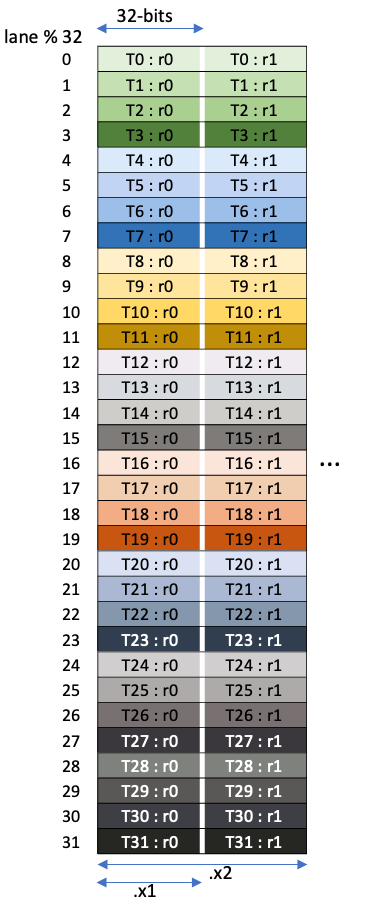}
    \caption{\texttt{tcgen05\{.ld,.st\}.32x32b.\{x1,x2\}}}
    \label{fig:tmem_fragment-32x32}
\end{subfigure}
\hfill
\begin{subfigure}[b]{0.45\textwidth}
    \centering
    \includegraphics[width=0.8\textwidth]{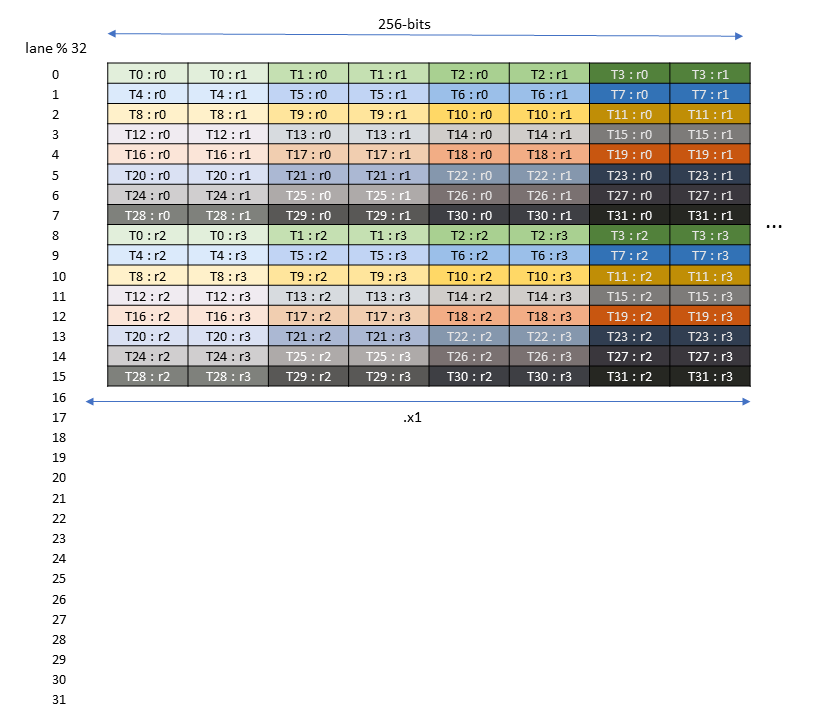}
    \caption{\texttt{tcgen05\{.ld,.st\}.16x256b.x1}}
    \label{fig:tmem-fragment-16x256}
\end{subfigure}
\caption{TMEM load/store instructions access specific offsets and assigns data to thread-local registers~\cite{NVIDIA:PTX:9.0}.}
\label{fig:tmem-fragments}
\end{figure}

Loading and storing data to and from Blackwell ``tensor memory'' (TMEM) uses instructions that access TMEM at a predefined set of very specific offsets, as shown in Figure~\ref{fig:tmem-fragments}. The physical TMEM addressing is a 2D grid of up to 512 ``cols" and 128 ``lanes". The images show the access pattern for one warp of each instruction, but each instruction actually uses four warps and the patterns above repeat every 32 lanes four times. In addressing physical TMEM, striding across the 32-bit ``cols" increments the TMEM address by 1 and striding down the ``lanes'' increments the TMEM address by 16384. With the physical addressing in mind, we define a layout that records all physical offsets that each instruction accesses:
\begin{center}
\begin{tabular}{l | r@{:}l}
\quad\quad\quad\quad \textbf{Instruction} & \multicolumn{2}{c}{$(\text{InstrCOL}, \text{InstrLANE}) \to \text{TMEM Offset}$} \\
\hline
\texttt{tcgen05\{.ld,.st\}.32x32b.x1} & $(1,128)$ & $(1,16384)$ \\
\texttt{tcgen05\{.ld,.st\}.32x32b.x2} & $(2,128)$ & $(1,16384)$ \\
\texttt{tcgen05\{.ld,.st\}.16x256b.x1} & $(8,(16,4))$ & $(1,(16384,32\cdot 16384))$ \\
\end{tabular}
\end{center}
These layouts are instruction-specific and can be used to determine exactly which offsets within TMEM each instruction will access.

Despite the 2D addressing and the use of TMEM as the accumulators of MMAs with shape $(M,N)$, the TMEM ``lanes" and ``cols" do not always correspond to logical matrix rows and columns. Like any addressing, the physical offsets can be wrapped and folded into any shape via an arbitrary \CuTe layout. Thus, given a data layout and a TMEM instruction, we wish to determine
\begin{compactenum}
\item whether the data layout contains all offsets that the instruction will access, and
\item where, in logical coordinates, the offsets accessed by the instruction are located within the data layout.
\end{compactenum}

In general, for a data layout $\v{A} : \Z(\v{A}) \to \Z_\alpha$ that maps logical coordinates to data offsets, and an instruction layout $\v{T} : \Z(\v{T}) \to \Z_\beta$ that maps instruction coordinates to data offsets, we wish to determine the existence and location of all offsets $\v{T}(i)$ in the image of $\v{A}$. This is stated as
\begin{align*}
\forall i \in \Z_{\abs{\v{T}}}, \exists c \in \Z(\v{A}) \ \ s.t. \ \ \v{T}(i) = \v{A}(c)
\end{align*}
and can be computed efficiently by computing the left-inverse of $\v{A}$ and checking
\begin{align*}
\v{A}(\v{A}^\dagger(\v{T}(i))) = \v{T}(i)
\end{align*}
That is,
\begin{compactitem}
\item All offsets $\v{T}(i)$ are in the domain of $\v{A}^\dagger$.
\item All coordinates $\v{A}^\dagger(\v{T}(i))$ are unique and in the domain of $\v{A}$.
\end{compactitem}
With these conditions, we can say that each offset $\v{T}(i)$ appears in the image of $\v{A}$ and is located at the coordinate $\v{A}^\dagger(\v{T}(i))$. The layout $\v{A}^\dagger \circ \v{T}$ is a layout that maps instruction coordinates to logical data coordinates and can be used via {\tt zipped\_divide}, for example, to partition the data layout $\v{A}$ into sublayouts that correspond to the offsets accessed by the instruction.

\subsection{Complement}
\label{sec:complement}

The \emph{complement} of a layout $\v{L} : \Z(\v{L}) \to \D$ is a layout $\v{L}^* : \Z(\v{L}^*) \to \D$ that satisfies the following conditions
\begin{align}
\label{eqn:comp:congruence}
\text{Weak congruence with codomain:} && \D \lesssim \v{L}^*, \\
\label{eqn:comp:disjoint}
\text{Disjoint images:} && \forall b \in \Z(\v{L}), \ \forall a \in \Z^{\v{L}^*}/\{0\}, \ &\v{L}(b) \neq \v{L}^*(a), \\
\label{eqn:comp:ordered}
\text{Ordered image:} && \forall a \in \Z_{\abs{\v{L}^*}}/\{0\}, \ &\v{L}^*(a-1) < \v{L}^*(a).
\end{align}
The complement of a layout is a layout that generates elements in the codomain of the original layout but not in the image. In composition, a layout $\v{L}$ is often used as an indirection into another layout, so the complement $\v{L}^*$ is a layout that points to elements omitted by $\v{L}$. The {\tt complement} operation enables the splitting of a layout via {\em logical divide} (Section~\ref{sec:logical_divide}) and enables repetition and extension of a layout via {\em logical product} (Section~\ref{sec:logical_product}).

In this section, we are only concerned with layouts that have a codomain $\D$ that is an {\tt HTuple}($\Z$): the integers or coordinates with $\D = \Z^D$ for some profile $D$. This enables the weak congruence condition Eq.~\ref{eqn:comp:congruence}. The disjoint images condition, Eq.~\ref{eqn:comp:disjoint}, ensures that the complement generates distinct values from the original layout. Note that the finite domain $\Z(\v{L})$ of the original layout is used, while the infinite extended domain $\Z^{\v{L}^*}$ of the complement is used. This provides that whether the complement shape is truncated or extended (e.g. through composition with a layout on the right), it will always generate distinct values. Finally, the ordered image condition, Eq.~\ref{eqn:comp:ordered}, establishes uniqueness of the complement layout.

Examples of complements in Table~\ref{tab:complements}.

\begin{table}[ht]
\begin{center}
\begin{tabular}{|>{$}c<{$}|>{$}c<{$}|c|}
\v{L} & \v{L}^* & Comment \\ \hline \hline
(4,8):(1,4) & 1:32 & Note that $\forall i \in \Z^+, \v{L}^*(i) > \max_{j \in \Z_\v{L}} \v{L}(j)$. \\ \hline
(4,8):(8,1) & 1:32 & Order of input doesn't matter. \\ \hline
(4,(4,2)):(4,(1,16)) & 1:32 & Hierarchy of input doesn't matter. \\ \hline
(4,8):(1,5) & 1:40 \\ \hline
(4,8):(1,8) & (2,1):(4,64) & Still disjoint, but ``filling holes" in the codomain. \\ \hline
((2,2),(2,4)):((0,1),(0,2)) & 1:8 & Stride-0 modes do not contribute. \\ \hline
((2,2),(2,4)):((0,2),(0,4)) & (2,1):(1,16) \\ \hline
(4,8):(e_0,e_1) & (1,1):(4e_0,8e_1) & Result domain congruent with codomain $\Z^{(\ast,\ast)}$. \\ \hline
(4,(4,2)):(e_1,(e_0,12e_1)) & (1,(3,1)):(4e_0,(4e_1,24e_1)) & Weak congruence and disjoint and ordered. \\ \hline
\end{tabular}
\end{center}
\caption{Examples of layout complements.}
\label{tab:complements}
\end{table}

\subsubsection{Application: Logical Product}
\label{sec:logical_product}

The logical product of two layouts $\v{A}$ and $\v{B}$ is a layout $\v{R}$ where ``each element of layout $\v{B}$ has been replaced with a uniquely shifted version of the layout $\v{A}$.''

Concretely, we define the logical product as a function of two layouts to produce a rank-2 layout
\begin{align*}
\v{A} \otimes \v{B} = (\v{A}, \v{A}^* \circ \v{B})
\end{align*}
where $\v{A}^*$ is the complement of $\v{A}$. Note that the first mode of the result is exactly the input layout $\v{A}$ and the second mode is compatible with the input layout $\v{B}$ in shape and order of elements. The first mode, the $\v{A}$ layout, is typically referred to as the ``tile" which is repeated across the ``grid" layout $\v{B}$. 

Consider the example of a row-major layout $(\textcolor{red}{3},\textcolor{red}{4}):(\textcolor{red}{4},\textcolor{red}{1})$ tile to be reproduced in a col-major layout $(\textcolor{blue}{2},\textcolor{blue}{5}):(\textcolor{blue}{1},\textcolor{blue}{2})$ grid:
\begin{align*}
\begin{array}{cc} (\textcolor{red}{3}, & \textcolor{red}{4}) \\ (\textcolor{red}{4}, & \textcolor{red}{1}) \end{array}
\ \otimes \
\begin{array}{cc} (\textcolor{blue}{2}, & \textcolor{blue}{5}) \\ (\textcolor{blue}{1}, & \textcolor{blue}{2}) \end{array}
\ =\
\left(
\begin{array}{cc} (\textcolor{red}{3}, & \textcolor{red}{4}) \\ (\textcolor{red}{4}, & \textcolor{red}{1}) \end{array}, \
\begin{array}{c} 1 \\ 12 \end{array}
\ \circ \
\begin{array}{cc}(\textcolor{blue}{2}, & \textcolor{blue}{5}) \\ (\textcolor{blue}{1},& \textcolor{blue}{2}) \end{array}
\right)
\ = \
\begin{array}{cccc}((\textcolor{red}{3}, & \textcolor{red}{4}), & (\textcolor{blue}{2}, & \textcolor{blue}{5})) \\ ((\textcolor{red}{4}, & \textcolor{red}{1}), & (\textcolor{blue}{12}, & \textcolor{blue}{24})) \end{array}
\end{align*}
where the layout $1:12$ is the complement of $(\textcolor{red}{3},\textcolor{red}{4}):(\textcolor{red}{4},\textcolor{red}{1})$. In the first mode of the result, the original tile is available and the second mode of the result repeats the original tile uniquely 10 times in a $2\times 5$ col-major order.

As another example, consider layout $(\textcolor{red}{4},\textcolor{red}{8}):(\textcolor{red}{20},\textcolor{red}{2})$ tile to be reproduced in a row-major layout $(\textcolor{blue}{3},\textcolor{blue}{2}):(\textcolor{blue}{2},\textcolor{blue}{1})$ grid:
\begin{align*}
\begin{array}{cc} (\textcolor{red}{4}, & \textcolor{red}{8}) \\ (\textcolor{red}{20}, & \textcolor{red}{2}) \end{array}
\ \otimes \
\begin{array}{cc} (\textcolor{blue}{3}, & \textcolor{blue}{2}) \\ (\textcolor{blue}{2}, & \textcolor{blue}{1}) \end{array}
\ = \
\left(
\begin{array}{cc} (\textcolor{red}{4}, & \textcolor{red}{8}) \\ (\textcolor{red}{20}, & \textcolor{red}{2}) \end{array}, \
\begin{array}{cc} (2, & 1) \\ (1, & 80) \end{array}
\ \circ \
\begin{array}{cc} (\textcolor{blue}{3}, & \textcolor{blue}{2}) \\ (\textcolor{blue}{2}, & \textcolor{blue}{1}) \end{array}
\right)
\ = \
\begin{array}{cccc}((\textcolor{red}{4}, & \textcolor{red}{8}), & (\textcolor{blue}{3}, & \textcolor{blue}{2})) \\ ((\textcolor{red}{20}, & \textcolor{red}{2}), & (\textcolor{blue}{80}, & \textcolor{blue}{1})) \end{array}
\end{align*}
where the layout $(2,1):(1,80)$ is the complement of $(\textcolor{red}{4},\textcolor{red}{8}):(\textcolor{red}{20},\textcolor{red}{2})$. Note that the grid is now interleaving the tiles to produce the smallest codomain of the result. The complement layout tells the result how to stride the layout $\v{B}$ and layout $\v{B}$ is telling the result how to order and shape the complement layout.

\paragraph{Related Products}

The logical product produces a rank-2 layout, where the first mode is the original tile and the second mode is interpreted as the grid, or "tiling", of that tile.

While the logical product has no restrictions on the rank or compatibility of $\v{A}$ and $\v{B}$, the operation {\tt blocked\_product} is often a more intuitive version which uses the rank of the inputs to construct the shape of the result. We might expect the product of a $3 {\times} 4$ layout with a $2 {\times} 5$ layout to produce a $6 {\times} 20$ layout. The {\tt blocked\_product} requires the rank of $\v{A}$ and $\v{B}$ to be the same,
\begin{align*}
\text{rank}(\v{A}) = \text{rank}(\v{B})
\end{align*}
so that it can apply a zip operation to combine logical row-modes with row-modes and logical col-modes with col-modes. For instance, rearranging the modes of the example above
\begin{align*}
\begin{array}{cccc}((\textcolor{red}{3}, & \textcolor{red}{4}), & (\textcolor{blue}{2}, & \textcolor{blue}{5})) \\ ((\textcolor{red}{4}, & \textcolor{red}{1}), & (\textcolor{blue}{12}, & \textcolor{blue}{24})) \end{array}
\ \Longrightarrow \
\begin{array}{cccc}((\textcolor{red}{3}, & \textcolor{blue}{2}), & (\textcolor{red}{4}, & \textcolor{blue}{5})) \\ ((\textcolor{red}{4}, & \textcolor{blue}{12}), & (\textcolor{red}{1}, & \textcolor{blue}{24})) \end{array}
\end{align*}
results in a $6 {\times} 20$ layout. Figure~\ref{fig:blockedproduct} plots the resulting layout with each $3 {\times} 4$ tile shaded. This is considered ``blocked'' because each $3 {\times} 4$ block of the resulting layout are precisely shifted versions of our original $3 {\times} 4$ tile.

\begin{figure}[ht]
\centering
\scalebox{0.5}{
\begin{tikzpicture}[x={(0cm,-1cm)},y={(1cm,0cm)},every node/.style={minimum size=1cm, outer sep=0pt}]
\node[fill={rgb,255:red,255;green,255;blue,255}] at (0,0) {\large 0};
\node[fill={rgb,255:red,255;green,255;blue,255}] at (0,1) {\large 1};
\node[fill={rgb,255:red,255;green,255;blue,255}] at (0,2) {\large 2};
\node[fill={rgb,255:red,255;green,255;blue,255}] at (0,3) {\large 3};
\node[fill={rgb,255:red,210;green,210;blue,255}] at (0,4) {\large 24};
\node[fill={rgb,255:red,210;green,210;blue,255}] at (0,5) {\large 25};
\node[fill={rgb,255:red,210;green,210;blue,255}] at (0,6) {\large 26};
\node[fill={rgb,255:red,210;green,210;blue,255}] at (0,7) {\large 27};
\node[fill={rgb,255:red,210;green,255;blue,210}] at (0,8) {\large 48};
\node[fill={rgb,255:red,210;green,255;blue,210}] at (0,9) {\large 49};
\node[fill={rgb,255:red,210;green,255;blue,210}] at (0,10) {\large 50};
\node[fill={rgb,255:red,210;green,255;blue,210}] at (0,11) {\large 51};
\node[fill={rgb,255:red,255;green,255;blue,210}] at (0,12) {\large 72};
\node[fill={rgb,255:red,255;green,255;blue,210}] at (0,13) {\large 73};
\node[fill={rgb,255:red,255;green,255;blue,210}] at (0,14) {\large 74};
\node[fill={rgb,255:red,255;green,255;blue,210}] at (0,15) {\large 75};
\node[fill={rgb,255:red,255;green,210;blue,210}] at (0,16) {\large 96};
\node[fill={rgb,255:red,255;green,210;blue,210}] at (0,17) {\large 97};
\node[fill={rgb,255:red,255;green,210;blue,210}] at (0,18) {\large 98};
\node[fill={rgb,255:red,255;green,210;blue,210}] at (0,19) {\large 99};
\node[fill={rgb,255:red,255;green,255;blue,255}] at (1,0) {\large 4};
\node[fill={rgb,255:red,255;green,255;blue,255}] at (1,1) {\large 5};
\node[fill={rgb,255:red,255;green,255;blue,255}] at (1,2) {\large 6};
\node[fill={rgb,255:red,255;green,255;blue,255}] at (1,3) {\large 7};
\node[fill={rgb,255:red,210;green,210;blue,255}] at (1,4) {\large 28};
\node[fill={rgb,255:red,210;green,210;blue,255}] at (1,5) {\large 29};
\node[fill={rgb,255:red,210;green,210;blue,255}] at (1,6) {\large 30};
\node[fill={rgb,255:red,210;green,210;blue,255}] at (1,7) {\large 31};
\node[fill={rgb,255:red,210;green,255;blue,210}] at (1,8) {\large 52};
\node[fill={rgb,255:red,210;green,255;blue,210}] at (1,9) {\large 53};
\node[fill={rgb,255:red,210;green,255;blue,210}] at (1,10) {\large 54};
\node[fill={rgb,255:red,210;green,255;blue,210}] at (1,11) {\large 55};
\node[fill={rgb,255:red,255;green,255;blue,210}] at (1,12) {\large 76};
\node[fill={rgb,255:red,255;green,255;blue,210}] at (1,13) {\large 77};
\node[fill={rgb,255:red,255;green,255;blue,210}] at (1,14) {\large 78};
\node[fill={rgb,255:red,255;green,255;blue,210}] at (1,15) {\large 79};
\node[fill={rgb,255:red,255;green,210;blue,210}] at (1,16) {\large 100};
\node[fill={rgb,255:red,255;green,210;blue,210}] at (1,17) {\large 101};
\node[fill={rgb,255:red,255;green,210;blue,210}] at (1,18) {\large 102};
\node[fill={rgb,255:red,255;green,210;blue,210}] at (1,19) {\large 103};
\node[fill={rgb,255:red,255;green,255;blue,255}] at (2,0) {\large 8};
\node[fill={rgb,255:red,255;green,255;blue,255}] at (2,1) {\large 9};
\node[fill={rgb,255:red,255;green,255;blue,255}] at (2,2) {\large 10};
\node[fill={rgb,255:red,255;green,255;blue,255}] at (2,3) {\large 11};
\node[fill={rgb,255:red,210;green,210;blue,255}] at (2,4) {\large 32};
\node[fill={rgb,255:red,210;green,210;blue,255}] at (2,5) {\large 33};
\node[fill={rgb,255:red,210;green,210;blue,255}] at (2,6) {\large 34};
\node[fill={rgb,255:red,210;green,210;blue,255}] at (2,7) {\large 35};
\node[fill={rgb,255:red,210;green,255;blue,210}] at (2,8) {\large 56};
\node[fill={rgb,255:red,210;green,255;blue,210}] at (2,9) {\large 57};
\node[fill={rgb,255:red,210;green,255;blue,210}] at (2,10) {\large 58};
\node[fill={rgb,255:red,210;green,255;blue,210}] at (2,11) {\large 59};
\node[fill={rgb,255:red,255;green,255;blue,210}] at (2,12) {\large 80};
\node[fill={rgb,255:red,255;green,255;blue,210}] at (2,13) {\large 81};
\node[fill={rgb,255:red,255;green,255;blue,210}] at (2,14) {\large 82};
\node[fill={rgb,255:red,255;green,255;blue,210}] at (2,15) {\large 83};
\node[fill={rgb,255:red,255;green,210;blue,210}] at (2,16) {\large 104};
\node[fill={rgb,255:red,255;green,210;blue,210}] at (2,17) {\large 105};
\node[fill={rgb,255:red,255;green,210;blue,210}] at (2,18) {\large 106};
\node[fill={rgb,255:red,255;green,210;blue,210}] at (2,19) {\large 107};
\node[fill={rgb,255:red,175;green,175;blue,175}] at (3,0) {\large 12};
\node[fill={rgb,255:red,175;green,175;blue,175}] at (3,1) {\large 13};
\node[fill={rgb,255:red,175;green,175;blue,175}] at (3,2) {\large 14};
\node[fill={rgb,255:red,175;green,175;blue,175}] at (3,3) {\large 15};
\node[fill={rgb,255:red,175;green,175;blue,255}] at (3,4) {\large 36};
\node[fill={rgb,255:red,175;green,175;blue,255}] at (3,5) {\large 37};
\node[fill={rgb,255:red,175;green,175;blue,255}] at (3,6) {\large 38};
\node[fill={rgb,255:red,175;green,175;blue,255}] at (3,7) {\large 39};
\node[fill={rgb,255:red,175;green,255;blue,175}] at (3,8) {\large 60};
\node[fill={rgb,255:red,175;green,255;blue,175}] at (3,9) {\large 61};
\node[fill={rgb,255:red,175;green,255;blue,175}] at (3,10) {\large 62};
\node[fill={rgb,255:red,175;green,255;blue,175}] at (3,11) {\large 63};
\node[fill={rgb,255:red,255;green,255;blue,175}] at (3,12) {\large 84};
\node[fill={rgb,255:red,255;green,255;blue,175}] at (3,13) {\large 85};
\node[fill={rgb,255:red,255;green,255;blue,175}] at (3,14) {\large 86};
\node[fill={rgb,255:red,255;green,255;blue,175}] at (3,15) {\large 87};
\node[fill={rgb,255:red,255;green,175;blue,175}] at (3,16) {\large 108};
\node[fill={rgb,255:red,255;green,175;blue,175}] at (3,17) {\large 109};
\node[fill={rgb,255:red,255;green,175;blue,175}] at (3,18) {\large 110};
\node[fill={rgb,255:red,255;green,175;blue,175}] at (3,19) {\large 111};
\node[fill={rgb,255:red,175;green,175;blue,175}] at (4,0) {\large 16};
\node[fill={rgb,255:red,175;green,175;blue,175}] at (4,1) {\large 17};
\node[fill={rgb,255:red,175;green,175;blue,175}] at (4,2) {\large 18};
\node[fill={rgb,255:red,175;green,175;blue,175}] at (4,3) {\large 19};
\node[fill={rgb,255:red,175;green,175;blue,255}] at (4,4) {\large 40};
\node[fill={rgb,255:red,175;green,175;blue,255}] at (4,5) {\large 41};
\node[fill={rgb,255:red,175;green,175;blue,255}] at (4,6) {\large 42};
\node[fill={rgb,255:red,175;green,175;blue,255}] at (4,7) {\large 43};
\node[fill={rgb,255:red,175;green,255;blue,175}] at (4,8) {\large 64};
\node[fill={rgb,255:red,175;green,255;blue,175}] at (4,9) {\large 65};
\node[fill={rgb,255:red,175;green,255;blue,175}] at (4,10) {\large 66};
\node[fill={rgb,255:red,175;green,255;blue,175}] at (4,11) {\large 67};
\node[fill={rgb,255:red,255;green,255;blue,175}] at (4,12) {\large 88};
\node[fill={rgb,255:red,255;green,255;blue,175}] at (4,13) {\large 89};
\node[fill={rgb,255:red,255;green,255;blue,175}] at (4,14) {\large 90};
\node[fill={rgb,255:red,255;green,255;blue,175}] at (4,15) {\large 91};
\node[fill={rgb,255:red,255;green,175;blue,175}] at (4,16) {\large 112};
\node[fill={rgb,255:red,255;green,175;blue,175}] at (4,17) {\large 113};
\node[fill={rgb,255:red,255;green,175;blue,175}] at (4,18) {\large 114};
\node[fill={rgb,255:red,255;green,175;blue,175}] at (4,19) {\large 115};
\node[fill={rgb,255:red,175;green,175;blue,175}] at (5,0) {\large 20};
\node[fill={rgb,255:red,175;green,175;blue,175}] at (5,1) {\large 21};
\node[fill={rgb,255:red,175;green,175;blue,175}] at (5,2) {\large 22};
\node[fill={rgb,255:red,175;green,175;blue,175}] at (5,3) {\large 23};
\node[fill={rgb,255:red,175;green,175;blue,255}] at (5,4) {\large 44};
\node[fill={rgb,255:red,175;green,175;blue,255}] at (5,5) {\large 45};
\node[fill={rgb,255:red,175;green,175;blue,255}] at (5,6) {\large 46};
\node[fill={rgb,255:red,175;green,175;blue,255}] at (5,7) {\large 47};
\node[fill={rgb,255:red,175;green,255;blue,175}] at (5,8) {\large 68};
\node[fill={rgb,255:red,175;green,255;blue,175}] at (5,9) {\large 69};
\node[fill={rgb,255:red,175;green,255;blue,175}] at (5,10) {\large 70};
\node[fill={rgb,255:red,175;green,255;blue,175}] at (5,11) {\large 71};
\node[fill={rgb,255:red,255;green,255;blue,175}] at (5,12) {\large 92};
\node[fill={rgb,255:red,255;green,255;blue,175}] at (5,13) {\large 93};
\node[fill={rgb,255:red,255;green,255;blue,175}] at (5,14) {\large 94};
\node[fill={rgb,255:red,255;green,255;blue,175}] at (5,15) {\large 95};
\node[fill={rgb,255:red,255;green,175;blue,175}] at (5,16) {\large 116};
\node[fill={rgb,255:red,255;green,175;blue,175}] at (5,17) {\large 117};
\node[fill={rgb,255:red,255;green,175;blue,175}] at (5,18) {\large 118};
\node[fill={rgb,255:red,255;green,175;blue,175}] at (5,19) {\large 119};
\draw[color=black,thick,shift={(-0.5,-0.5)}] (0,0) grid (6,20);

\node at (0,-1) {\Large{\texttt{0}}};
\node at (1,-1) {\Large{\texttt{1}}};
\node at (2,-1) {\Large{\texttt{2}}};
\node at (3,-1) {\Large{\texttt{3}}};
\node at (4,-1) {\Large{\texttt{4}}};
\node at (5,-1) {\Large{\texttt{5}}};
\node at (-1,0) {\Large{\texttt{0}}};
\node at (-1,1) {\Large{\texttt{1}}};
\node at (-1,2) {\Large{\texttt{2}}};
\node at (-1,3) {\Large{\texttt{3}}};
\node at (-1,4) {\Large{\texttt{4}}};
\node at (-1,5) {\Large{\texttt{5}}};
\node at (-1,6) {\Large{\texttt{6}}};
\node at (-1,7) {\Large{\texttt{7}}};
\node at (-1,8) {\Large{\texttt{8}}};
\node at (-1,9) {\Large{\texttt{9}}};
\node at (-1,10) {\Large{\texttt{10}}};
\node at (-1,11) {\Large{\texttt{11}}};
\node at (-1,12) {\Large{\texttt{12}}};
\node at (-1,13) {\Large{\texttt{13}}};
\node at (-1,14) {\Large{\texttt{14}}};
\node at (-1,15) {\Large{\texttt{15}}};
\node at (-1,16) {\Large{\texttt{16}}};
\node at (-1,17) {\Large{\texttt{17}}};
\node at (-1,18) {\Large{\texttt{18}}};
\node at (-1,19) {\Large{\texttt{19}}};
\end{tikzpicture}
}
\caption{The {\tt blocked\_product} of row-major $(3,4):(4,1)$ tile with col-major $(2,5):(1,2)$ tiling.}
\label{fig:blockedproduct}
\end{figure}

The operation {\tt raked\_product} is similar to the blocked product except the modes are zipped in reverse:
\begin{align*}
\begin{array}{cccc}((\textcolor{red}{3}, & \textcolor{red}{4}), & (\textcolor{blue}{2}, & \textcolor{blue}{5})) \\ ((\textcolor{red}{4}, & \textcolor{red}{1}), & (\textcolor{blue}{12}, & \textcolor{blue}{24})) \end{array}
\ \Longrightarrow \
\begin{array}{cccc}((\textcolor{blue}{2}, & \textcolor{red}{3}), & (\textcolor{blue}{5}, & \textcolor{red}{4})) \\ ((\textcolor{blue}{12}, & \textcolor{red}{4}), & (\textcolor{blue}{24}, & \textcolor{red}{1})) \end{array}
\end{align*}
which results in the $3 {\times} 4$ tile being ``raked'' over the $2 {\times} 5$ tile rather than ``blocked''. Figure~\ref{fig:rakedproduct} plots the resulting layout with each $3 {\times} 4$ tile shaded.

\begin{figure}[ht]
\centering
\scalebox{0.5}{
\begin{tikzpicture}[x={(0cm,-1cm)},y={(1cm,0cm)},every node/.style={minimum size=1cm, outer sep=0pt}]

\node[fill={rgb,255:red,255;green,255;blue,255}] at (0,0) {\large 0};
\node[fill={rgb,255:red,210;green,210;blue,255}] at (0,1) {\large 24};
\node[fill={rgb,255:red,210;green,255;blue,210}] at (0,2) {\large 48};
\node[fill={rgb,255:red,255;green,255;blue,210}] at (0,3) {\large 72};
\node[fill={rgb,255:red,255;green,210;blue,210}] at (0,4) {\large 96};
\node[fill={rgb,255:red,255;green,255;blue,255}] at (0,5) {\large 1};
\node[fill={rgb,255:red,210;green,210;blue,255}] at (0,6) {\large 25};
\node[fill={rgb,255:red,210;green,255;blue,210}] at (0,7) {\large 49};
\node[fill={rgb,255:red,255;green,255;blue,210}] at (0,8) {\large 73};
\node[fill={rgb,255:red,255;green,210;blue,210}] at (0,9) {\large 97};
\node[fill={rgb,255:red,255;green,255;blue,255}] at (0,10) {\large 2};
\node[fill={rgb,255:red,210;green,210;blue,255}] at (0,11) {\large 26};
\node[fill={rgb,255:red,210;green,255;blue,210}] at (0,12) {\large 50};
\node[fill={rgb,255:red,255;green,255;blue,210}] at (0,13) {\large 74};
\node[fill={rgb,255:red,255;green,210;blue,210}] at (0,14) {\large 98};
\node[fill={rgb,255:red,255;green,255;blue,255}] at (0,15) {\large 3};
\node[fill={rgb,255:red,210;green,210;blue,255}] at (0,16) {\large 27};
\node[fill={rgb,255:red,210;green,255;blue,210}] at (0,17) {\large 51};
\node[fill={rgb,255:red,255;green,255;blue,210}] at (0,18) {\large 75};
\node[fill={rgb,255:red,255;green,210;blue,210}] at (0,19) {\large 99};
\node[fill={rgb,255:red,175;green,175;blue,175}] at (1,0) {\large 12};
\node[fill={rgb,255:red,175;green,175;blue,255}] at (1,1) {\large 36};
\node[fill={rgb,255:red,175;green,255;blue,175}] at (1,2) {\large 60};
\node[fill={rgb,255:red,255;green,255;blue,175}] at (1,3) {\large 84};
\node[fill={rgb,255:red,255;green,175;blue,175}] at (1,4) {\large 108};
\node[fill={rgb,255:red,175;green,175;blue,175}] at (1,5) {\large 13};
\node[fill={rgb,255:red,175;green,175;blue,255}] at (1,6) {\large 37};
\node[fill={rgb,255:red,175;green,255;blue,175}] at (1,7) {\large 61};
\node[fill={rgb,255:red,255;green,255;blue,175}] at (1,8) {\large 85};
\node[fill={rgb,255:red,255;green,175;blue,175}] at (1,9) {\large 109};
\node[fill={rgb,255:red,175;green,175;blue,175}] at (1,10) {\large 14};
\node[fill={rgb,255:red,175;green,175;blue,255}] at (1,11) {\large 38};
\node[fill={rgb,255:red,175;green,255;blue,175}] at (1,12) {\large 62};
\node[fill={rgb,255:red,255;green,255;blue,175}] at (1,13) {\large 86};
\node[fill={rgb,255:red,255;green,175;blue,175}] at (1,14) {\large 110};
\node[fill={rgb,255:red,175;green,175;blue,175}] at (1,15) {\large 15};
\node[fill={rgb,255:red,175;green,175;blue,255}] at (1,16) {\large 39};
\node[fill={rgb,255:red,175;green,255;blue,175}] at (1,17) {\large 63};
\node[fill={rgb,255:red,255;green,255;blue,175}] at (1,18) {\large 87};
\node[fill={rgb,255:red,255;green,175;blue,175}] at (1,19) {\large 111};
\node[fill={rgb,255:red,255;green,255;blue,255}] at (2,0) {\large 4};
\node[fill={rgb,255:red,210;green,210;blue,255}] at (2,1) {\large 28};
\node[fill={rgb,255:red,210;green,255;blue,210}] at (2,2) {\large 52};
\node[fill={rgb,255:red,255;green,255;blue,210}] at (2,3) {\large 76};
\node[fill={rgb,255:red,255;green,210;blue,210}] at (2,4) {\large 100};
\node[fill={rgb,255:red,255;green,255;blue,255}] at (2,5) {\large 5};
\node[fill={rgb,255:red,210;green,210;blue,255}] at (2,6) {\large 29};
\node[fill={rgb,255:red,210;green,255;blue,210}] at (2,7) {\large 53};
\node[fill={rgb,255:red,255;green,255;blue,210}] at (2,8) {\large 77};
\node[fill={rgb,255:red,255;green,210;blue,210}] at (2,9) {\large 101};
\node[fill={rgb,255:red,255;green,255;blue,255}] at (2,10) {\large 6};
\node[fill={rgb,255:red,210;green,210;blue,255}] at (2,11) {\large 30};
\node[fill={rgb,255:red,210;green,255;blue,210}] at (2,12) {\large 54};
\node[fill={rgb,255:red,255;green,255;blue,210}] at (2,13) {\large 78};
\node[fill={rgb,255:red,255;green,210;blue,210}] at (2,14) {\large 102};
\node[fill={rgb,255:red,255;green,255;blue,255}] at (2,15) {\large 7};
\node[fill={rgb,255:red,210;green,210;blue,255}] at (2,16) {\large 31};
\node[fill={rgb,255:red,210;green,255;blue,210}] at (2,17) {\large 55};
\node[fill={rgb,255:red,255;green,255;blue,210}] at (2,18) {\large 79};
\node[fill={rgb,255:red,255;green,210;blue,210}] at (2,19) {\large 103};
\node[fill={rgb,255:red,175;green,175;blue,175}] at (3,0) {\large 16};
\node[fill={rgb,255:red,175;green,175;blue,255}] at (3,1) {\large 40};
\node[fill={rgb,255:red,175;green,255;blue,175}] at (3,2) {\large 64};
\node[fill={rgb,255:red,255;green,255;blue,175}] at (3,3) {\large 88};
\node[fill={rgb,255:red,255;green,175;blue,175}] at (3,4) {\large 112};
\node[fill={rgb,255:red,175;green,175;blue,175}] at (3,5) {\large 17};
\node[fill={rgb,255:red,175;green,175;blue,255}] at (3,6) {\large 41};
\node[fill={rgb,255:red,175;green,255;blue,175}] at (3,7) {\large 65};
\node[fill={rgb,255:red,255;green,255;blue,175}] at (3,8) {\large 89};
\node[fill={rgb,255:red,255;green,175;blue,175}] at (3,9) {\large 113};
\node[fill={rgb,255:red,175;green,175;blue,175}] at (3,10) {\large 18};
\node[fill={rgb,255:red,175;green,175;blue,255}] at (3,11) {\large 42};
\node[fill={rgb,255:red,175;green,255;blue,175}] at (3,12) {\large 66};
\node[fill={rgb,255:red,255;green,255;blue,175}] at (3,13) {\large 90};
\node[fill={rgb,255:red,255;green,175;blue,175}] at (3,14) {\large 114};
\node[fill={rgb,255:red,175;green,175;blue,175}] at (3,15) {\large 19};
\node[fill={rgb,255:red,175;green,175;blue,255}] at (3,16) {\large 43};
\node[fill={rgb,255:red,175;green,255;blue,175}] at (3,17) {\large 67};
\node[fill={rgb,255:red,255;green,255;blue,175}] at (3,18) {\large 91};
\node[fill={rgb,255:red,255;green,175;blue,175}] at (3,19) {\large 115};
\node[fill={rgb,255:red,255;green,255;blue,255}] at (4,0) {\large 8};
\node[fill={rgb,255:red,210;green,210;blue,255}] at (4,1) {\large 32};
\node[fill={rgb,255:red,210;green,255;blue,210}] at (4,2) {\large 56};
\node[fill={rgb,255:red,255;green,255;blue,210}] at (4,3) {\large 80};
\node[fill={rgb,255:red,255;green,210;blue,210}] at (4,4) {\large 104};
\node[fill={rgb,255:red,255;green,255;blue,255}] at (4,5) {\large 9};
\node[fill={rgb,255:red,210;green,210;blue,255}] at (4,6) {\large 33};
\node[fill={rgb,255:red,210;green,255;blue,210}] at (4,7) {\large 57};
\node[fill={rgb,255:red,255;green,255;blue,210}] at (4,8) {\large 81};
\node[fill={rgb,255:red,255;green,210;blue,210}] at (4,9) {\large 105};
\node[fill={rgb,255:red,255;green,255;blue,255}] at (4,10) {\large 10};
\node[fill={rgb,255:red,210;green,210;blue,255}] at (4,11) {\large 34};
\node[fill={rgb,255:red,210;green,255;blue,210}] at (4,12) {\large 58};
\node[fill={rgb,255:red,255;green,255;blue,210}] at (4,13) {\large 82};
\node[fill={rgb,255:red,255;green,210;blue,210}] at (4,14) {\large 106};
\node[fill={rgb,255:red,255;green,255;blue,255}] at (4,15) {\large 11};
\node[fill={rgb,255:red,210;green,210;blue,255}] at (4,16) {\large 35};
\node[fill={rgb,255:red,210;green,255;blue,210}] at (4,17) {\large 59};
\node[fill={rgb,255:red,255;green,255;blue,210}] at (4,18) {\large 83};
\node[fill={rgb,255:red,255;green,210;blue,210}] at (4,19) {\large 107};
\node[fill={rgb,255:red,175;green,175;blue,175}] at (5,0) {\large 20};
\node[fill={rgb,255:red,175;green,175;blue,255}] at (5,1) {\large 44};
\node[fill={rgb,255:red,175;green,255;blue,175}] at (5,2) {\large 68};
\node[fill={rgb,255:red,255;green,255;blue,175}] at (5,3) {\large 92};
\node[fill={rgb,255:red,255;green,175;blue,175}] at (5,4) {\large 116};
\node[fill={rgb,255:red,175;green,175;blue,175}] at (5,5) {\large 21};
\node[fill={rgb,255:red,175;green,175;blue,255}] at (5,6) {\large 45};
\node[fill={rgb,255:red,175;green,255;blue,175}] at (5,7) {\large 69};
\node[fill={rgb,255:red,255;green,255;blue,175}] at (5,8) {\large 93};
\node[fill={rgb,255:red,255;green,175;blue,175}] at (5,9) {\large 117};
\node[fill={rgb,255:red,175;green,175;blue,175}] at (5,10) {\large 22};
\node[fill={rgb,255:red,175;green,175;blue,255}] at (5,11) {\large 46};
\node[fill={rgb,255:red,175;green,255;blue,175}] at (5,12) {\large 70};
\node[fill={rgb,255:red,255;green,255;blue,175}] at (5,13) {\large 94};
\node[fill={rgb,255:red,255;green,175;blue,175}] at (5,14) {\large 118};
\node[fill={rgb,255:red,175;green,175;blue,175}] at (5,15) {\large 23};
\node[fill={rgb,255:red,175;green,175;blue,255}] at (5,16) {\large 47};
\node[fill={rgb,255:red,175;green,255;blue,175}] at (5,17) {\large 71};
\node[fill={rgb,255:red,255;green,255;blue,175}] at (5,18) {\large 95};
\node[fill={rgb,255:red,255;green,175;blue,175}] at (5,19) {\large 119};
\draw[color=black,thick,shift={(-0.5,-0.5)}] (0,0) grid (6,20);

\node at (0,-1) {\Large{\texttt{0}}};
\node at (1,-1) {\Large{\texttt{1}}};
\node at (2,-1) {\Large{\texttt{2}}};
\node at (3,-1) {\Large{\texttt{3}}};
\node at (4,-1) {\Large{\texttt{4}}};
\node at (5,-1) {\Large{\texttt{5}}};
\node at (-1,0) {\Large{\texttt{0}}};
\node at (-1,1) {\Large{\texttt{1}}};
\node at (-1,2) {\Large{\texttt{2}}};
\node at (-1,3) {\Large{\texttt{3}}};
\node at (-1,4) {\Large{\texttt{4}}};
\node at (-1,5) {\Large{\texttt{5}}};
\node at (-1,6) {\Large{\texttt{6}}};
\node at (-1,7) {\Large{\texttt{7}}};
\node at (-1,8) {\Large{\texttt{8}}};
\node at (-1,9) {\Large{\texttt{9}}};
\node at (-1,10) {\Large{\texttt{10}}};
\node at (-1,11) {\Large{\texttt{11}}};
\node at (-1,12) {\Large{\texttt{12}}};
\node at (-1,13) {\Large{\texttt{13}}};
\node at (-1,14) {\Large{\texttt{14}}};
\node at (-1,15) {\Large{\texttt{15}}};
\node at (-1,16) {\Large{\texttt{16}}};
\node at (-1,17) {\Large{\texttt{17}}};
\node at (-1,18) {\Large{\texttt{18}}};
\node at (-1,19) {\Large{\texttt{19}}};
\end{tikzpicture}
}
\caption{The {\tt raked\_product} of row-major $(3,4):(4,1)$ tile with col-major $(2,5):(1,2)$ tiling.}
\label{fig:rakedproduct}
\end{figure}


Many other versions of logical product can be produced that additionally group and rearrange modes to produce a convenient result.

\subsubsection{Application: Logical Divide}
\label{sec:logical_divide}

The logical division of two layouts $\v{A}$ and $\v{B}$ is a layout $\v{R}$ where ``layout $\v{A}$ is split into two: the elements that are pointed to by $\v{B}$ and those elements that are not pointed to by $\v{B}$.''.

Concretely, we define logical division as a function of two layouts to produce a rank-2 layout
\begin{align*}
\v{A} \oslash \v{B} = \v{A} \circ (\v{B}, \v{B}^*_{\abs{\v{A}}}) = \v{A} \circ \v{B}^\bigstar
\end{align*}
where $\v{B}^*_{\abs{\v{A}}}$ is the complement of $\v{B}$ taken with respect to the size of layout $\v{A}$. Because this operation intends to preserve all elements of $\v{A}$ and simply split layout $\v{A}$ into two, we require that $\v{B}^\bigstar = (\v{B},\v{B}^*_{\abs{\v{A}}})$ is surjective onto $\Z_{\abs{A}}$. This often involves an extension of the size of $\v{B}^*$ so that $\abs{\v{B}^\bigstar} \geq \abs{\v{A}}$. Furthermore, we require the complement $\v{B}^*$ to ``complete'' the layout $\v{B}$ in the sense that $\v{B}^\bigstar = (\v{B},\v{B}^*_{\abs{A}})$ has a reflexive generalized inverse $\v{B}^+$ which satisfies:
\begin{align}
\v{B}^\bigstar \v{B}^+ \v{B}^\bigstar = \v{B}^\bigstar \\
\v{B}^+ \v{B}^\bigstar \v{B}^+ = \v{B}^+
\end{align}

In analogy with the logical product operation, the first mode of the result layout $\v{A} \circ \v{B}$ is typically referred to as the ``tile" which is repeated across the ``grid", or ``tiling", layout $\v{A} \circ \v{B}^*_{\abs{\v{A}}}$. The $\v{B}$ argument is referred to as the ``tiler", which can be extended beyond simply being a layout as in Section~\ref{sec:tiler}.

As an immediate example, it is common to ask for some set of logical elements from another layout. For example, to extract every third element from a layout we can use composition:
\begin{align*}
\begin{array}{c} \textcolor{red}{24} \\ \textcolor{red}{3} \end{array}
\ \circ \
\begin{array}{c} \textcolor{blue}{8} \\ \textcolor{blue}{3} \end{array}
\ &= \
\begin{array}{c} \textcolor{blue}{8} \\ \textcolor{blue}{9} \end{array} \\
\begin{array}{ccc} (\textcolor{red}{6}, & \textcolor{red}{2}, & \textcolor{red}{2}) \\ (\textcolor{red}{2}, & \textcolor{red}{1}, & \textcolor{red}{20}) \end{array}
\ \circ \
\begin{array}{c} \textcolor{blue}{8} \\ \textcolor{blue}{3} \end{array}
\ &= \
\begin{array}{ccc} (\textcolor{blue}{2}, & \textcolor{blue}{2}, & \textcolor{blue}{2}) \\ (\textcolor{blue}{6}, & \textcolor{blue}{1}, & \textcolor{blue}{20}) \end{array}
\end{align*}
In both cases above, a layout of 24 elements is composed with the layout $8:3$, which yields a layout of every third element. But what about the "rest" of the elements from the left-hand side layout? Logical divide splits the left-hand side layout into two pieces: the part that gets ``hit" by the right-hand side and the ``rest":
\begin{align*}
\begin{array}{c} \textcolor{red}{24} \\ \textcolor{red}{3} \end{array}
\ \oslash \
\begin{array}{c} \textcolor{blue}{8} \\ \textcolor{blue}{3} \end{array}
\ &=& \
\begin{array}{c} \textcolor{red}{24} \\ \textcolor{red}{3} \end{array}
\ \circ \
\begin{array}{cc} (\textcolor{blue}{8}, & 3) \\ (\textcolor{blue}{3}, & 1) \end{array}
\ &=& \
\begin{array}{cc} (\textcolor{blue}{8}, & 3) \\ (\textcolor{blue}{9}, & 3) \end{array} \\
\begin{array}{ccc} (\textcolor{red}{6}, & \textcolor{red}{2}, & \textcolor{red}{2}) \\ (\textcolor{red}{2}, & \textcolor{red}{1}, & \textcolor{red}{20}) \end{array}
\ \oslash \
\begin{array}{c} \textcolor{blue}{8} \\ \textcolor{blue}{3} \end{array}
\ &=& \
\begin{array}{ccc} (\textcolor{red}{6}, & \textcolor{red}{2}, & \textcolor{red}{2}) \\ (\textcolor{red}{2}, & \textcolor{red}{1}, & \textcolor{red}{20}) \end{array}
\ \circ \
\begin{array}{cc} (\textcolor{blue}{8}, & 3) \\ (\textcolor{blue}{3}, & 1) \end{array}
\ &=& \
\begin{array}{cccc} ((\textcolor{blue}{2}, & \textcolor{blue}{2}, & \textcolor{blue}{2}), & 3) \\ ((\textcolor{blue}{6}, & \textcolor{blue}{1}, & \textcolor{blue}{20}), & 2) \end{array}
\end{align*}
where $3:1$ is the complement of $8:3$ under the size of 24. The first mode of the result again contains every third element of the left-hand side layout as requested and the second mode of the result is the number of those ``tiles`` that are contained in the original layout (3 in this case) and the strides between those tiles. The complement is used to determine which elements are missing and in what order to keep them.

\paragraph{Related Divides}

Similar to composition, it is very common to want to apply logical divide by-mode. The same notation and strategy is used
\begin{align*}
\v{A} \oslash \langle \v{B},\v{C} \rangle = (\v{A}_0,\v{A}_1) \oslash \langle \v{B},\v{C} \rangle = (\v{A}_0 \oslash \v{B}, \v{A}_1 \oslash \v{C}).
\end{align*}
The intuition for this is to be able to perform these operations across columns and rows independently and then potentially rearrange those modes to construct new layouts, similar to the layout products. For example, returning to the tilers in Section~\ref{sec:tiler}, using the same tilers with logical divide rather than composition yields the $2 {\times} 2$ layout of the $4 {\times} 8$ sublayouts that the tilers point to. This is shown in Figure~\ref{fig:tilers_divide} where one could imagine a $2 {\times} 2$ layout which maps to each individually shaded tile of data.

\begin{figure}[ht]
\centering
\begin{subfigure}{0.22\textwidth}
\centering
\resizebox{\linewidth}{!}{
\begin{tikzpicture}[x={(0cm,-1cm)},y={(1cm,0cm)},every node/.style={minimum size=1cm, outer sep=0pt}]
\foreach \i in {0,1,2,3} {
\foreach \j in {0,1,2,3,4,5,6,7} {
\node[fill=black!80!white] at (\i,\j) {};
}}
\foreach \i in {4,5,6,7} {
\foreach \j in {0,1,2,3,4,5,6,7} {
\node[fill=black!60!white] at (\i,\j) {};
}}
\foreach \i in {0,1,2,3} {
\foreach \j in {8,9,10,11,12,13,14,15} {
\node[fill=black!30!white] at (\i,\j) {};
}}
\foreach \i in {4,5,6,7} {
\foreach \j in {8,9,10,11,12,13,14,15} {
\node[fill=black!10!white] at (\i,\j) {};
}}
\draw[color=black,thick,shift={(-0.5,-0.5)}] (0,0) grid (8,16);
\end{tikzpicture}}
\caption{$\langle 4,8 \rangle \equiv \langle 4:1,8:1 \rangle$}
\end{subfigure}\quad
\begin{subfigure}{0.22\textwidth}
\centering
\resizebox{\linewidth}{!}{
\begin{tikzpicture}[x={(0cm,-1cm)},y={(1cm,0cm)},every node/.style={minimum size=1cm, outer sep=0pt}]
\foreach \i in {0,1,4,5} {
\foreach \j in {0,1,2,3,4,5,6,7} {
\node[fill=black!80!white] at (\i,\j) {};
}}
\foreach \i in {2,3,6,7} {
\foreach \j in {0,1,2,3,4,5,6,7} {
\node[fill=black!60!white] at (\i,\j) {};
}}
\foreach \i in {0,1,4,5} {
\foreach \j in {8,9,10,11,12,13,14,15} {
\node[fill=black!30!white] at (\i,\j) {};
}}
\foreach \i in {2,3,6,7} {
\foreach \j in {8,9,10,11,12,13,14,15} {
\node[fill=black!10!white] at (\i,\j) {};
}}
\draw[color=black,thick,shift={(-0.5,-0.5)}] (0,0) grid (8,16);
\end{tikzpicture}}
\caption{$\langle (2,2):(1,4),8:1 \rangle$}
\end{subfigure}\quad
\begin{subfigure}{0.22\textwidth}
\centering
\resizebox{\linewidth}{!}{
\begin{tikzpicture}[x={(0cm,-1cm)},y={(1cm,0cm)},every node/.style={minimum size=1cm, outer sep=0pt}]
\foreach \i in {0,1,4,5} {
\foreach \j in {0,2,4,6,8,10,12,14} {
\node[fill=black!80!white] at (\i,\j) {};
}}
\foreach \i in {2,3,6,7} {
\foreach \j in {0,2,4,6,8,10,12,14} {
\node[fill=black!60!white] at (\i,\j) {};
}}
\foreach \i in {0,1,4,5} {
\foreach \j in {1,3,5,7,9,11,13,15} {
\node[fill=black!30!white] at (\i,\j) {};
}}
\foreach \i in {2,3,6,7} {
\foreach \j in {1,3,5,7,9,11,13,15} {
\node[fill=black!10!white] at (\i,\j) {};
}}
\draw[color=black,thick,shift={(-0.5,-0.5)}] (0,0) grid (8,16);
\end{tikzpicture}}
\caption{$\langle (2,2):(1,4),8:2 \rangle$}
\end{subfigure}\quad
\begin{subfigure}{0.22\textwidth}
\centering
\resizebox{\linewidth}{!}{
\begin{tikzpicture}[x={(0cm,-1cm)},y={(1cm,0cm)},every node/.style={minimum size=1cm, outer sep=0pt}]
\foreach \i in {0,2,4,6} {
\foreach \j in {0,2,4,6,8,10,12,14} {
\node[fill=black!80!white] at (\i,\j) {};
}}
\foreach \i in {1,3,5,7} {
\foreach \j in {0,2,4,6,8,10,12,14} {
\node[fill=black!60!white] at (\i,\j) {};
}}
\foreach \i in {0,2,4,6} {
\foreach \j in {1,3,5,7,9,11,13,15} {
\node[fill=black!30!white] at (\i,\j) {};
}}
\foreach \i in {1,3,5,7} {
\foreach \j in {1,3,5,7,9,11,13,15} {
\node[fill=black!10!white] at (\i,\j) {};
}}
\draw[color=black,thick,shift={(-0.5,-0.5)}] (0,0) grid (8,16);
\end{tikzpicture}}
\caption{$\langle 4:2,8:2 \rangle$}
\end{subfigure}
\caption{Examples of tilers that split an $8 {\times} 16$ layout into $4 {\times} 8$ tiles with a $2 {\times} 2$ tiling.}
\label{fig:tilers_divide}
\end{figure}

For instance, given a layout $(\textcolor{red}{8},\textcolor{red}{16}):(\textcolor{red}{20},\textcolor{red}{1})$, we would like to extract a ``tile" where we want 4 consecutive elements down each column and every other element across each row. The by-mode logical divide produces:
\begin{align*}
\begin{array}{cc} (\textcolor{red}{8}, & \textcolor{red}{16})\\
                  (\textcolor{red}{20},& \textcolor{red}{1}) \end{array}
\ \oslash \
\left\langle
\begin{array}{c} \textcolor{blue}{4} \\ \textcolor{blue}{1} \end{array},
\begin{array}{c} \textcolor{blue}{8} \\ \textcolor{blue}{2} \end{array}
\right\rangle
\ = \
\begin{array}{cc} (\textcolor{red}{8}, & \textcolor{red}{16}) \\ (\textcolor{red}{20}, & \textcolor{red}{1}) \end{array}
\ \circ \
\left\langle
\begin{array}{cc} (\textcolor{blue}{4}, & \textcolor{black}{2}) \\ (\textcolor{blue}{1}, & \textcolor{black}{4}) \end{array},
\begin{array}{cc} (\textcolor{blue}{8}, & \textcolor{black}{2}) \\ (\textcolor{blue}{2}, & \textcolor{black}{1}) \end{array}
\right\rangle
\ = \
\begin{array}{cccc}((\textcolor{blue}{4}, & \textcolor{black}{2}), & (\textcolor{blue}{8}, & \textcolor{black}{2})) \\ ((\textcolor{blue}{20}, & \textcolor{black}{80}), & (\textcolor{blue}{2}, & \textcolor{black}{1})) \end{array}
\end{align*}
where the ``tile" modes are colored blue and the ``rest" modes are colored black. Note that the resulting layout remains $8 {\times} 16$ and the elements pointed to by the tiler have been permuted to first $4 {\times} 8$block.

The operation {\tt zipped\_divide} performs by-mode logical divide then zips together like-modes:
\begin{align*}
\begin{array}{cccc}((\textcolor{blue}{4}, & \textcolor{blue}{8}), & (\textcolor{black}{2}, & \textcolor{black}{2})) \\ ((\textcolor{blue}{20}, & \textcolor{blue}{2}), & (\textcolor{black}{80}, & \textcolor{black}{1})) \end{array}
\end{align*}
which creates a ``tile" mode (blue) that is precisely the sublayout pointed to by the tiler and a ``rest" mode (black) that is the sublayout that iterates over tiles. The operation {\tt zipped\_divide} will always return a rank-2 layout result where the first mode is precisely the composition with the tiler and the second is the composition with the complement.

A very common pattern in software is tiling a data layout into a grid of tiles and then selecting a particular tile for each processing element. In the Python reference implementation (PyCuTe)~\cite{NVlabs:PyCuTe}, this looks like
\begin{python}
data    = Tensor(MyPtr, MyLayoutMxN)   # Tensor:                Crd -> Offset
tiler   = (4,8)                        # Tiler:           TileCoord -> Crd
data_vt = zipped_divide(data, tiler)   # Compose: (TileCrd,GridCrd) -> Offset
tile    = data_vt[:, blk_id]           # Slice for a 1D block id
tile    = data_vt[:, (blk_x, blk_y)]   # Slice for a 2D block id, equivalently
\end{python}
As is evident, this pattern is very similar to the thread-value partitioning of Section~\ref{sec:layouttv}, but only the tile-mode is actually specified and the missing ``Grid" mode is computed as the complement of the tiler. This ``completed'' tiler with shape $(\text{Tile}, \text{Grid})$ is then composed with the data to produce the grid of tiles, which can be sliced in the second mode by a block identifier to extract a particular tile for a processing element.

Many other versions of logical divide can be produced that additionally group and rearrange modes to produce a convenient result.

\section{Conclusion}

This paper has presented \CuTe, a mathematical framework for representing and manipulating tensor layouts that addresses the increasing complexity of modern GPU architectures. Through its hierarchical layout representation, \CuTe provides a generic approach to writing and managing the intricate data layouts and partitioning patterns required by specialized tensor instructions on contemporary hardware. Through its rich algebraic operations, \CuTe provides a principled approach to manipulating layouts and generating new layouts in the development of high-performance linear algebra kernels. Together, the representation and algebra enable support for complex layouts and partitioning, separation of concerns, and static analysis and optimization.

The expressiveness of \CuTe's layout representation has proven instrumental in several key applications. \CuTe facilitates representing diverse algorithms such as general tensor-tensor contractions (GETT) and convolutions (CONV) as instances of general matrix multiplication (GEMM), promoting code reuse and algorithmic uniformity. For instance, \CuTe provides for matrices to be indexed using $(m,n)$ coordinates regardless of their concrete data layout and allows generic algorithms to remain orthogonal to specific layouts, while algorithms or instructions that require specific layouts can statically inspect, verify, and reason about their tensor arguments.

The \CuTe layout algebra enables the static analysis and transformation of tensor layouts to implement and validate generic algorithms. For instance, it enables the derivation of maximal vectorization opportunities for \texttt{COPY} operations, validation of input and output layouts for matrix multiply-accumulate (\texttt{MMA}) and \texttt{COPY} instructions, and derivation of optimal shared memory layouts complete with bank conflict-avoiding swizzling patterns. These capabilities transform layout management from an error-prone manual process into a systematic, verifiable methodology. To support study, experimentation, and verification of the definitions in this paper, a pure-Python reference implementation, PyCuTe~\cite{NVlabs:PyCuTe}, is publicly available at \url{https://github.com/NVlabs/CuTe}.

The practical impact of \CuTe is evidenced by its successful deployment in production systems, most notably as the foundation of NVIDIA's CUTLASS v3 and v4 libraries. Performance evaluations in CUTLASS and applications such as FlashAttention demonstrate that \CuTe's abstractions introduce no performance overhead while significantly accelerating software development. The framework's adoption has validated its dual promise: maintaining the performance characteristics of hand-optimized code while substantially reducing development complexity and time.

Looking forward, \CuTe's mathematical foundations and algebraic approach position it as a sustainable framework for adapting to future architectural innovations. By separating layout concerns from algorithmic logic and providing powerful compile-time reasoning capabilities, \CuTe establishes a methodology for tensor-centric programming that can accommodate the demands of a wide range of computing systems and applications.

\bibliographystyle{unsrturl} 
\bibliography{CuTeWhitepaper}

\end{document}